\documentclass[12pt]{MUNThesis}

\usepackage{IEEEtrantools}
\usepackage[left=3.3cm, right=2.3cm, top=3cm, bottom=3cm]{geometry}
\usepackage{graphicx}
\usepackage{subfigure}
\usepackage[strict]{chngpage}
\usepackage{url}
\usepackage{fancyhdr}
\usepackage{footmisc}
\usepackage{algorithm}
\usepackage{algpseudocode}
\usepackage{cite}
\usepackage{amsmath,amssymb}
\usepackage{hyperref}
\hypersetup{
    bookmarks=false,         
    unicode=false,          
    pdftoolbar=true,        
    pdfmenubar=true,        
    pdffitwindow=false,     
    pdftitle={},    
    pdfauthor={},     
    pdfsubject={},   
    pdfcreator={},   
    pdfproducer={},  
    pdfkeywords={} {} {}, 
    pdfnewwindow=true,      
    colorlinks=false,       
    linkcolor=red,          
    citecolor=green,        
    filecolor=magenta,      
    urlcolor=cyan           
}

\usepackage[duplicate,sectionbib]{chapterbib}

\usepackage{nomencl}
\title{Optimal Link Adaptation for Multicarrier Communication Systems}

\author{Ebrahim E. Bedeer-Mohamed}

\deg{Doctor of Philosophy}

\fac{Faculty of Engineering and Applied Science\\ Memorial University of Newfoundland}

\date{}

\copyrightyear{}

\begin{document}

\setcounter{secnumdepth}{2} \setcounter{tocdepth}{2}

\pagenumbering{roman} \setcounter{page}{1}

\newpage
\begin{center}
\vfill
{\Huge
Optimal Link Adaptation for Multicarrier Communication Systems
}
\\
\vspace{5cm}
{\Large by} \vspace{0.25cm}\\
{\LARGE \copyright Ebrahim Bedeer}
\vfill
\end{center}

\newpage
\doublespacing
\setlength{\topmargin}{-.5in}
\chapter*{Abstract}

\addcontentsline{toc}{chapter}{Abstract}
Link adaptation is the terminology used to describe techniques that improve multicarrier communication systems performance by dynamically adapting the transmission parameters, i.e., transmit power and number of bits per subcarrier, to the changing quality of the wireless link. The research literature has focused on single objective optimization techniques to optimize the multicarrier communication systems performance, e.g., maximizing the throughput/capacity or minimizing the transmit power subject to a set of constraints. In this dissertation, we adopt a novel optimization concept, namely multiobjective optimization, where our objective is to simultaneously optimize the conflicting and incommensurable throughput and power objectives.

More specifically, in Chapters \ref{ch:TVT_GA} and \ref{ch:WCL}, we propose novel algorithms that jointly maximize the multicarrier system throughput and minimize its total transmit power subject to quality-of-service, total transmit power, and maximum allocated bits per subcarrier constraints. The proposed algorithms require prior knowledge about the importance of the competing objective functions in terms of pre-determined weighting coefficients, or they can adapt the weighting coefficients during the solution process while meeting the constraints, in order to reduce the computational complexity. Simulation results show significant performance gains in terms of the achieved throughput and transmit power when compared to single optimization approaches, at the cost of no additional complexity.

Motivated by the obtained results, in Chapter \ref{ch:TW} the problem is extended to the cognitive radio environment where the multicarrier unlicensed/secondary user, with limited sensing capabilities, needs to satisfy additional constraints for the leaked interference to existing licensed/primary users. In Chapter \ref{ch:TVT_MOOP}, a multiobjective optimization problem is formulated to balance between the SU capacity and the leaked interference to existing primary users, where the effect of the imperfect channel-state-information on the links from the secondary user transmitter to the primary users receivers is considered. Simulation results show improvements of the energy efficiency of the secondary user when compared to its counterparts of the works in the literature, with reduced computational complexity.

In Chapter \ref{ch:TVT_EE} we investigate the optimal link adaptation problem to optimize the energy efficiency of secondary users while considering the effect of imperfect channel-state-information on the links between the secondary user transmitter and receiver pairs and the limited sensing capabilities of the secondary user. The proposed link adaptation algorithm guarantees minimum required rate for the secondary user and statistical interference constraints to the existing primary users.

Finally, conclusions and possible extensions to the optimal link adaptation problem is discussed in Chapter \ref{ch:conc}.

\clearpage
\thispagestyle{plain}
\par\vspace*{.35\textheight}{\centering \textit{
To my parents ... \\
To my two sisters ... \\
To my wife and my kids, Noor-Eddin, Omar, and Ali ...
}\par} 

\chapter*{Acknowledgements}
\addcontentsline{toc}{chapter}{Acknowledgments}
As the formal part of my education comes to an end, it is a great pleasure to acknowledge several individuals who have had contributed to who I am today.

I would like to offer my sincere thanks to my supervisor Dr. Octavia A. Dobre and my co-supervisor Dr. Mohamed H. Ahmed for their valuable guidance and advice. I would like to thank them for allowing me to change my research topic after one and a half year from starting my Ph.D. program and for the flexibility they gave me to choose the research topic presented in this dissertation. 
The financial support provided by my supervisors, the Faculty of Engineering and Applied Science, the School of Graduate Studies, Defence Research and Development Canada (DRDC), and Natural Science and Engineering Research Council of Canada (NSERC) is duly acknowledged.

A significant part of my education was in Egypt and my foundations were laid in schools at the city of Tanta.
I would like to thank my elementary school teachers, my math teachers at secondary and high schools, my B.Sc. project supervisor, and my M.Sc. thesis supervisor.

The final word of acknowledgement is reserved to my parents for their unconditional support, to my two sisters for their love, to my wife for her patience and unwavering believe in me, and to my kids for bringing joy and hope to my life. 


\renewcommand{\contentsname}{Table of Contents}
\tableofcontents{}
\addcontentsline{toc}{chapter}{Table of Contents}
\listoffigures{}
\addcontentsline{toc}{chapter}{List of Figures}
\clearpage
\addcontentsline{toc}{chapter}{List of Abbreviations}
\printnomenclature[4.0cm]
\nomenclature{CR}{Cognitive Radio}
\nomenclature{SU}{Secondary User}
\nomenclature{OFDM}{Orthogonal Frequency Division Multiplexing}
\nomenclature{BER}{Bit Error Rate}
\nomenclature{AWGN}{Additive White Gaussian Noise}
\nomenclature{QAM}{Quadrature Amplitude Modulation}
\nomenclature{KKT}{Karush-Kuhn-Tucker}
\nomenclature{MCM}{Multicarrier Modulation}
\nomenclature{PU}{Primary User}
\nomenclature{RM}{Rate Maximization}
\nomenclature{MM}{Margin Maximization}
\nomenclature{SNR}{Signal-to-Noise Ratio}
\nomenclature{CCI}{Co-Channel Interference}
\nomenclature{ACI}{Adjacent Channel Interference}
\nomenclature{CSI}{Channel-State-Information}
\nomenclature{MOOP}{Multiobjective Optimization}
\nomenclature{GA}{Genetic Algorithm}
\nomenclature{FM}{Fading Margin}
\nomenclature{QoS}{Quality-of-Service}
\nomenclature{EE}{Energy-Efficiency}
\nomenclature{LMMSE}{Linear Minimum Mean Square Error}



\doublespacing
\clearpage
\pagenumbering{arabic}

\chapter{Introduction and Overview}
\section{Background}

``See if you can hear anything, Mr. Kemp.'' Guglielmo Marconi asked his assistant at noon on Thursday December 12, 1901, heralding the success of the first transatlantic wireless communication at the Signal Hill in St. John's, Newfoundland. This was preceded by the first ever wireless transmission when Marconi was able to ring a wireless alarm across his room in the summer of 1894. Since that time, the realm of wireless communication is one of the fastest expanding in the world.

Unlike wired channels that are stationary and predictable, wireless channels are extremely random and the transmission path between the transmitter and the receiver can vary from simple line-of-sight to one that is severely obstructed by buildings, mountains, and foliage. Due to multiple reflections from these objects, the electromagnetic waves travel along different paths of varying lengths. The interaction between these waves causes multipath fading which yields drastic problems for single carrier communication systems \cite{rappaport1996wireless}.


Multicarrier communication systems (MCM) provide numerous advantages over single carrier systems due to their ability to cope with the severe idiosyncrasies of the wireless channel \cite{wang2000wireless, fazel2008multi}. For example, frequency-selective fading that occurs due to multipath propagation, is a major performance-limiting challenge for single carrier communication systems. It arises when the channel coherence bandwidth is smaller than the signal bandwidth; therefore, different frequency components of the signal experience independent fading, and, thus, the received signal spectrum is distorted.
MCM systems overcome this problem by dividing the wideband signal into a number of narrowband subcarriers of equal bandwidth. Each subcarrier bandwidth is smaller than the channel coherence bandwidth; hence, it experiences frequency-flat fading and avoids the need of complex equalizers.

Due to their efficient digital implementations, MCM spurred widespread interest in various single user and multiple access communication standards. MCM systems, such as Discrete Multi-tone (DMT) has been applied to high speed asynchronous digital subscriber line (ADSL) modems, while orthogonal frequency division multiplexing (OFDM), its wireless counterpart, has been adopted in various wireless standards, such as digital cable television systems, IEEE 802.11 wireless LAN standard, IEEE 802.15 personal area network standard, IEEE 802.16 WiMAX standard, IEEE 802.20 mobile broadband wireless access standard, and the downlink of the 3GPP LTE and LTE-A fourth generation mobile broadband standard \cite{fu2010multicarrier}. Recently, it is also considered as the physical layer modulation of interest for CR systems
due to its flexibility in adjusting its transmission parameters to meet surrounding environment constraints, adaptivity in allocating vacant radio resources, and underlying sensing and spectrum shaping capabilities \cite{mahmoud2009ofdm}.


In wireless communication, the radio spectrum is the most scarce resource due to the ceaselessly demands of spectrum by new applications and services. However, this spectrum scarcity happens while most of the allocated spectrum is underutilized as reported by many jurisdictions \cite{fcc2002spectrum}. This paradox occurs only due to the inefficiency of traditional static spectrum allocation policies.
CR \cite{Mitola1999} provides a solution to the spectrum utilization inefficiency problem by allowing unlicensed/secondary users (SUs) to \textit{underlay}, \textit{overlay}, or \textit{interweave} their signals with licensed/primary users (PUs) \cite{kolodzy2005cognitive, srinivasa2007cognitive, goldsmith2009breaking}.
The underlay approach allows concurrent transmission of PUs and SUs as in ultra-wide band systems. SUs spread their transmission over a wide bandwidth; hence, their interference is below an acceptable noise floor to PUs. The overlay approach also allows concurrent transmission of PUs and SUs with a premise that SUs use part of their power to assist/relay PUs transmission. The interweave approach allows SUs to opportunistically access voids in PUs frequency bands/time slots under the condition that no harmful interference occurs to PUs. In our research, we consider the interweave CR systems.

In the past few years, the concept of energy aware communications has spurred the interest of the research community due to various environmental and economical reasons \cite{bolla2011energy}. It becomes indispensable for wireless communication systems to shift their resource allocation problems from optimizing traditional metrics such as throughput and transmit power to environmental-friendly energy metric.
Considering an adequate energy efficiency metric---that considers the transmit power consumption, circuitry power, and signaling overhead---is of momentous importance such that optimal resource allocations in cognitive radio systems reduce the energy consumption of SUs.

\section{Link Adaptation Algorithms}

In conventional MCM systems, all subcarriers employ the same signal constellation and transmit power; hence, the overall performance is dominated by subcarriers with worst channel conditions (i.e., deep fade). The performance of MCM systems can be significantly improved by dynamically loading/allocating different bits and/or powers per each subcarrier according to the channel quality or the wireless standard specifications \cite{hughes1988ensemble, chow1995practical, campello1999practical, papandreou2005new, liu2009adaptive, leke1997maximum, wyglinski2005bit, krongold2000computationally, sonalkar2000efficient, bedeer2011partial, bedeer2012jointVTC, bedeer2012EBERGC, bedeer2012UBERGC, bedeer2012novelICC, bedeer2012adaptiveRWS, bedeer2013resource, bedeer2013adaptive, bedeer2013novel, bedeer2014rateCONF, bedeer2014multiobjective, bedeer2013joint, 
bedeer2014energy, bedeer2015systematic, bedeer2015rate}. Broadly speaking, the link adaptation problem for MCM systems in non-CR environment focuses on optimizing the transmit power and constellation size. Consequently, the optimal link adaptation can be categorized into two main classes: \textit{rate maximization} (RM) and \textit{margin maximization} (MM) \cite{hughes1988ensemble, chow1995practical, campello1999practical, papandreou2005new, liu2009adaptive, leke1997maximum, wyglinski2005bit, krongold2000computationally, sonalkar2000efficient}. For the former, the objective is to maximize the achievable data rate \cite{leke1997maximum, wyglinski2005bit, krongold2000computationally, sonalkar2000efficient}, while for the latter the objective is to maximize the achievable system margin \cite{hughes1988ensemble, chow1995practical, campello1999practical, papandreou2005new, liu2009adaptive} (i.e., minimizing the total transmit power given a target data rate or a target bit error rate (BER)).


Most of the algorithms for loading bits and power are variant of two main types: greedy algorithms \cite{hughes1988ensemble, chow1995practical, campello1999practical, papandreou2005new, wyglinski2005bit, sonalkar2000efficient, song2002joint} and water-filling based algorithms \cite{liu2009adaptive, leke1997maximum, krongold2000computationally, goldfeld2002minimum}. Greedy algorithms seek to find the global optimum by repeatedly determining the local optimum at each stage \cite{mehlhorn2008algorithms}, i.e., decisions at each stage are based on local conditions only with no considerations of any future states. Hence, greedy algorithms are not guaranteed to find global optimum. In MCM systems, greedy algorithms incrementally load an integer number of bits to subcarriers, initially nulled, that requires the least amount of transmit power until the power constraint or the average BER is reached; or unload an integer number of bits from subcarriers, initially loaded with the maximum allowed constellation size, until the average BER is reached. Greedy algorithms in MCM systems provide near optimal allocation at the cost of high complexity. On the other hand, water-filling based algorithms formulate the loading problem as a constrained optimization problem that can be solved by classical optimization methods \cite{cover2004elements}. The water-filling based algorithms maximize the capacity on all subcarriers by loading power on each subcarrier in proportional to the subcarrier channel gain, while the total transmit power is kept within a fixed constraint. The capacity on each subcarrier is related to its power through the Shannon's capacity formula. Typically, water-filling based algorithms allocate a non-integer number of bits per each subcarrier; hence, it is generally followed by a rounding-off step to allocate an integer number of bits to the transmitted symbols across all subcarriers, which compromises performance for lower complexity.


The foregoing research in \cite{hughes1988ensemble, chow1995practical, campello1999practical, papandreou2005new, liu2009adaptive} pored on maximizing the margin. In \cite{hughes1988ensemble}, Hughes-Hartog proposed a greedy algorithm to maximize the margin by successively allocating bits to subcarriers requiring the minimum incremental power until the total target date rate is reached. The algorithm converges very slowly as it requires extensive sorting; hence, it is very complex and not suitable for practical implementations. Chow \textit{et al.} \cite{chow1995practical} proposed a practical iterative bit loading algorithm to maximize the margin that offers significant implementation advantages over Hughes-Hartog algorithm. The algorithm uses the channel capacity approximation to compute the initial bit allocation across all subcarriers assuming uniform power loading. Then, it iteratively changes the allocated bits to achieve the optimal margin and the target data rate.
Papandreou and Antonakopoulos \cite{papandreou2005new} presented an efficient bit loading algorithm to minimize the transmit power that achieves the same bit allocation as the discrete optimal bit-filling and bit-removal techniques\footnote{In the bit-removal allocation techniques, all subcarriers are initially loaded with the maximum allowed constellation size that is decrementally decreased to meet the constraints. On the contrary, in the bit-filling allocation techniques, all subcarriers are initially nulled and constellation size is incrementally increased until meeting the constraints.}, but with faster convergence.
The algorithm exploits the differences between the subchannel gain-to-noise ratios in order to determine an initial bit allocation and then performs a multiple bit insertion or removal loading procedure to achieve the requested target rate. In \cite{liu2009adaptive}, Liu \textit{et al.} proposed a low complexity power loading algorithm that aims to minimize the transmit power while guaranteeing a target BER. Closed-form expressions for the optimal BER and power distributions were derived. Noteworthy, the reduced complexity of the proposed algorithm comes as a result of assuming uniform bit allocation across all subcarriers.

On the other hand, in \cite{leke1997maximum, wyglinski2005bit, krongold2000computationally, sonalkar2000efficient} the authors focused on maximizing the rate. Leke and Cioffi \cite{leke1997maximum} proposed a finite granularity algorithm that maximizes the data rate for a given margin. Subcarriers with signal-to-noise ratios (SNR) below a predefined threshold are nulled, and then remaining subcarriers are identified and the available power is distributed either optimally using a water-filling approach or suboptimally by assuming equal power to maximize the data rate. Krongold \textit{et al.} \cite{krongold2000computationally} presented a computationally efficient algorithm to maximize the throughput using a look-up table search and the Lagrange multiplier bisection method \cite{ramchandran1993best}. The algorithm converges faster to the optimal solution when compared to other allocation schemes. In \cite{wyglinski2005bit}, Wyglinski \textit{et al.} proposed an incremental bit loading algorithm to maximize the throughput while guaranteeing a target mean BER. The algorithm loads all subcarriers with the highest possible constellation size, and then calculate the BER per subcarrier depending on the channel state condition. The average BER is calculated and checked against the target BER. If the average BER meets the target BER, the final bit allocation is reached; otherwise, the signal constellation on the worst performance subcarrier is decreased and the process repeats. The algorithm nearly achieves the optimal solution given in \cite{fox1966discrete} but with lower complexity, which is the result of assuming uniform power allocations across all subcarriers.

Song \textit{et~al.} \cite{song2002joint} proposed an iterative joint bit loading and power allocation algorithm based on \textit{statistical} channel conditions to meet a target BER, i.e., the algorithm loads bits and power per subcarrier based on long-term frequency domain channel conditions, rather than \textit{instantaneous} channel conditions as in \cite{hughes1988ensemble, chow1995practical, campello1999practical, papandreou2005new, liu2009adaptive, leke1997maximum, wyglinski2005bit, krongold2000computationally, sonalkar2000efficient}. The algorithm marginally improves the  performance when compared to conventional MCM systems. The authors conclude that their algorithm is not meant to compete with its counterparts that adapt according to the instantaneous channel conditions. In \cite{fischer1996new}, Fischer and Huber proposed a low complexity loading algorithm to minimize the BER given a maximum allowed power and minimum throughput constraints. The authors claimed that in order to minimize the average BER, all subcarriers should experience the same BER so that the average BER is not dominated by the worst subcarrier. Goldfeld \textit{et~al.} \cite{goldfeld2002minimum} formulated an optimization problem to minimize the aggregate BER with a constraint on the total transmit power. Unfortunately, the problem was too complex to solve; hence, they resorted to a sub-optimal power loading algorithm that assumes uniform constellation size across all subcarriers.

\section{Link Adaptation Algorithms in the CR Environment}

The water-filling algorithms, which have been proven to be optimal for the link adaptation problem in non-CR environment, are no longer considered as optimal solutions in the CR environment. This is due to the fact that several new parameters need to be considered. For example, the interference from the PU to the SU, the interference from the SU to the PU, and the predefined threshold on the interference from the SU to the PU. The water-filling solutions for optimal link adaptation in the CR environment are found to load power to each subcarrier inversely proportional to the spectral distance between the subcarrier and the PU location, while still proportional to the SU subcarrier channel gain \cite{wang2007power, wang2009general, bansal2008optimal, zhang2008optimal, zhang2010efficient, kang2009optimal, attar2008interference, zhao2010power, Tang2010optimal, ngo2009resource, bansal2011adaptive, hasan2009energy}.



In \cite{wang2007power}, Wang \textit{et al.} proposed a novel iterative partitioned water-filling power allocation algorithm to maximize the  SU capacity, where the  SU power budget (constraints on the interference to the PUs are converted to constraints on the  SU transmitted power) and the peak transmission power per  SU subcarrier are considered as constraints. The authors considered only the effect of co-channel interference. The work in \cite{wang2007power} was generalized in \cite{wang2009general}, where the effect of adjacent channel interference is further considered. Bansal \textit{et al.} in \cite{bansal2008optimal} investigated the optimal power allocation problem in CR networks to maximize the  SU downlink transmission capacity under a constraint on the instantaneous interference to PUs. The proposed algorithm was complex and several suboptimal algorithms were developed to decrease the computational complexity. In \cite{zhang2008optimal}, Zhang optimized the  SU transmit power to maximize its ergodic capacity with constraints on the PU average capacity loss and the  SU average transmit power. The authors assumed perfect channel-state-information (CSI) between the PU and the SU receivers, as well as between the PU transmitter and receiver. Zhang and Leung \cite{zhang2010efficient} proposed a low complexity suboptimal algorithm for an OFDM-based CR system in which SUs may access both nonactive and active PU frequency bands, as long as the total co-channel interference (CCI) and adjacent channel interference (ACI) are within acceptable limits. The complexity reduction is the results of two validated approximations: 1) adjacent channel interference from SU to PUs is mainly limited to a few subcarriers adjacent to the PUs frequency bands and 2) the bandwidth of the PUs is typically much larger than that of a SU subcarrier. The proposed suboptimal algorithm shows significant improvement over its counterparts that use only nonactive PU frequency bands. Kang \textit{et al.}, in \cite{kang2009optimal}, studied the problem of optimal power allocation to achieve the ergodic, delay-sensitive, and outage capacities of a SU under a constrained average/peak  SU transmit power and interference to the PUs, with no interference from the PUs to the  SU taken into consideration. The ergodic capacity is defined as the maximum achievable rate averaged over all fading blocks. The delay-limited capacity is defined as the maximum constant transmission rate achievable over each fading block, which can be zero for severe fading scenarios. Thus, for such scenarios, the outage capacity, defined as the maximum constant rate that can be maintained over fading blocks with a given outage probability, is a good choice. It was shown that under the same threshold value, average interference constraints are more flexible over peak interference constraints to maximize the SU capacities. Attar \textit{et al.} in \cite{attar2008interference} maximized the total throughput (of both the  SU and PU) under a constraint of threshold interference to each user. In \cite{zhao2010power}, Zhao and Kwak maximized the throughput of the  SU while keeping the interference to PUs below a certain threshold. The mutual interference between the SU and the PUs were comprehensively modeled into constraints on the transmit power of the SU. A low-complexity iterative power loading algorithm and a suboptimal iterative bit loading algorithm were proposed to solve the modeled optimization problem. Ngo \textit{et al.}, in \cite{ngo2009resource}, proposed a practically optimal joint subcarrier assignment and power allocation algorithm to maximize the weighted sum rate of all secondary users of an OFDM-based CR network, while satisfying tolerable interference levels to PUs. The optimization problem was solved in the dual domain, where the duality gap tends to zero as the number of subcarriers goes to infinity. In \cite{bansal2011adaptive}, Bansal \textit{et al.} developed an optimal power allocation algorithm for OFDM-based CR systems with different statistical interference constraints imposed by different PUs. Since the interference constraints are met in a statistical manner, the SU transmitter does not require instantaneous CSI feedback from the PU receivers. Hasan \textit{et al.} \cite{hasan2009energy} presented a novel solution to maximize the SU capacity while taking into account the availability of subcarriers, i.e., the activity of PUs in the licensed bands, and limiting the interference leaked to~PUs.

\section{Energy Efficient Link Adaptation Algorithms}

As discussed earlier, the existing research has focused on optimizing the transmission rate of SUs while limiting the interference introduced to PUs to predefined thresholds. Recently, optimizing the energy-efficiency (EE)---defined as the total energy consumed to deliver one bit, or  the number of bits per unit energy \cite{amin2012cooperative, wang2012optimal,  oto2012energy}
---has received increasing attention due to steadily rising energy costs and environmental concerns \cite{amin2012cooperative, wang2012optimal,  oto2012energy, xie2012energy, wangenergy, mao2013energy, mao2013energy2}.

Wang   \textit{et al.} in \cite{wang2012optimal} optimized the EE of an OFDM-based CR network subject to power budget and interference constraints; however, this comes at the expense of deteriorating the rate of the SU.
Mao  \textit{et al.} in \cite{mao2013energy2} optimized the EE of OFDM-based CR systems subject to controlled interference leakage to PUs.
The authors proposed a so called waterfilling factors aided search to solve the non-convex EE optimization problem.
In \cite{han2011energy}, Han  \textit{et al.} proposed a novel channel management scheme that switches between different operations modes in order to maximize the EE of a CR sensor network.
Oto and Akan in \cite{oto2012energy} found the optimal packet size that maximizes the EE of a CR sensor networks while maintaining acceptable interference levels to the licensed PUs. In \cite{xie2012energy}, Xie \textit{et al.} investigated the problem of maximizing the EE of heterogeneous macrocells and femtocells cognitive networks.
The resource allocation problem is formulated as a Stackelberg game where the solution is obtained using a gradient-based iterative algorithm.
Wang \textit{et al.} in \cite{wangenergy} optimized the EE of an OFDM-based CR system subject to PUs interference constraints and different SUs rates.
In \cite{mao2013energy}, Mao \textit{et al.} optimized the EE of CR MIMO broadcast channels while guaranteeing certain interference threshold at the PUs receivers.
The authors transformed the non-convex optimization problem into an equivalent one-dimensional problem with a quasi-concave objective function that was solved using a golden search.
The same authors optimized the EE of an OFDM-based CR systems subject to controlled interference leakage to PUs in \cite{mao2013energy2}.



\section{Motivation and Outline}

As discussed earlier, the link adaptation algorithms in the literature focused on a single objective optimization, i.e., maximizing the throughput/capacity or minimizing the transmit power. However, in emerging wireless communication systems including CR systems, different requirements are needed. For example, minimizing the transmit power is prioritized when operating in interference-prone shared spectrum environments or in the proximity to other frequency-adjacent users. On the other hand, maximizing the throughput is favoured if sufficient guard bands exist to separate users. So, instead of solving different optimization problems for different applications, we adopt a multiobjective optimization (MOOP) approach that formulate a general optimization problem to balance between competing objectives.
In Chapter \ref{ch:TVT_GA}, we consider the MOOP problem that jointly maximizes the throughput and minimizes the transmit power of multicarrier systems subject to the quality-of-service (QoS), total transmit power, and maximum allowed bits per subcarrier constraints. The competing objective functions are linearly combined through weighting coefficients that represent the prior information about the preferences/importance of each objective. Novel algorithms are proposed to solve the MOOP problem and simulation results show significant performance improvements in terms of the achieved throughput and transmit power when compared to single optimization approaches, at the cost of no additional complexity. Chapter \ref{ch:WCL} proposes an evolutionary algorithm that adapts the preferences during the solution process in order to reduce the computational complexity.

Most of the optimal link adaptation algorithms in the CR-environment tend to assume practically unrealistic assumptions such as perfect sensing capabilities of the SU and perfect CSI on the links between the SU transmitter and the PUs receiver.  In Chapter \ref{ch:TW}, we consider the coexistence between an SU and multiple PUs and investigate the MOOP problem that simultaneously maximizes the SU throughput and minimizes its transmit power while considering the following: 1) total transmit power constraint, 2) maximum allowed CCI to the co-channel PUs constraint, 3) maximum allowed ACI to frequency adjacent PUs constraints, 4) QoS for the SU constraint, 5) maximum allocated bits per subcarrier constraint, and 6) imperfect sensing capabilities of the SU.  Chapter \ref{ch:TVT_MOOP} extends the MOOP formulation to balance between the SU transmission rate, CCI to co-channel PU, and ACI to adjacent PUs while assuming imperfect CSI on the links between the SU transmitter and the PUs receivers. Simulation results show that the MOOP approach provides improvements in the energy efficiency of the SU when compared to the works in the literature, with reduced computational effort.

Finally, Chapter \ref{ch:TVT_EE} proposes an energy-efficient power loading algorithm that considers the imperfect CSI on the links between the SU transmitter and receiver pairs subject to statistical constraints on the leaked interference to the PUs receivers and minimum supported rate for the SU.

\section{Contributions}

This dissertation presents the following novel contributions to the optimal link adaptation problem for MCM systems.

\begin{itemize}
  \item We propose a novel optimization framework for the optimal link adaptation problem for MCM systems. More specifically, we adopt a MOOP approach that jointly optimizes conflicting and incommensurable throughput/capacity, power, and interference objectives.
  \item We model the MOOP problems to guarantee certain QoS, maximum transmit power, maximum allocated bits per subcarrier, and certain interference thresholds to the PUs receivers. Some of the formulated MOOP problems are non-convex and we introduce approximate convex optimization problems, where the global optimal solution is guaranteed.
  \item We formulate the interference leaked to the PUs receivers with different degree of channel knowledge of the links between the SU transmitter and the PUs receivers.
  \item We consider the effect of imperfect spectrum sensing of the SU while formulating the MOOP problem in the CR environment.
  \item We optimize the EE of the SU while considering the channel sensing errors on the links between the SU transmitter and receiver pair and a minimum SU supported rate.
  \item We propose low complexity algorithms to solve the formulated optimization problems.
  \item We setup various simulation scenarios to investigate the performance of the proposed algorithms.
  \item We show that the adopted MOOP approach achieves significant performance improvements in terms of the achieved throughput and transmit power, when compared with other works in the literature that separately maximized the throughput (while constraining the transmit power) or minimized the transmit power (while constraining the throughput), at the cost of no additional complexity. Additionally, the MOOP improves the EE of the multicarrier systems.
  \item We illustrate that the interference constraints at the PUs receivers can be severely violated due to 1) assuming that the SU has perfect spectrum sensing capabilities and 2) imperfect CSI knowledge on the links between the SU transmitter and the PUs receivers. We additionally quantify the performance loss associated with the imperfect CSI knowledge on the links between the SU transmitter and the PUs receivers.
\end{itemize}

\bibliographystyle{IEEEtran}
\bibliography{IEEEabrv,mybib_file} 


\chapter{} \label{ch:TVT_GA}
\section{Abstract}
This paper investigates the problem of bit and power allocation for orthogonal frequency division multiplexing (OFDM) systems. Unlike all the proposed works in the literature that have focused on single objective optimizations,
in this paper we adopt the concept of multiobjective optimization to approach the bit and power allocation problem in order to meet the requirements of emerging wireless systems, i.e., achieving higher throughput without considerably increasing the transmit power. More specifically, we propose to simultaneously maximize the throughput and minimize the transmit power  of an OFDM system subject to a set of constraints. The formulated optimization problem is not convex and we use an evolutionary algorithm, i.e., genetic algorithm, in order to obtain the solution. To obtain closed-form expressions for the solution and reduce the complexity, we propose an approximate convex optimization problem where the global optimality of the Pareto solutions is guaranteed. Simulation results show that the proposed multiobjective optimization approach provides significant performance improvements over single objective optimization techniques presented in the literature, without incurring additional complexity.
\section{Introduction}
Orthogonal frequency division multiplexing (OFDM) is recognized as a robust and efficient transmission technique, as evidenced by its consideration for diverse communication systems and adoption by several wireless standards \cite{fazel2008multi, hwang2009ofdm, fu2010multicarrier}. The performance of OFDM systems can be improved  by dynamically adapting various transmission parameters, i.e., transmit power and number of bits per subcarrier, to the changing quality of the wireless link \cite{leke1997maximum, wyglinski2005bit, krongold2000computationally, sonalkar2000efficient, hughes1988ensemble, chow1995practical, liu2009adaptive, campello1999practical, papandreou2005new, mahmood2010efficient, wang2010efficient, goldsmith1998adaptive, chung2001degrees, song2002statistical, bedeer2011partial, bedeer2012jointVTC, bedeer2012EBERGC, bedeer2012UBERGC, bedeer2012novelICC, bedeer2012adaptiveRWS, bedeer2013resource, bedeer2013adaptive, bedeer2013novel, bedeer2014rateCONF, bedeer2014multiobjective, bedeer2013joint, 
bedeer2014energy, bedeer2015systematic, bedeer2015rate}.
The bit and power allocation problems can be categorized into two main classes: \textit{rate maximization} (RM) and \textit{margin maximization} (MM) \cite{hughes1988ensemble, chow1995practical, campello1999practical, papandreou2005new, liu2009adaptive, leke1997maximum, wyglinski2005bit, krongold2000computationally, sonalkar2000efficient}. For the former, the objective is to maximize the achievable data rate \cite{leke1997maximum, wyglinski2005bit, krongold2000computationally, sonalkar2000efficient}, while for the latter the objective is to maximize the achievable system margin \cite{hughes1988ensemble, chow1995practical, campello1999practical, papandreou2005new, liu2009adaptive, mahmood2010efficient, wang2010efficient} (i.e., minimizing the total transmit power given a target data rate or a target bit error rate (BER)). Modern wireless communication systems are required to satisfy conflicting objectives (e.g., increasing the OFDM system throughput without considerably increasing the transmit power) that do not optimize the RM and MM, separately.

To date, most of the research literature has focused on the \emph{single} objective function of maximizing either the RM or MM problems separately. For example, Leke and Cioffi \cite{leke1997maximum} proposed a finite granularity optimal algorithm that maximizes the throughput for a given power budget. The algorithm identifies and nulls subcarriers with signal-to-noise ratios (SNRs) below a predefined threshold, and optimally distributes the available power over the remaining subcarriers using a water-filling approach. In order to reduce the complexity of the proposed algorithm, the authors proposed a suboptimal algorithm that allocates equal power per subcarrier.
In \cite{wyglinski2005bit}, Wyglinski \textit{et al.} proposed an incremental bit loading algorithm to maximize the throughput while guaranteeing a target mean BER. The algorithm nearly achieves the optimal solution given in \cite{fox1966discrete} but with lower complexity, which is a result of employing uniform power
allocations across all subcarriers.
On the other hand, Hughes-Hartog \cite{hughes1988ensemble} proposed a greedy algorithm to maximize the margin by successively allocating bits to subcarriers requiring the minimum incremental power until the total target data rate is reached. The algorithm converges very slowly as it requires extensive subcarrier sorting; hence, it is very complex and not suitable for practical implementations. In \cite{chow1995practical}, Chow \textit{et al.}  proposed a suboptimal iterative bit loading algorithm that minimizes the transmit power subject to a target throughput. The algorithm calculates the initial bit allocations using the channel capacity approximation. Then, it iteratively adjusts the allocated bits to meet the target throughput.
Liu \textit{et al.} \cite{liu2009adaptive} proposed a low complexity power loading algorithm that aims to minimize the transmit power while guaranteeing a target BER. Closed-form expressions for the optimal power distributions were derived. The reduced complexity of the proposed algorithm comes as a result of assuming uniform bit allocation across all subcarriers.
Song \textit{et~al.} \cite{song2002statistical} proposed a statistical loading algorithm for multicarrier modulation (MCM) systems, i.e., the algorithm jointly loads bits and powers per subcarrier based on the fading statistics rather than the instantaneous channel conditions as in \cite{hughes1988ensemble, chow1995practical, campello1999practical, papandreou2005new, liu2009adaptive, leke1997maximum, wyglinski2005bit, krongold2000computationally, sonalkar2000efficient, goldsmith1998adaptive, chung2001degrees, mahmood2010efficient, wang2010efficient}.
The algorithm attains a marginal performance improvement when compared to conventional MCM systems. The authors conclude that their algorithm is not meant to compete with algorithms that adapt according to the instantaneous channel conditions.

In emerging wireless communication systems, different and flexible requirements are needed. For example, minimizing the transmit power is prioritized for battery operated devices, when operating in interference-limited shared spectrum environments, or in the proximity of other frequency-adjacent users. On the other hand, maximizing the throughput is favoured if high date rate is required and/or if sufficient guard bands exist to separate users.
This motivates us to formulate a multiobjective optimization (MOOP) problem that optimizes the conflicting and incommensurable throughput and power objectives. Recently, MOOP has attracted researchers' attention due to its flexible and superior performance \cite{elmusrati2008applications, devarajan2012energy, elmusrati2007multiobjective, sun2009modified, bedeer2013joint}. Jointly maximizing the throughput and minimizing the transmit power provides significant performance improvements in terms of the achieved throughput and transmit power, when compared with other works in the literature that separately maximize the throughput (while constraining the transmit power) or minimize the transmit power (while constraining the throughput); this is verified through the results presented in Section \ref{sec_Ch_1:sim}.


In this paper, we adopt a MOOP approach that simultaneously minimizes the OFDM system transmit power and maximizes its throughput subject to constraints on the quality-of-service (QoS), total transmit power, and maximum allocated bits per subcarrier. The QoS constraint is set to limit the average BER to a certain threshold. This constraint is not convex, and, hence, the formulated MOOP is not convex and the global optimality of the Pareto solutions is not guaranteed. The solution of this problem is found using an evolutionary algorithm, i.e., genetic algorithm. We noticed that the constraint on the average BER can be transformed to a constraint on the BER per subcarrier. This helps us to formulate an approximate convex MOOP problem where the global optimality of the Pareto solutions is guaranteed and closed-form expressions for the optimal bit and power allocations can be reached. Simulation results illustrate that the proposed algorithms are suprior to existing allocation algorithms in the literature, without incurring additional complexity.

The remainder of the paper is organized as follows. Section \ref{sec_Ch_1:opt} formulates and solves the optimization problems. Simulation results are presented in Section \ref{sec_Ch_1:sim}, while conclusions are drawn in Section \ref{sec_Ch_1:conc}.

Throughout this paper we use bold-faced lower case letters for vectors, e.g., $\mathbf{x}$, and light-faced letters for scalar quantities, e.g., $x$. $[.]^{\rm{T}}$ denotes the transpose operation, $\nabla$ represents the gradient, $\lfloor x \rfloor$ is the largest integer not greater than $x$, $\lfloor x \rceil$ is the nearest integer to $x$,  $[x,y]^-$ represents $\rm{min}(x,y)$, and $\bar{\bar{\mathbb{X}}}$ is the cardinality of the set $\mathbb{X}$.
\section{Optimization Problems: Formulation and Solution} \label{sec_Ch_1:opt}

\subsection{MOOP Principles}
The bit and power allocation problems in the literature are usually formulated as single objective optimization problems (minimizing cost function or maximizing utility function) and other functions are treated as constraint. Examples are minimizing the total transmit power subject to QoS and total transmit power, and maximizing the throughput/capacity subject to total transmit power and QoS constraints. Hence, the general form of the single objective bit and power allocation optimization problems can be written as
\begin{IEEEeqnarray}{c}
\underset{\mathbf{x}}{\rm{min}} \quad f(\mathbf{x}) \nonumber \\
{\rm{subject \: to}} \quad \mathbf{x} \in \mathcal{S},
\end{IEEEeqnarray}
where we have a single objective function $f(\mathbf{x})$: $\mathbf{R}^n \rightarrow \mathbf{R}$ and the decision variable $\mathbf{x} = \{x_1, x_2, ..., x_n\}^{\rm{T}}$ belongs to the non-empty feasible set $\mathcal{S}$, which is a subset of the decision variable space $\mathbf{R}^n$. We assume that the feasible region is formed by a set of inequality constraints, i.e., $\mathcal{S} = \{\mathbf{x} \in \mathbf{R}^n | \: \mathbf{g}(\mathbf{x}) = \{g_1(\mathbf{x}), g_2(\mathbf{x}), ..., g_C(\mathbf{x}) \}^{\rm{T}} \leq 0 \}$ and $C$ is the number of the inequality constraints.

In this paper, we propose a new formulation to the bit and power allocation problem, which is based on MOOP concepts, and prove that this new formulation provides superior performance over traditional formulations, i.e., single objective optimization. The MOOP formulation to the bit and power resource allocation problem can be written as
\begin{IEEEeqnarray}{c}
\underset{\mathbf{x}}{\rm{min}} \quad \{f_1(\mathbf{x}), f_2(\mathbf{x}), ..., f_O(\mathbf{x})\} \nonumber \\
{\rm{subject \: to}} \quad \mathbf{x} \in \mathcal{S},
\end{IEEEeqnarray}
where we have $O\:(\geq 2)$ objective functions. We denote the vector of the objective functions $\mathbf{f}(\mathbf{x}) = \{f_1(\mathbf{x}), f_2(\mathbf{x}), ..., f_O(\mathbf{x})\}^{\rm{T}}$ and we need to minimize all the objectives in $\mathbf{f}(\mathbf{x})$ simultaneously\footnote{If a function $f_o(\mathbf{x}), o = 1, ..., O,$ is to be maximized, i.e., $\underset{\mathbf{x}}{\rm{max}}  \: f_o(\mathbf{x})$, we transform it into an equivalent minimization problem, i.e., $\underset{\mathbf{x}}{\rm{min}}  \: \{- f_o(\mathbf{x})\}$.}. If there is no conflict between the objective functions, then an optimal solution can be found where every objective function attains its optimum. However, such a case is not common in practice, as  the objective functions are conflicting, i.e., there is no single optimal solution for all objective functions in $\mathbf{f}(\mathbf{x})$. Also, the objective functions in $\mathbf{f}(\mathbf{x})$ are usually incommensurable, i.e., of different units. The MOOP approach tries to search for non-dominant solutions $\mathbf{x}^*$, called Pareto optimal solutions, that can best compromise between different conflicting objectives. Mathematically, a decision vector $\mathbf{x}^* \in \mathcal{S}$ is Pareto optimal if there is not any other decision vector $\mathbf{x} \in \mathcal{S}$ such that $f_o(\mathbf{x}) \leq f_o(\mathbf{x}^*), \forall \: o = 1, ..., O,$ and $f_{o'}(\mathbf{x}) < f_{o'}(\mathbf{x}^*)$ for at least one index $\forall \: o' = 1, ..., O$ \cite{miettinen1999nonlinear}. If the objective functions $\mathbf{f}(\mathbf{x})$ and the feasible region $\mathcal{S}$ are convex, then the obtained Pareto optimal solution is referred to as a global Pareto optimal solution; otherwise, it is referred to as a local Pareto optimal solution \cite{miettinen1999nonlinear}. Furthermore, a decision vector $\mathbf{x}^* \in \mathcal{S}$ is weak Pareto optimal if there does not exist another decision vector $\mathbf{x} \in \mathcal{S}$ such that $f_o(\mathbf{x}) < f_o(\mathbf{x}^*), \forall \: o = 1, ..., O$.
In other words, a weak Pareto optimal solution is the solution for which there are no possible alternative solutions that cause every objective function to gain/improve. Note that the Pareto optimal set is a subset of the weakly Pareto optimal set \cite{miettinen1999nonlinear}.
Moving from a Pareto optimal solution to another one necessitates trading off; this is a basic concept in MOOP. Different methods exist to approach the MOOP tradeoff and they may produce worse/better results for the competing objective functions. Choosing the most efficient method is out of the scope of this work and we adopt the simple weighting sum method to explore the tradeoff and to show the effectiveness of the MOOP approach when compared to single objective approaches. 
In the weighting sum method, the competing objective functions are linearly combined through weighting coefficients that represent the preference/importance of each objective \cite{miettinen1999nonlinear}. Accordingly, the MOOP is formulated as
\begin{IEEEeqnarray}{c} \label{eq:MOOP_standard}
\underset{\mathbf{x}}{\rm{min}} \quad \alpha_1 \: f_1(\mathbf{x}) + \alpha_2 \: f_2(\mathbf{x}) +  ... + \alpha_O \: f_O(\mathbf{x}) \nonumber \\
{\rm{subject \: to}} \quad \mathbf{x} \in \mathcal{S},
\end{IEEEeqnarray}
where $\alpha_o \geq 0, o = 1, ..., O,$ are the tradeoff factors (weighting coefficients) that satisfy $\sum_{o = 1}^{O} \alpha_o = 1$. By changing the weighting parameters, the Pareto optimal set can be obtained through solving the MOOP problem in \eqref{eq:MOOP_standard}.

\subsection{Optimization Problem Formulation and Solution}


---\textit{Optimization problem}:
The new proposed formulation of the bit and power allocation problem that jointly minimizes the transmit power and maximizes the throughput can be written as
\begin{IEEEeqnarray}{c}
\underset{\mathbf{b, p}}{\rm{min}} \quad  \left\{\sum_{i = 1}^{N} p_i, \:  -\sum_{i = 1}^{N} b_i \right\} \nonumber \\
{\rm{subject \: to}} \quad {\rm{BER}}_{\rm{av}}(\mathbf{b,p}) \leq \rm{BER}_{\rm{th}}, \nonumber \\
\qquad \sum_{i = 1}^{N}p_i \leq P_{\rm{th}}, \nonumber \\
\qquad b_i  \leq  b_{i,\rm{max}},
\end{IEEEeqnarray}
where the decision variables $\mathbf{b} = [b_1, b_2, ..., b_N]^{\rm{T}}$ and $\mathbf{p} = [p_1, p_2, ..., p_N]^{\rm{T}}$ are the allocated bits and powers per each subcarrier, respectively. ${\rm{BER}}_{\rm{av}}$ and $\rm{BER}_{\rm{th}}$ are the average (over the total number of subcarriers) and the threshold values of the BER, respectively. $P_{\rm{th}}$ and $b_{i,\rm{max}}$ are the threshold value of the total transmit power and the maximum allocated bits per subcarrier, respectively. The minus sign associated with $\sum_{i = 1}^{N} b_i$ is added to reflect the throughput maximization. The constraint on the total transmit power is considered to meet the transmit power amplifier limitations and the constraint on the maximum allocated bit per subcarrier is added as it is not practical for some wireless applications to load a very high number of bits per subcarrier. ${\rm{BER}}_{\rm{av}}(\mathbf{b,p})$ is calculated as
\begin{IEEEeqnarray}{c}
{\rm{BER}}_{\rm{av}}(\mathbf{b,p}) = \frac{\sum_{i=1}^{N}b_i \: {\rm{BER}}_i(\mathbf{b,p})}{\sum_{i=1}^{N}b_i},
\end{IEEEeqnarray}
where ${\rm{BER}}_i$ is the BER per subcarrier $i$, $i$ = 1, ..., $N$. An approximate expression for the BER per subcarrier $i$ in the case of $M$-ary quadrature amplitude modulation (QAM) is given by\footnote{This expression is tight within 1 dB for BER $\leq$ $10^{-3}$ \cite{chung2001degrees}.} \cite{liu2009adaptive,chung2001degrees}
\begin{IEEEeqnarray}{rCl}
{\rm{BER}}_i &{} \approx  {}& 0.2 \: \exp\left ( - 1.6 \: \gamma_i \: \frac{p_i}{2^{b_i} - 1} \right ), \label{eq:BER}
\end{IEEEeqnarray}
where $\gamma_i$ is the channel-to-noise ratio for subcarrier~$i$. As mentioned earlier, we adopt the weighting sum method to solve the MOOP problem. Accordingly, the MOOP is formulated as
\begin{IEEEeqnarray}{c}
\mathcal{OP}1: \quad \underset{\mathbf{b, p}}{\rm{min}} \quad \mathbf{f}_{\rm{MOOP}}(\mathbf{b,p}) = \: \frac{\alpha}{u_{\rm{p}}} \sum_{i = 1}^{N}p_i - \frac{1-\alpha}{u_{\rm{b}}}\sum_{i = 1}^{N}b_i, \nonumber \\
{\rm{subject \: to}} \quad \frac{0.2 \: \sum_{i=1}^{N} b_i \: \exp\left ( \frac{- 1.6 \: \gamma_i p_i}{2^{b_i} - 1} \right ) }{\sum_{i=1}^{N} b_i} \leq \rm{BER}_{\rm{th}}, \nonumber \\
 \sum_{i = 1}^{N}p_i  \leq P_{\rm{th}}, \nonumber \\
 b_i   \leq  b_{i,\rm{max}}, \label{eq:MOOP_f}
\end{IEEEeqnarray}
where $\alpha$ ($0 < \alpha < 1$) is a constant whose value indicates the relative importance of one objective function relative to the other (i.e., a higher value of $\alpha$ favors minimizing the transmit power, whereas a lower value of $\alpha$ favors maximizing the throughput) and $u_{\rm{p}}$ and $u_{\rm{b}}$ are normalization factors used such that the two objectives are approximately within the same range. As such, $\alpha$ and $1-\alpha$ reflect the true preferences about each objective. We assume that the resource allocation entity of the OFDM system chooses the proper value of $\alpha$ depending on the application and/or the surrounding environment.
Further, as the minimum value of each objective is zero, we choose $u_{\rm{p}}$ and $u_{\rm{b}}$ to be equal to the maximum value of each objective, i.e., $P_{\rm{th}}$ and $N b_{i,\rm{max}}$, respectively, such that both objectives are in the range of [0,1].

The MOOP problem in \eqref{eq:MOOP_f} is non-convex as the constraint on the average BER is not convex for both $\mathbf{p}$ and $\mathbf{b}$. Hence, solving the problem using any gradient-based or numerical method can lead to a local optimum and not necessary to the global optimum with very large computational complexity, depending on the initial starting point. One way to overcome this difficulty is to use a gradient-based method and try many initial starting points; then, we select the solution that achieves the lowest objective function value. However, this is complex and may not be of interest especially in practical applications. Another possible way to approach the problem in \eqref{eq:MOOP_f} is to adopt  gradient-free algorithms, e.g., genetic algorithms (GAs) where we start with an initial set of points (population) and not with a single starting point, and, hence, it is less likely that GAs get trapped in a local optimum \cite{deb2001multi}.

---\textit{GA solution to the formulated problem}:
In GAs, a population of potential solutions, termed as chromosomes/individuals, is evolved over successive generations using a set of genetic operators called selection, crossover, and mutation. The selection operator selects the relatively fit individuals, based on their fitness value, to be part of the reproduction process of the new generation. In the reproduction process, new generations (children) are created using crossover and mutation operators. In the crossover operator, new children are created by blending genetic information between current individuals (i.e., parents) in order to explore the search space. On the other hand, the mutation operator changes one of more genes of the parents in order to maintain diversity and avoid premature convergence. The reproduction process repeats until meeting a certain stopping/convergence criteria \cite{deb2001multi}. It is worthy to mention that beside crossover and mutation operators, some individuals from the current generation with best fitness function values (i.e., lowest as the optimization problem in \eqref{eq:MOOP_f} is a minimization problem) are passed to the next generation and they are called elite children \cite{deb2001multi}.

In this paper, we adopt the real-coded GA proposed by Deep \emph{et al.} in \cite{deep2009real} in order to solve the MOOP problem in \eqref{eq:MOOP_f}.
Most of the real-coded GAs round off the real value of the decision variables to the nearest integer in order to meet the integer restriction of the integer variables. The real-coded GA in \cite{deep2009real} uses a truncation methods that ensures randomness in the generated solutions and avoids the possibility that the same integer value is generated when a real value lies between the same two consecutive integers. The truncation method works as follows. If the decision variable $b_i$ is integer, then it is accepted. Otherwise, it is equal to either $\lfloor b_i \rfloor$ or  $\lfloor b_i \rfloor + 1$ with equal probability. This truncation method increases the possibility to find the optimal solution \cite{deep2009real}.

We choose a tournament selection as it converges faster to the optimal solution even with lower complexity when compared to other selection schemes \cite{deep2009real, goldberg1991comparative}.
In the tournament selection, a number of individuals are chosen randomly from a given population, the best individual from this group is selected for further processing, and then the process repeats.
The selection of the best individual is done as follows: 1) a feasible solution with the lowest objective function value is chosen when compared to other feasible solutions, 2) a feasible solution is chosen when compared to infeasible solutions, and 3) an infeasible solution with the lowest constraint violation is chosen when compared to other infeasible solutions. This can be defined mathematically as \cite{goldberg1991comparative}
\begin{IEEEeqnarray}{c}
\mathbf{f}_{\rm{fitness}}(\mathbf{b,p}) =  \left\{\begin{matrix}\mathbf{f}_{\rm{MOOP}}(\mathbf{b,p}), \hfill \rm{for \: feasible} \: (\mathbf{b,p}),\\
\mathbf{f}_{\rm{worst}} + \sum_{c = 1}^{C = N +2} |\phi_c(\mathbf{b,p})|, \quad \hfill \rm{for \: infeasible} \: (\mathbf{b,p}),
\end{matrix}\right.
\end{IEEEeqnarray}
where $\mathbf{f}_{\rm{fitness}}(\mathbf{b,p})$ is the fitness function value, $\mathbf{f}_{\rm{worst}}$ is the objective function value of the worst feasible solution in a given population, and $\phi_c(\mathbf{b,p})$ is the left hand side of the inequality constraints in \eqref{eq:MOOP_f}. As can be seen, the fitness function of feasible solutions equals to their objective function value. On the other hand, for infeasible solutions, the fitness function depends on the constraints violations as well as the current population, i.e., the value of $\mathbf{f}_{\rm{worst}}$. In case there were no feasible solutions for a given population,  $\mathbf{f}_{\rm{worst}}$ is set to 0.

We use the Laplace crossover operator due to its superiority over other crossover techniques \cite{deep2009real, deep2007new}.
Laplace crossover generates two offsprings $w^{(1)}_k$ and $w^{(2)}_k$ from a pair of parents $z^{(1)}_k$ and $z^{(2)}_k, k = 1, ..., K,$ where $K$ is the size of the decision variables, as follows. First, uniform random numbers $\nu_k$ and $r_k$ between 0 and 1 are generated. Based on the Laplace inverse cumulative distribution function, a random number $\beta_k, k = 1, ..., K,$ that satisfies the Laplace distribution is generated as:
\begin{IEEEeqnarray}{c}
\beta_k =  \left\{\begin{matrix} a - \xi \log_e(\nu_k), \qquad r_k \leq \frac{1}{2},\\
a + \xi \log_e(\nu_k), \qquad r_k > \frac{1}{2},
\end{matrix}\right.
\end{IEEEeqnarray}
where $a$ and $\xi > 0$ are the location and the scale parameters for the Laplace distribution function \cite{kotz2001laplace}; these are chosen adaptively to distribute the children based on the spread of the parents \cite{deep2007new}. Finally, the children are generated as
\begin{IEEEeqnarray}{c}
w^{(1)}_k = z^{(1)}_k + \beta_k |z^{(1)}_k - z^{(2)}_k|, \nonumber\\
w^{(2)}_k = z^{(2)}_k + \beta_k |z^{(1)}_k - z^{(2)}_k|.
\end{IEEEeqnarray}

We use the power mutation operator that is superior when compared to the mutation operators \cite{deep2009real, deep2007newmu}.
A child is created in the vicinity of a parent solution through the following steps. First, a random number $s$ that follows the power-law distribution is generated \cite{kotz2001laplace}. Then, the muted solution is calculated as
\begin{IEEEeqnarray}{c}
w_k =  \left\{\begin{matrix} z_k - s(z_k - z^{(l)}), \qquad t < r,\\
z_k + s(z^{(u)} - z_k), \qquad t \geq r,
\end{matrix}\right.
\end{IEEEeqnarray}
where $z^{(l)}$ and $z^{(u)}$ are the lower and upper bounds on the decision variable $z_k$, respectively, $r$ is uniformly distributed random number between 0 and 1, and $t = \frac{z_k - z^{(l)}}{z^{(u)} - z_k}$.

The proposed GA algorithm to solve $\mathcal{OP}1$ in \eqref{eq:MOOP_f} is outlined as follows:

\floatname{algorithm}{}
\begin{algorithm}
\renewcommand{\thealgorithm}{}
\caption{\textbf{\small{Proposed Algorithm to Solve $\mathcal{OP}1$}}}
\begin{algorithmic}[1]
\small
\State \textbf{INPUT} $\gamma_i$, $\rm{BER}_{\rm{th}}$, $u_{\rm{p}}$, $u_{\rm{b}}$, $\alpha$, $P_{\rm{th}}$, $b_{i,\rm{max}}$, population size, number of generations.
\State Create random population.
\State Check the stopping criteria (i.e., change in objective function value is less than a certain threshold or maximum number of generations is reached). If satisfied, stop; otherwise, proceed to next step.
        \algstore{myalg}
  \end{algorithmic}
\end{algorithm}

\floatname{algorithm}{}
\begin{algorithm}
 \renewcommand{\thealgorithm}{}
  \caption{\textbf{\small{Proposed Algorithm to Solve $\mathcal{OP}1$} (continued)}}
  \begin{algorithmic}
      \algrestore{myalg}
      \small
\State Create new population.
\begin{itemize}
\item Compute the fitness value for each member in the current population and select individuals, based on their fitness value, using a tournament selection.
\item Select the elite individuals (i.e., members of lower fitness value) with a certain probability and pass them to the next population.
\item Apply the crossover operator to the selected parents from the old population in order to produce children.
\item Apply the power mutation operator to the selected parents from the old population in order to produce muted children.
\item Replace the current population with the produced children in order to form the next generation.
\end{itemize}
\State Go to step 3.
\State \textbf{OUTPUT} $b_i^*$ and $p_i^*$, $i = 1, ..., N$.
\end{algorithmic}
\end{algorithm}

\subsection{Approximate MOOP Problem}

According to results in \cite{willink1997optimization} and numerical results presented in Section \ref{sec_Ch_1:sim}, the constraint on the BER per subcarrier is an acceptable substitute to the constraint on the average BER. To avoid the computational complexity of the proposed algorithm to solve $\mathcal{OP}1$ and in order to obtain closed-form solutions, we consider the approximate problem of $\mathcal{OP}1$, where the constraint on the average BER is replaced with a constraint on the BER per subcarrier.
The new optimization problem is formulated as\footnote{The optimization problem with discrete constraints for the number of the allocated bits per subcarrier is a mixed integer nonlinear programming problem that can be solved by the branch and bound algorithm \cite{floudas1995nonlinear}. However, this will be significantly complex and not tractable for large number of subcarriers. In the rest of the paper, we assume continuous values for the number of bits per subcarrier in order to obtain a low complexity solution, and then discretize the number of allocated bits per subcarrier. 
}

\begin{IEEEeqnarray}{c}
\underset{\mathbf{b, p}}{\rm{min}} \quad  \left\{\sum_{i = 1}^{N} p_i, \: -\sum_{i = 1}^{N} b_i \right\} \nonumber \\
{\rm{subject \: to}} \quad {\rm{BER}}_i(\mathbf{b,p}) = 0.2 \: \exp\left ( \frac{- 1.6 \: \gamma_i p_i}{2^{b_i} - 1} \right ) \leq \rm{BER}_{\rm{th},\it{i}}, \nonumber \\
 										    \sum_{i = 1}^{N}p_i \leq P_{\rm{th}},   \nonumber \\
                                            b_i  \leq b_{i,\rm{max}}, \quad i = 1, ..., N. \label{eq:ineq_const_op2}
\end{IEEEeqnarray}

The optimization problem in \eqref{eq:ineq_const_op2} is not convex due to the constraint on the BER per subcarrier, and hence, the global optimality of the Pareto set of solutions is not guaranteed. An important remark that helps to resolve the non-convexity issue is that the constraint on the BER per subcarrier, i.e., ${\rm{BER}}_i(\mathbf{b,p}) \leq \rm{BER}_{\rm{th},\it{i}}$, which is the source of the non-convexity, is always active\footnote{An inequality constraints $g_c, c = 1, ..., C,$ is said to be active at a point $\mathbf{x}^*$ if $g_c(\mathbf{x}^*) = 0$, and it is said to be inactive at a point  $\mathbf{x}^*$ if $g_c(\mathbf{x}^*) < 0$.} and it can be relaxed in order to obtain a convex problem equivalent to the optimization problem in \eqref{eq:ineq_const_op2}. We can prove that the constraint on the BER per subcarrier is always active by contradiction, as follows. Let us assume that the optimal bit and power allocations ($b_i^*,p_i^*$) exist at a value for the BER per subcarrier that is not at the boundary, i.e., at ${\rm{BER}}_{{\rm{th}},i}^* < {\rm{BER}}_{{\rm{th}},i}$. In this case, a new solution could be obtained at ${\rm{BER}}_{{\rm{th}},i}^{\rm{new}}$, ${\rm{BER}}_{{\rm{th}},i}^*  < {\rm{BER}}_{{\rm{th}},i}^{\rm{new}} \leq   {\rm{BER}}_{{\rm{th}},i}$,  where the power could be decreased, i.e., $p_{i}^{\rm{new}} < p_i^*$ or the rate can be increased, i.e., $b_{i}^{\rm{new}} > b_i^*$ without violating the BER constraint. Clearly, this results in a lower objective function value in \eqref{eq:ineq_const_op2}, and hence, the allocation of the bit and power $(b_{i}^*, p_{i}^*)$ that is  at ${\rm{BER}}_{{\rm{th}},i}^* < {\rm{BER}}_{{\rm{th}},i}$ cannot be an optimal solution. This can be mathematically proved by applying the Karush-Khun-Tucker (KKT) conditions to the problem in \eqref{eq:ineq_const_op2}.

Since the constraint on the BER per subcarrier (the source of the non-convexity of the problem in \eqref{eq:ineq_const_op2}) is always active, we can relate $p_i$ and $b_i$ from (\ref{eq:BER}) as follows
\begin{IEEEeqnarray}{c}
p_i =  \frac{\Gamma_i}{\gamma_i} \: (2^{b_i} - 1),
\label{eq:sub_P_b}
\end{IEEEeqnarray}
where $\Gamma_i = \frac{- \ln(5 \rm{BER}_{\rm{th},\it{i}})}{1.6}$ is the signal-to-noise ratio (SNR) gap that represents the difference between the maximum achieved rate and the practical achievable transmission rate \cite{goldsmith1998adaptive}. The BER constraint can be removed from the optimization problem in \eqref{eq:ineq_const_op2} after substituting $p_i$, $i = 1, ..., N$, from (\ref{eq:sub_P_b}). Hence, we formulate a new optimization problem $\mathcal{OP}2$ as follows
\begin{IEEEeqnarray}{c}
\mathcal{OP}2: \mathbf{f}_{\rm{MOOPapprox}}(\mathbf{b}) = \frac{\alpha}{u_{\rm{p}}} \sum_{i = 1}^{N} \frac{\Gamma_i}{\gamma_i} \: (2^{b_i} - 1) - \frac{1-\alpha}{u_{\rm{b}}}\sum_{i = 1}^{N}b_i, \nonumber \\
{\rm{subject \: to}} \quad
G_{\varrho}(\mathbf{b}) =  \left\{\begin{matrix}
b_i - b_{i,\rm{max}} \leq 0, \hfill \varrho = i = 1, ..., N, \\
\sum_{i = 1}^{N}  \frac{\Gamma_i}{\gamma_i} (2^{b_i} - 1) - P_{\rm{th}} \leq 0, \quad \varrho = N+1. \label{eq:ineq_const_op5} \\
\end{matrix}\right.
\end{IEEEeqnarray}

One can easily show that $\mathcal{OP}2$ is a convex optimization problem, and hence, the obtained Pareto optimal solution is guaranteed to be a global optimum. Applying the method of Lagrangian multipliers, the inequality constraints in (\ref{eq:ineq_const_op5}) are transformed to equality constraints by adding non-negative slack variables, $y^2_{\varrho}$, $\varrho = 1, ..., N+1$ \cite{Boyd2004convex}. Hence, the constraints are rewritten as
\begin{IEEEeqnarray}{rCl}
\mathcal{G}_{\varrho}(\mathbf{b}, \mathbf{y}) & = & G_{\varrho}(\mathbf{b}) + y_{\varrho}^2 = 0, \qquad \varrho = 1, ..., N+1,
\label{eq:eq_const_op5}
\end{IEEEeqnarray}
where $\mathbf{y} = [y_1^2, ..., y_{N+1}^2]^T$ is the vector of slack variables. The Lagrange function $\mathcal{L}$ is expressed as
\begin{IEEEeqnarray}{rCl}
\mathcal{\mathcal{L}}(\mathbf{b}, \mathbf{y},\boldsymbol\lambda) &{} = {}& \mathbf{f}_{\rm{MOOPapprox}}(\mathbf{b}) + \sum_{\varrho = 1}^{N+1} \lambda_{\varrho} \: \mathcal{G}_{\varrho}(\mathbf{b}, \mathbf{y}), \nonumber \\
 &{} = {}& \frac{\alpha}{u_{\rm{p}}} \sum_{i = 1}^{N} \frac{\Gamma_i}{\gamma_i} (2^{b_i} - 1) - \frac{1-\alpha}{u_{\rm{b}}} \sum_{i = 1}^{N}b_i \nonumber \\
& + & \sum_{i = 1}^{N} \lambda_i \: [b_i - b_{i,\rm{max}} + y^2_i], \label{eq:L_op5} \nonumber \\
& + &  \lambda_{N+1}\: \big[\sum_{i = 1}^{N} \frac{\Gamma_i}{\gamma_i} (2^{b_i} - 1) - P_{\rm{th}} + y_{N+1}^2\big],
\end{IEEEeqnarray}
where $\boldsymbol\lambda = [\lambda_1, ..., \lambda_{N+1}]^T$  is the vector of the Lagrange multipliers. A stationary point can be found when $\nabla \mathcal{\mathcal{L}}(\mathbf{b}, \mathbf{y},\boldsymbol\lambda) = 0$, which yields

\begin{subequations}
\label{eq:lag_op5}
\begin{IEEEeqnarray}{RCL}
\frac{\partial \mathcal{L}}{\partial b_i} &{}  = {}& \frac{\alpha}{u_{\rm{p}}} \ln(2) \frac{\Gamma_i}{\gamma_i} 2^{b_i}  - \frac{1-\alpha}{u_{\rm{b}}} + \lambda_i + \ln(2) \lambda_{N+1} \frac{\Gamma_i}{\gamma_i} 2^{b_i}  = 0 \label{eq:eq2_op5} \IEEEeqnarraynumspace\\
\frac{\partial \mathcal{L}}{\partial \lambda_i} &{}  = {}& b_i - b_{i,\rm{max}} +  y_i^2 = 0, \IEEEeqnarraynumspace \label{eq:eq33_op5}\\
\frac{\partial \mathcal{L}}{\partial \lambda_{N+1}} &{}  = {}& \sum_{i = 1}^{N} \frac{\Gamma_i}{\gamma_i} (2^{b_i} - 1) - P_{\rm{th}} + y_{N+1}^2 = 0, \IEEEeqnarraynumspace \label{eq:eq3_op5}\\
\frac{\partial \mathcal{L}}{\partial y_i} &{}  = {}&  2\lambda_i y_i = 0, \label{eq:eq4_op5} \\
\frac{\partial \mathcal{L}}{\partial y_{N+1}} &{}  = {}&  2\lambda_{N+1} y_{N+1} = 0. \label{eq:eq44_op5}
\end{IEEEeqnarray}
\end{subequations}

It can be seen that (\ref{eq:eq2_op5})--(\ref{eq:eq44_op5}) represent $3N+2$ equations in the $3N+2$ unknown elements of the vectors $\mathbf{b}, \mathbf{y}$,  and $\boldsymbol\lambda$. Equation (\ref{eq:eq4_op5}) implies that either $\lambda_i$ = 0 or $y_i$ = 0, while (\ref{eq:eq44_op5}) implies that either $\lambda_{N+1}$ = 0 or $y_{N+1}$ = 0. Accordingly, four possible cases exist, as follows:

--- \textit{Case \MakeUppercase{\romannumeral 1}}: Choosing $\lambda_{N+1} = 0$ ($y_{N+1} \neq 0$, i.e., inactive power constraint) and $\lambda_i = 0$ ($y_i \neq 0$, i.e., inactive maximum bit constraint) gives the optimal values of $b_i^*$ as
\begin{IEEEeqnarray}{c}
b_i^* = \Big\lfloor \log_2\big(\frac{\frac{1-\alpha}{u_{\rm{b}}} }{\frac{\alpha}{u_{\rm{p}}} \ln(2)} \frac{\gamma_i}{\Gamma_i}\big) \Big\rceil, \label{eq:b_optimal}
\end{IEEEeqnarray}
and from (\ref{eq:sub_P_b}), the optimal power allocation $p^*_i$ is given by
\begin{IEEEeqnarray}{c}
p_i^* = \frac{\Gamma_i}{\gamma_i} \Big(2^{\big\lfloor \log_2\big( \frac{\frac{1-\alpha}{u_{\rm{b}}} }{\frac{\alpha}{u_{\rm{p}}} \ln(2)} \frac{\gamma_i}{\Gamma_i}\big) \big\rceil}  - 1\Big).  \label{eq:p_optimal} \IEEEeqnarraynumspace
\end{IEEEeqnarray}
Since we consider $M$-ary QAM, $b_i$ should be greater than 2. From (\ref{eq:b_optimal}), to have $b_i \geq 2$, the channel-to-noise ratio per subcarrier, $\gamma_i$, must satisfy the condition
\setlength{\arraycolsep}{0.0em}
\begin{IEEEeqnarray}{c}
\gamma_i \geq \gamma_{{\rm{th}},i}^{\rm{min}} =  \frac{\frac{\alpha}{u_{\rm{p}}} \ln(2)}{\frac{1-\alpha}{u_{\rm{b}}}} \Gamma_i \: 2^2, \qquad i = 1, ..., N. \label{eq:condition}
\end{IEEEeqnarray}

--- \textit{Case \MakeUppercase{\romannumeral 2}}: Choosing $\lambda_{N+1} = 0$ ($y_{N+1} \neq 0$, i.e., inactive power constraint) and $y_i = 0$ (i.e., active maximum bit constraint) leads to the optimal bit allocation $b_i^* = b_{i,\rm{max}}$ if and only if $\gamma_i \geq \gamma_{{\rm{th}},i}^{\rm{max}} =  \frac{\frac{\alpha}{u_{\rm{p}}} \ln(2)}{\frac{1-\alpha}{u_{\rm{b}}}} \Gamma_i \: 2^{b_{i,\rm{max}}}, i = 1, ..., N$ (the proof is provided in Appendix A) and $p_i^*$ is calculated according to  (\ref{eq:sub_P_b}). It is worthy to mention that limiting the allocated bits to the maximum value $b_{i,\rm{max}}$, when $\gamma_i \geq \gamma_{{\rm{th}},i}^{\rm{max}}$, reduces the transmit power on the corresponding subcarriers, and, hence, the total transmit power decreases (i.e., the power constraint is still inactive).

Given that $\gamma_{{\rm{th}},i}^{\rm{max}} \geq \gamma_{{\rm{th}},i}^{\rm{min}}$, the optimal solution of cases \emph{\MakeUppercase{\romannumeral 1}} and \emph{\MakeUppercase{\romannumeral 2}}, in case of $\gamma_i \geq \gamma_{{\rm{th}},i}^{\rm{min}}$, is joined as
\begin{IEEEeqnarray}{c}
b_i^* = \Big[ \big\lfloor\log_2\big(\frac{\frac{1-\alpha}{u_{\rm{b}}}}{\frac{\alpha}{u_{\rm{p}}} \: \ln(2)} \frac{\gamma_i}{\Gamma_i}\big)\big\rceil, b_{i,\rm{max}} \Big]^-, \label{eq:b_optimal_5}
\end{IEEEeqnarray}
\begin{IEEEeqnarray}{c}
p_i^* = \frac{\Gamma_i}{\gamma_i} \left(2^{\big[ \big\lfloor\log_2\big(\frac{\frac{1-\alpha}{u_{\rm{b}}}}{\frac{\alpha}{u_{\rm{p}}} \: \ln(2)} \frac{\gamma_i}{\Gamma_i}\big)\big\rceil, b_{i,\rm{max}} \big]^-} - 1\right). \label{eq:p_optimal_5}
\end{IEEEeqnarray}

--- \textit{Case \MakeUppercase{\romannumeral 3}}: Choosing $y_{N+1} = 0$ (i.e., active power constraint) and $\lambda_i = 0$ ($y_i \neq 0$,  i.e., inactive maximum bit constraint) gives the optimal values of $b_i^*$ as
\begin{IEEEeqnarray}{c}
b_i^* = \left\lfloor \log_2\Big( \frac{\frac{1-\alpha}{u_{\rm{b}}}}{(\frac{\alpha}{u_{\rm{p}}} + \lambda_{N+1}) \ln(2)} \frac{\gamma_i}{\Gamma_i}\Big) \right\rceil, \label{eq:b_optimal_2}
\end{IEEEeqnarray}
where $\lambda_{N+1}$ is calculated to satisfy the active power constraint in \eqref{eq:eq3_op5} ($y_{N+1} = 0$). Hence, the value of $\lambda_{N+1}$ is found to be
\begin{IEEEeqnarray}{c}
\lambda_{N+1} = \bar{\bar{\mathbb{N}}}_a \frac{\frac{1-\alpha}{u_{\rm{b}}}}{\ln 2} \frac{1}{p_{th} + \sum_{i \in \mathbb{N}_a}^{} \frac{\Gamma_i}{\gamma_i}} - \frac{\alpha}{u_{\rm{p}}}, \IEEEeqnarraynumspace \label{eq:lambda}
\end{IEEEeqnarray}
where $\bar{\bar{\mathbb{N}}}_a$ is the cardinality of the set of active subcarriers $\mathbb{N}_a$.
 Finally, the optimal bit $b_i^*$ and power $p_i^*$ allocations in case of active power constraint is given as

\begin{IEEEeqnarray}{c}
b_i^* = \big\lfloor\log_2\big(\frac{P_{\rm{th}} + \sum_{i \in \mathbb{N}_a}^{}\frac{\Gamma_i}{\gamma_{i}}}{\bar{\bar{\mathbb{N}}}_a} \frac{\gamma_i}{\Gamma_i}\big)\big\rceil. \label{eq:b_optimal_3}
\end{IEEEeqnarray}
\begin{IEEEeqnarray}{c}
p_i^* = \frac{\Gamma_i}{\gamma_i} \left(2^{\big\lfloor\log_2\big(\frac{P_{\rm{th}} + \sum_{i \in \mathbb{N}_a}^{}\frac{\Gamma_i}{\gamma_{i}}}{\bar{\bar{\mathbb{N}}}_a} \frac{\gamma_i}{\Gamma_i}\big)\big\rceil} - 1\right). \label{eq:p_optimal_3}
\end{IEEEeqnarray}

--- \textit{Case \MakeUppercase{\romannumeral 4}}: Choosing $y_{N+1} = 0$ (i.e., active power constraint) and $y_i = 0$ (i.e., active maximum bit constraint) leads to the optimal bit allocation $b_i^* = b_{i,\rm{max}}$ if and only if $\gamma_i \geq  \gamma_{{\rm{th}},i}^{\rm{max}}$ and $p_i^*$ is calculated according to  (\ref{eq:sub_P_b}). To find the bit and power allocation for the rest of subcarriers, the set of active subcarriers $\mathbb{N}_a$ is updated to exclude subcarriers with $b_i^* = b_{i,\rm{max}}$, and the Lagrangian multiplier $\lambda_{N+1}$ is calculated based on the new power budget $P_{\rm{th}} - \sum_{i_{\rm{max}}}^{} p_{i_{\rm{max}}}^*$, where $i_{\rm{max}}$ denotes subcarriers loaded with the maximum allocated bits per subcarrier $b_{i,\rm{max}}$.

The optimal solution of case \emph{\MakeUppercase{\romannumeral 4}}, in case of $\gamma_{{\rm{th}},i}^{\rm{max}} \geq \gamma_i \geq \gamma_{{\rm{th}},i}^{\rm{min}}$, is expressed as
\begin{IEEEeqnarray}{c}
b_i^* =  \big\lfloor\log_2\big(\frac{P_{\rm{th}} - \sum_{i_{\rm{max}}}^{} p_{i_{\rm{max}}}^* + \sum_{i \in \mathbb{N}_a}^{}\frac{\Gamma_i}{\gamma_i}}{\bar{\bar{\mathbb{N}}}_a} \frac{\gamma_i}{\Gamma_i} \big)\big\rceil \label{eq:b_optimal_55}
\end{IEEEeqnarray}
\begin{IEEEeqnarray}{c}
p_i^* = \frac{\Gamma_i}{\gamma_i} \left(2^{\big\lfloor\log_2\big(\frac{P_{\rm{th}} - \sum_{i_{\rm{max}}}^{} p_{i_{\rm{max}}}^* + \sum_{i_m \in \mathbb{N}_a}^{}\frac{\Gamma_i}{\gamma_i}}{\bar{\bar{\mathbb{N}}}_a} \frac{\gamma_i}{\Gamma_i}\big)\big\rceil} - 1\right). \label{eq:p_optimal_55} \IEEEeqnarraynumspace
\end{IEEEeqnarray}

The optimal solution of cases \emph{\MakeUppercase{\romannumeral 3}} and \emph{\MakeUppercase{\romannumeral 4}} is joined as
\begin{IEEEeqnarray}{c}
b_i^* =  \big\lfloor\log_2\big(\frac{P_{\rm{th}} - \sum_{i_{\rm{max}}}^{} p_{i_{\rm{max}}}^* + \sum_{i \in \mathbb{N}_a}^{}\frac{\Gamma_i}{\gamma_i}}{\bar{\bar{\mathbb{N}}}_a} \frac{\gamma_i}{\Gamma_i} \big)\big\rceil \label{eq:b_optimal_555}
\end{IEEEeqnarray}
\begin{IEEEeqnarray}{c}
p_i^* = \frac{\Gamma_i}{\gamma_i} \left(2^{\big\lfloor\log_2\big(\frac{P_{\rm{th}} - \sum_{i_{\rm{max}}}^{} p_{i_{\rm{max}}}^* + \sum_{i_m \in \mathbb{N}_a}^{}\frac{\Gamma_i}{\gamma_i}}{\bar{\bar{\mathbb{N}}}_a} \frac{\gamma_i}{\Gamma_i}\big)\big\rceil} - 1\right). \label{eq:p_optimal_555} \IEEEeqnarraynumspace
\end{IEEEeqnarray}
where for $\gamma_i \geq \gamma_{{\rm{th}},i}^{\rm{min}}$, $i_{\rm{max}}$ denotes the subcarriers loaded with $b_{i,\rm{max}}$; otherwise, $i_{\rm{max}}$ includes no subcarriers.

The obtained solution ($\mathbf{p}^*,\mathbf{b}^*$) represents a global minimum as the KKT conditions \cite{Boyd2004convex} are satisfied (see Appendix A for proof), and $\mathcal{OP}2$ is convex.

The proposed algorithm to solve $\mathcal{OP}2$ can be stated as follows:
\floatname{algorithm}{}
\begin{algorithm}
\renewcommand{\thealgorithm}{}
\caption{\textbf{\small{Proposed Algorithm to Solve $\mathcal{OP}2$}}}
\begin{algorithmic}[1]
\small
\State \textbf{INPUT} $\gamma_i$, $\rm{BER}_{\rm{th}}$, $u_{\rm{p}}$, $u_{\rm{b}}$, $\alpha$, $P_{\rm{th}}$, $b_{i,\rm{max}}$
\For{$i$ = 1, ..., $N$}
\If{$\gamma_i \geq \gamma_{{\rm{th}},i}^{\rm{min}}$}
\State  $b_i^*$ and $p_i^*$ are given by (\ref{eq:b_optimal_5}) and (\ref{eq:p_optimal_5}), respectively.
\Else
\State Null the corresponding subcarrier $i$.
        \algstore{myalg}
  \end{algorithmic}
\end{algorithm}

\floatname{algorithm}{}
\begin{algorithm}
 \renewcommand{\thealgorithm}{}
  \caption{\textbf{Proposed Algorithm to solve $\mathcal{OP}2$} (continued)}
  \begin{algorithmic}
      \algrestore{myalg}
\small
\EndIf
\EndFor
\If {$\sum_{i = 1}^{N}p_i \geq P_{\rm{th}}$}
\State For subcarriers with $\gamma_i \geq \gamma_{{\rm{th}},i}^{\rm{max}}$, $b_i^* = b_{i,\rm{max}}$ and $p_i^*$ is calculated from \eqref{eq:sub_P_b}.
\State  For subcarriers with $\gamma_i < \gamma_{{\rm{th}},i}^{\rm{max}}$, $b_i^*$ and $p_i^*$ are given by (\ref{eq:b_optimal_555}) and (\ref{eq:p_optimal_555}), respectively.
\EndIf
\State If the condition on the total transmit power is violated due to rounding, decrement the number of bits on the subcarrier that has the largest $\Delta p_{i}(b_{i}) = p_{i}(b_{i}) - p_{i}(b_{i} - 1)$ until satisfied.
\State \textbf{OUTPUT} $b^*_{i}$ and $p^*_{i}$, $i$ = 1, ..., $N$.
\end{algorithmic}
\end{algorithm}

According to the MOOP problem analysis, the optimal solution belongs to one of the four cases, \textit{case \MakeUppercase{\romannumeral 1}} to \textit{case \MakeUppercase{\romannumeral 4}}. So, the proposed algorithm starts by assuming that the optimal solution belongs to either \textit{case \MakeUppercase{\romannumeral 1}} or \textit{case \MakeUppercase{\romannumeral 2}}, where the optimal bit and power allocations for both cases are given by (\ref{eq:b_optimal_5}) and (\ref{eq:p_optimal_5}), respectively.
Based on this assumption and if the power constraint is violated, the optimal bit allocation is given by $b_i^* = b_{i,\rm{max}}$ and the optimal power is calculated according to \eqref{eq:sub_P_b} for subcarriers with $\gamma_i \geq \gamma_{{\rm{th}},i}^{\rm{min}}$. Otherwise, the optimal bit and power allocations  are given by (\ref{eq:b_optimal_555}) and (\ref{eq:p_optimal_555}), respectively.
The purpose of step 13 is to guarantee that the total transmit power constraint will not be violated due to rounding the allocated bits to the nearest integer. If violated, the subcarrier corresponding to the largest power reduction when the number of bits is decremented by 1 bit is chosen, and the number of bits is decreased by 1 bit on that subcarrier. The process repeats until the total transmit power constraint is satisfied.

The computational complexity of the proposed algorithm to solve $\mathcal{OP}2$ can be analyzed as follows. Steps 2 to 8 requires a complexity of $\mathcal{O}(N)$; steps 9 to 11 requires a complexity of  $\mathcal{O}(N^2)$; and step 13 requires a computational complexity of $\mathcal{O}(N^2)$. This can be explained as follows: First, step 13 finds the subcarrier $i$ with the maximum $\Delta p_{i}(b_{i}) = p_{i}(b_{i}) - p_{i}(b_{i} - 1)$ due to rounding, which is of complexity of  $\mathcal{O}(N)$. Second, step 13 decrements the allocated bits on $i'$ until the power constraint is satisfied. In the worst case, this process will be repeated $N$ times and hence, the computational complexity is of $\mathcal{O}(N)$ if all allocated bits are rounded up to the nearest integer. Hence, the worst case computational complexity  of the proposed algorithm to solve $\mathcal{OP}2$ is calculated as $\mathcal{O}(N) + \mathcal{O}(N^2) + \mathcal{O}(N^2) = \mathcal{O}(N^2)$.

\section{Numerical Results} \label{sec_Ch_1:sim}

This section investigates the performance of the proposed algorithms, and compares their performance with bit and power loading algorithms presented in the literature. The computational complexity of the proposed algorithms is also compared to the other schemes.

\subsection{Simulation Setup}
We consider an OFDM system with $N$ = 64 subcarriers and bandwidth of 1.25 MHz \cite{standard}. The average BER threshold $\rm{BER}_{th}$ is set to $10^{-4}$ and the BER threshold per subcarrier, ${\rm{BER}}_{{\rm{th}},i}$, $i = 1, ..., N$, is additionally set to $10^{-4}$. A Rayleigh fading channel with average channel power gain of 1
is considered. Representative results are presented in this section, which were obtained through Monte Carlo trials for $10^{4}$ channel realizations. Unless otherwise mentioned, $b_{i,\rm{max}} = 6$ and equal importance is considered for the transmit power and the throughput objectives, i.e., $\alpha = 0.5$.
For GA, the population size is set to 100 individuals, the maximum number of generations is 1500, and the change in the objective function threshold is $10^{-12}$. The number of the elite children is set to $0.05 \times \min(\max(10 (2 N), 40), 100) = 5$ children \cite{deep2009real}, the crossover probability is set to 0.8, i.e., the number of crossover children is $0.8\times(100-5) = 76$, and the number of the mutation children is $100 - 5 - 76 = 19$ children.

\subsection{Performance of the Proposed Algorithms}
\begin{figure}[!t]
	\centering
		\includegraphics[width=0.75\textwidth]{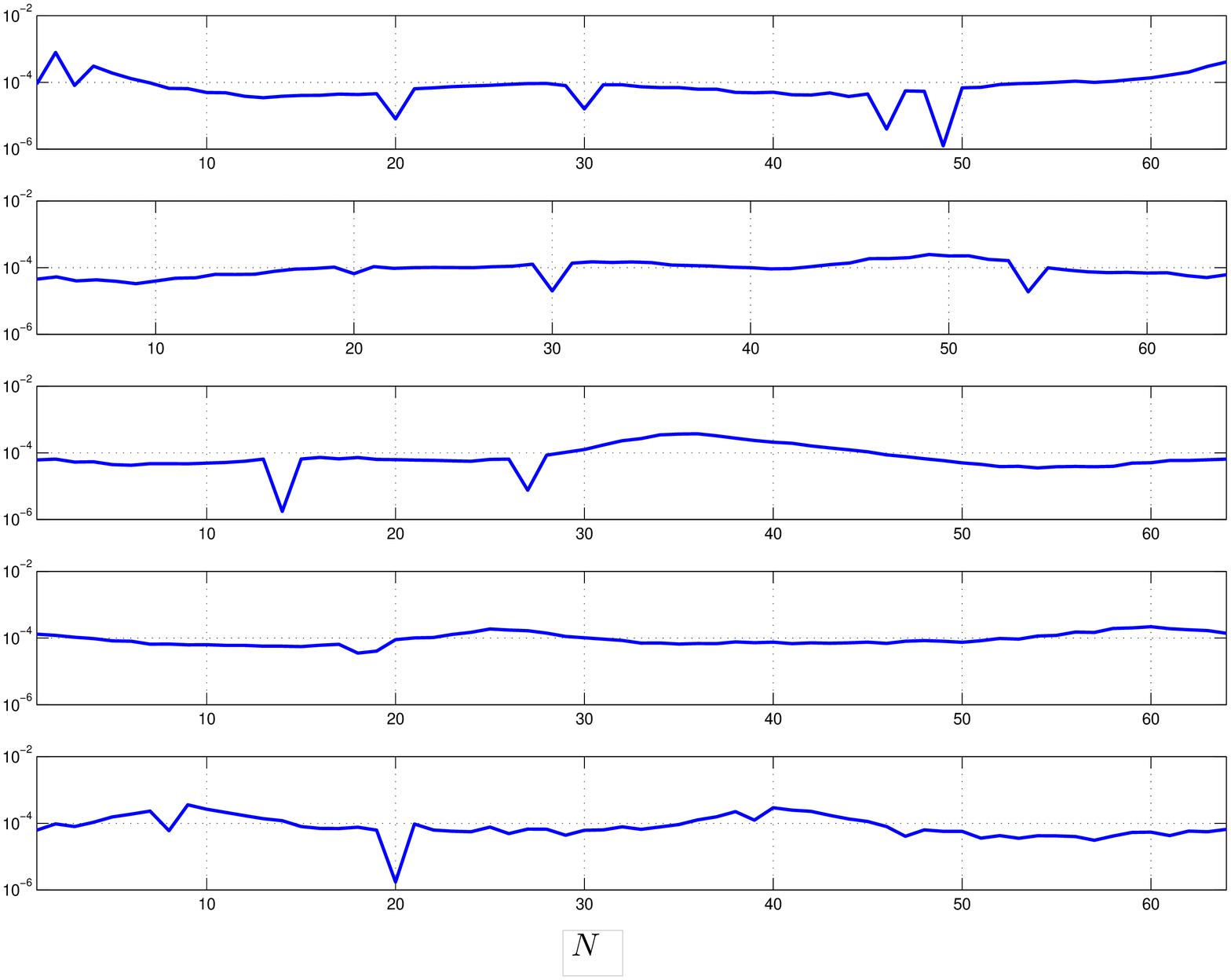}
	\caption{BER per subcarrier allocation of $\mathcal{OP}1$ for random channel realizations at $\alpha = 0.5$ and $\gamma_{\rm{av}} = 30$ dB.}
	\label{fig:real_new}
\end{figure}
Fig. \ref{fig:real_new} illustrates the BER per subcarrier resulting from the proposed algorithm to solve $\mathcal{OP}1$ using GA for different channel realizations for $\gamma_{\rm{av}} = 30$ dB\footnote{The average channel gain $\gamma_{\rm{av}}$ is calculated by averaging the instantaneous channel gain values per subcarrier over the total number of subcarriers and the total number of channel realizations, respectively.} and $\alpha = 0.5$. As can be seen, the resulting BER per subcarrier fluctuates around $10^{-4}$, and hence, the approximation optimization problem $\mathcal{OP}2$ is an acceptable reformulation for $\mathcal{OP}1$.

\begin{figure}[!t]
	\centering
		\includegraphics[width=0.75\textwidth]{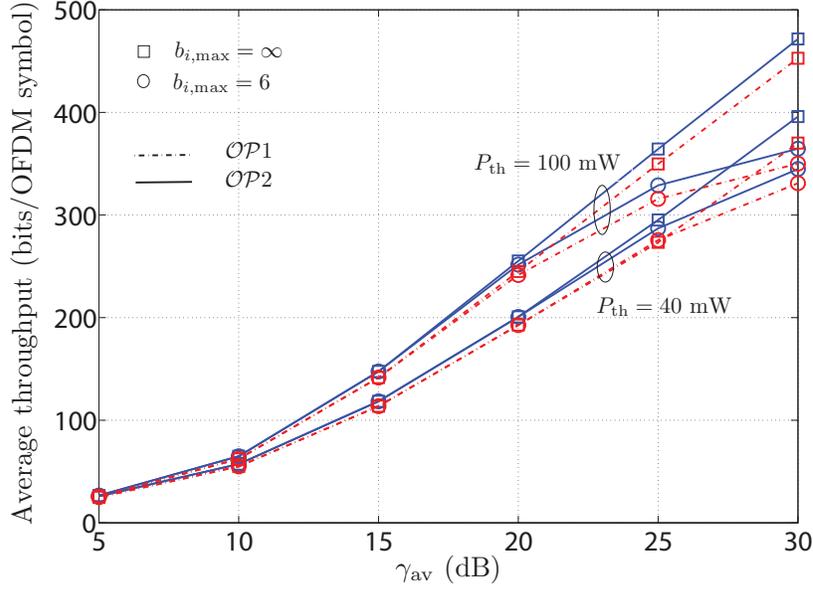}
	\caption{Average throughput of $\mathcal{OP}1$ and $\mathcal{OP}2$ as a function of $\gamma_{\rm{av}}$ at $\alpha = 0.5$.}
	\label{fig:Th_b_p}
\end{figure}
\begin{figure}[!t]
	\centering
		\includegraphics[width=0.75\textwidth]{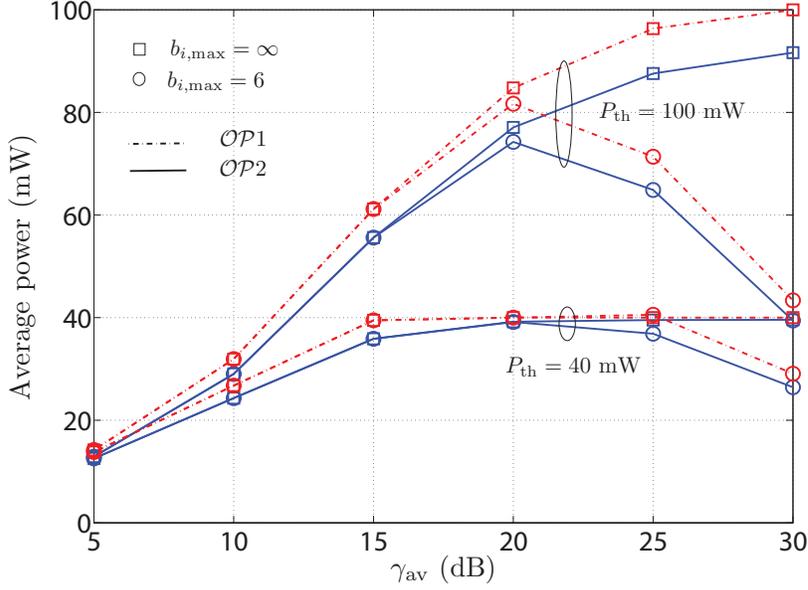}
	\caption{Average power of $\mathcal{OP}1$ and $\mathcal{OP}2$ as a function of $\gamma_{\rm{av}}$ at $\alpha = 0.5$.}
	\label{fig:Power_b_p}
\end{figure}

Fig. \ref{fig:Th_b_p} depicts the average throughput of $\mathcal{OP}1$ and $\mathcal{OP}2$ as a function of $\gamma_{\rm{av}}$, for different values of $P_{\rm{th}}$, $b_{i,\rm{max}}$ and $\alpha = 0.5$. As expected, for both $\mathcal{OP}1$ and $\mathcal{OP}2$, increasing the value of $\gamma_{\rm{av}}$ increases the average throughput. Additionally, the achieved throughput is higher at $P_{\rm{th}} = 100$ mW when compared to the throughput achieved at $P_{\rm{th}} = 40$ mW. For both $b_{i,\rm{max}} = 6$ and $\infty$, the same throughput is achieved for lower $\gamma_{\rm{av}}$ values; this is because for lower $\gamma_{\rm{av}}$, the proposed algorithms tend to allocate a number of bits lower than $b_{i,\rm{max}} = 6$. However, for high $\gamma_{\rm{av}}$, i.e., when the proposed algorithms allocate a higher number of bits, the throughput reduces for the constraint of $b_{i,\rm{max}} = 6$. The slightly reduced throughput of $\mathcal{OP}1$ compared to $\mathcal{OP}2$ is due to the fact that the average value of the BER for $\mathcal{OP}1$ is less than $10^{-4}$ (the average BER is $10^{-4}$, $9.96\times 10^{-5}$, $9.01\times 10^{-5}$, $8.46\times10^{-5}$, $8.96\times10^{-5}$, and $9.42\times10^{-5}$ for $\gamma_{\rm{av}} = 5, 10, 15, 20, 25$, and $30$ dB, respectively) while the average value of the BER for $\mathcal{OP}2$ is always equal to $10^{-4}$. The improvements of the average BER come at the expense of reduced throughput and increased transmit power (as shown in Fig. \ref{fig:Power_b_p}).

In Fig. \ref{fig:Power_b_p}, the average power of $\mathcal{OP}1$ and $\mathcal{OP}2$ is plotted as a function of $\gamma_{\rm{av}}$, for different values of $P_{\rm{th}}$, $b_{i,\rm{max}}$ and $\alpha = 0.5$. Similar to the discussion of Fig. \ref{fig:Th_b_p}, the average power increases with $\gamma_{\rm{av}}$. Additionally, the average power of $\mathcal{OP}1$ is slightly higher when compared to $\mathcal{OP}2$; this is because the average BER of $\mathcal{OP}1$ is less than $10^{-4}$ and the average BER of $\mathcal{OP}2$ is equal to $10^{-4}$. For both $\mathcal{OP}1$ and $\mathcal{OP}2$ at $b_{i,\rm{max}} = 6$, it is worthy to mention that the average power drops at higher values of $\gamma_{\rm{av}}$. This can be explained as follows: while higher average channel gains $\gamma_{\rm{av}}$ imply an increase in the throughput, this is actually limited by the constraint of $b_{i,\rm{max}} = 6$ and it will not increase beyond certain values. Hence, the improvements of the channel gain $\gamma_{\rm{av}}$ translate into a reduction in the transmit power. As the performance of the proposed algorithm to solve $\mathcal{OP}1$ is comparable to its counterpart of $\mathcal{OP}2$, in the rest of this section we focus our discussion on the performance of $\mathcal{OP}2$.

\begin{figure}[!t]
	\centering
		\includegraphics[width=0.75\textwidth]{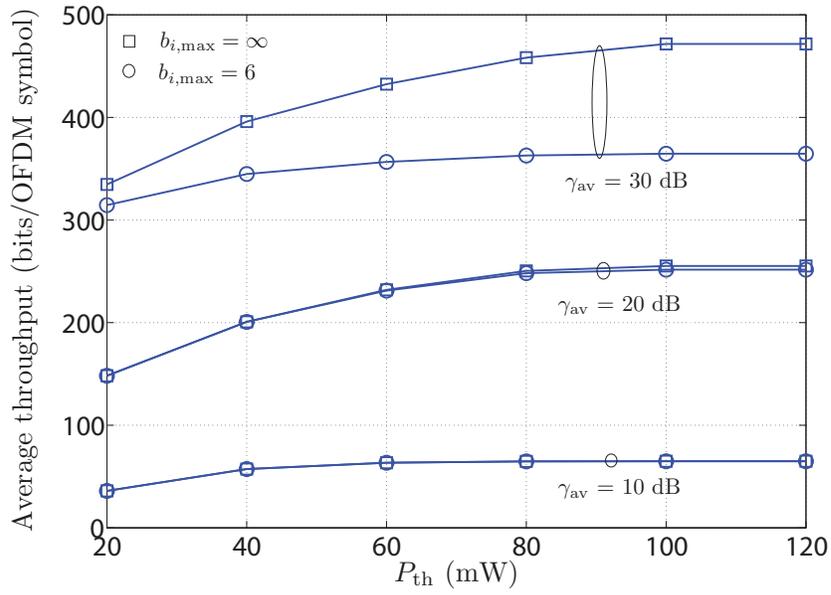}
	\caption{Average throughput of $\mathcal{OP}2$ as a function of $P_{\rm{th}}$ at $\alpha = 0.5$.}
	\label{fig:Th_P}
\end{figure}

\begin{figure}[!t]
	\centering
		\includegraphics[width=0.75\textwidth]{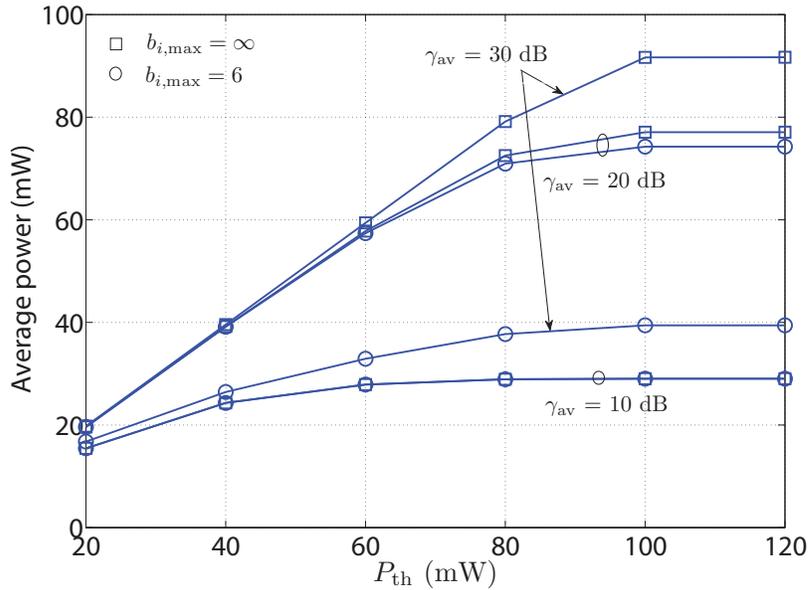}
	\caption{Average power of $\mathcal{OP}2$ as a function of $P_{\rm{th}}$ at $\alpha = 0.5$.}
	\label{fig:Power_P}
\end{figure}
Figs. \ref{fig:Th_P} and \ref{fig:Power_P} show the average throughput and power of $\mathcal{OP}2$, respectively, as a function of the power threshold $P_{\rm{th}}$, for different values of $\gamma_{\rm{av}}$ and $b_{i,\rm{max}}$, and with $\alpha = 0.5$. As can be seen, the average throughput and power increase as $P_{\rm{th}}$ increases, and saturates for higher values of $P_{\rm{th}}$. This can be explained as follows. For lower values of $P_{\rm{th}}$, the total transmit power, and hence, the throughput are restricted by this threshold value, while increasing $P_{\rm{th}}$ results in a corresponding increase in both the  throughput and total transmit power. For higher values of $P_{\rm{th}}$, the total transmit power is always less than the threshold value, and thus, it is as if the constraint on the total transmit power is actually inactive/relaxed. In this case, the proposed algorithm essentially minimizes the transmit power by keeping it constant; consequently, the average throughput remains constant. At $\gamma_{\rm{av}} = 10$ and $20$ dB, the average throughput and power exhibit the same performance for both $b_{i,\rm{max}} = 6$ and $\infty$. This is as at low $\gamma_{\rm{av}}$, the allocated bits are lower than the maximum value $b_{i,\rm{max}} = 6$, and hence, it is as if the maximum bit constraint is relaxed. However, for $\gamma_{\rm{av}} = 30$ dB the maximum allocated bits are limited by $b_{i,\rm{max}} = 6$ and the improvement in the average channel gain $\gamma_{\rm{av}}$ translates into a reduction in the average transmit power, and, thus, throughput.

\begin{figure}[!t]
	\centering
		\includegraphics[width=0.75\textwidth]{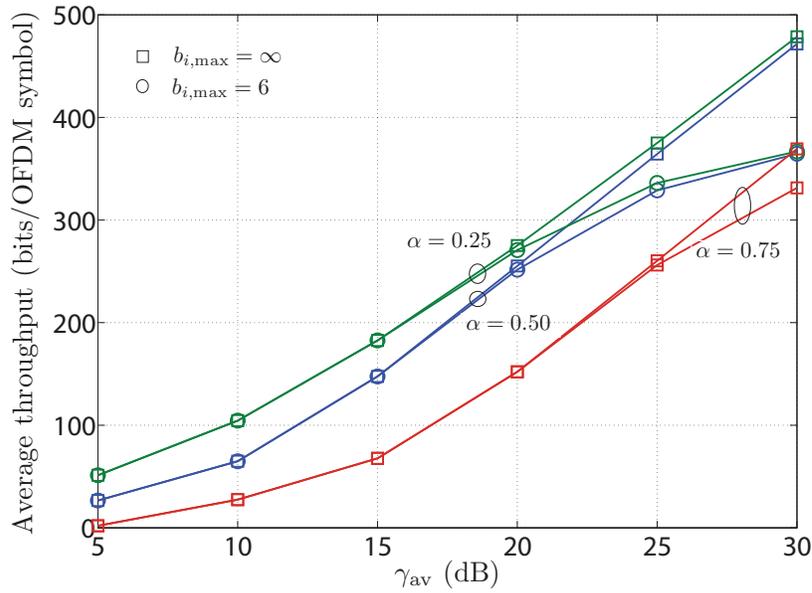}
	\caption{Average throughput of $\mathcal{OP}2$ as a function of $\gamma_{\rm{av}}$ at $P_{\rm{th}} = 100$ mW.}
	\label{fig:Th_alpha}
\end{figure}

\begin{figure}[!t]
	\centering
		\includegraphics[width=0.75\textwidth]{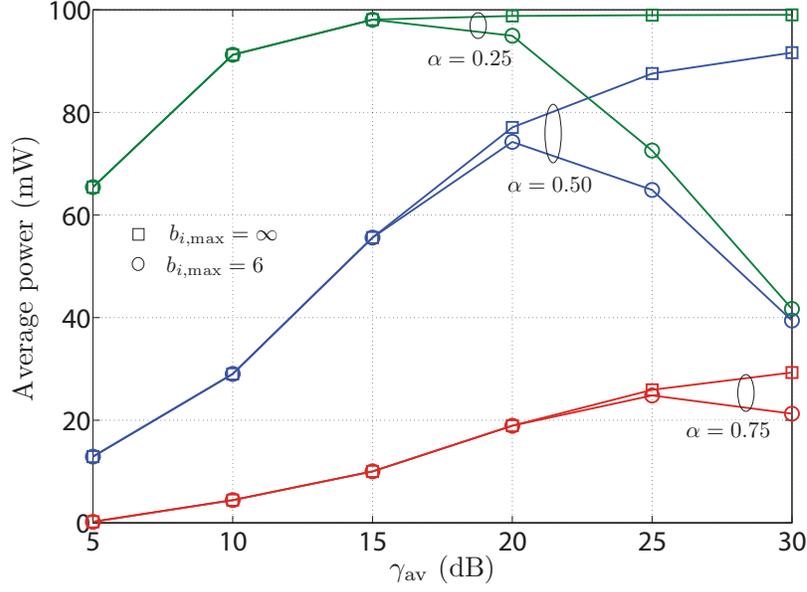}
	\caption{Average power of $\mathcal{OP}2$ as a function of $\gamma_{\rm{av}}$ at $P_{\rm{th}} = 100$ mW.}
	\label{fig:Power_alpha}
\end{figure}
The effect of $\alpha$ on the average throughput and power of $\mathcal{OP}2$ is depicted in Figs. \ref{fig:Th_alpha} and \ref{fig:Power_alpha}, respectively, for $P_{\rm{th}} = 100$ mW. One can see that increasing the value of $\alpha$, decreases the average throughput and power. This can be explained as follows. By increasing $\alpha$, more weight is given to the transmit power minimization, whereas less weight is given to the throughput maximization according to the problem formulation in \eqref{eq:ineq_const_op5}. As discussed earlier, the throughput is limited by the constraints on the maximum allowed bits per subcarrier, and hence, the power drops at higher values of $\gamma_{\rm{av}}$. It is worthy to mention that for lower values of $\alpha = 0.25$ and for $b_{i,\rm{max}} = \infty$, the average power is limited by $P_{\rm{th}} = 100$ mW.

\begin{figure}[!t]
	\centering
		\includegraphics[width=0.75\textwidth]{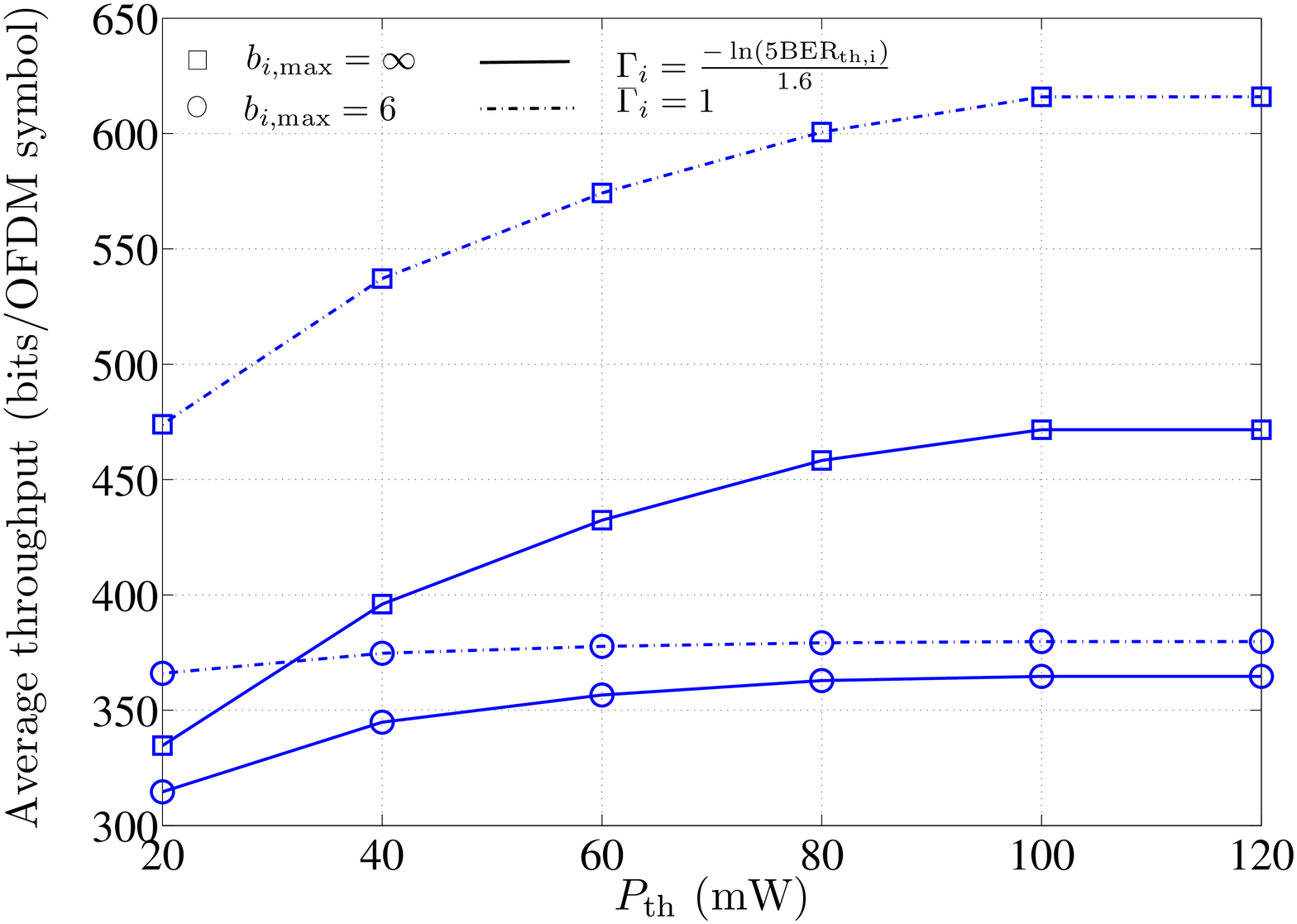}
	\caption{Effect of perfect coding on the average throughput of $\mathcal{OP}2$.}
	\label{fig:Th_P_max_rate}
\end{figure}

\begin{figure}[!t]
	\centering
		\includegraphics[width=0.75\textwidth]{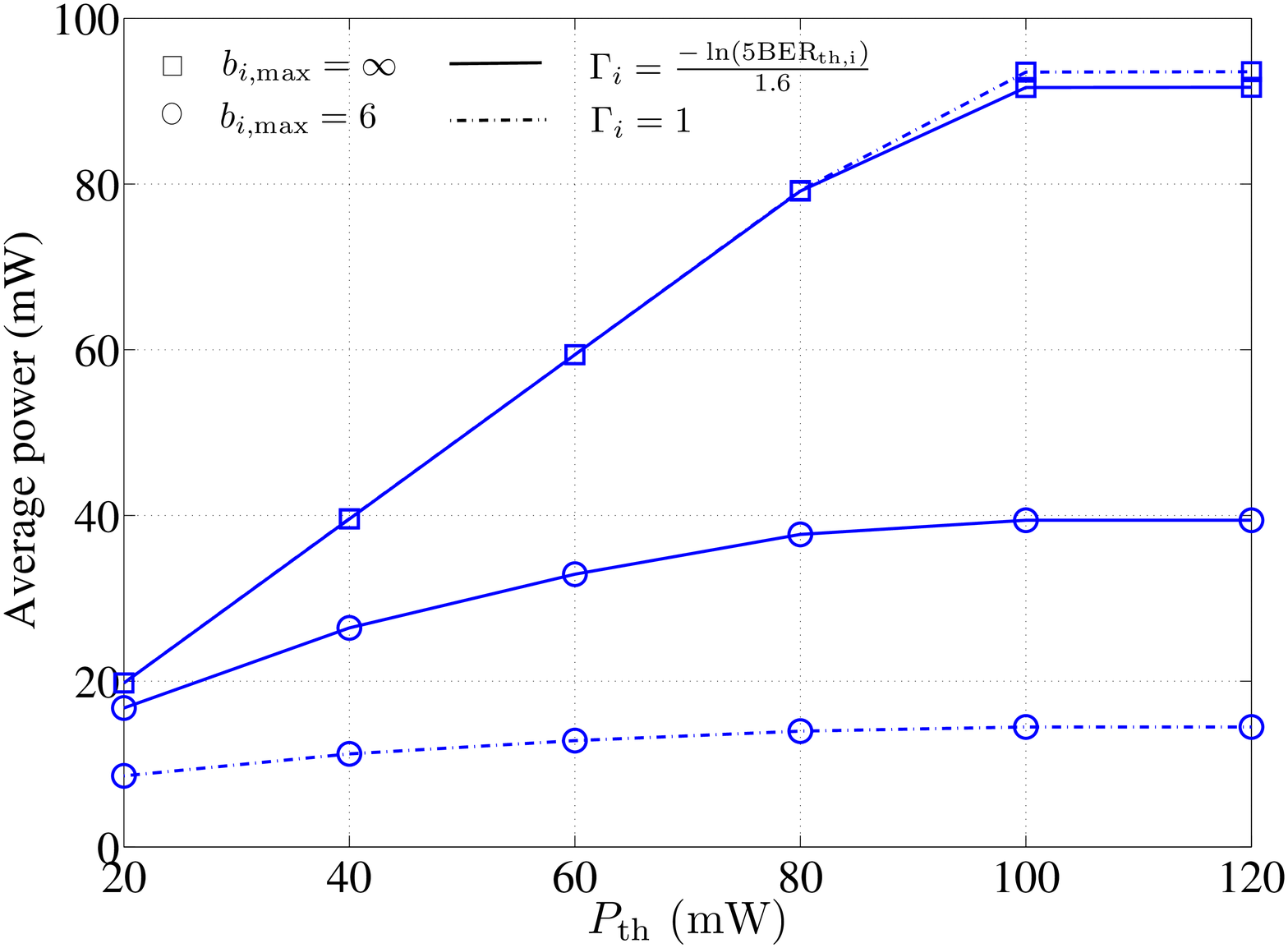}
	\caption{Effect of perfect coding on the average power of $\mathcal{OP}2$.}
	\label{fig:Power_P_max_rate}
\end{figure}
Figs. \ref{fig:Th_P_max_rate} and \ref{fig:Power_P_max_rate} characterize the gap between the performance of the proposed algorithm to solve $\mathcal{OP}2$ and its counterpart of the coded OFDM system at $\gamma_{\rm{av}} = 30$ dB. We assume ideal coding scheme, i.e., we set $\Gamma_i = 1$ (i.e., maximum possible rate is achieved), and the obtained performance in this case represents an upper bound of the performance when using any other practical coding scheme. For $b_{i,\rm{max}} = \infty$, introducing coding improves the achievable rate considerably at the same power levels. However, for $b_{i,\rm{max}} = 6$ the throughput cannot increase beyond this constraint (see Fig. \ref{fig:Th_P_max_rate}), and hence, the transmit power reduces considerably in the coded case (see Fig. \ref{fig:Power_P_max_rate}).

\subsection{Performance and Complexity Comparison}
\begin{figure}
	\centering
		\includegraphics[width=0.75\textwidth]{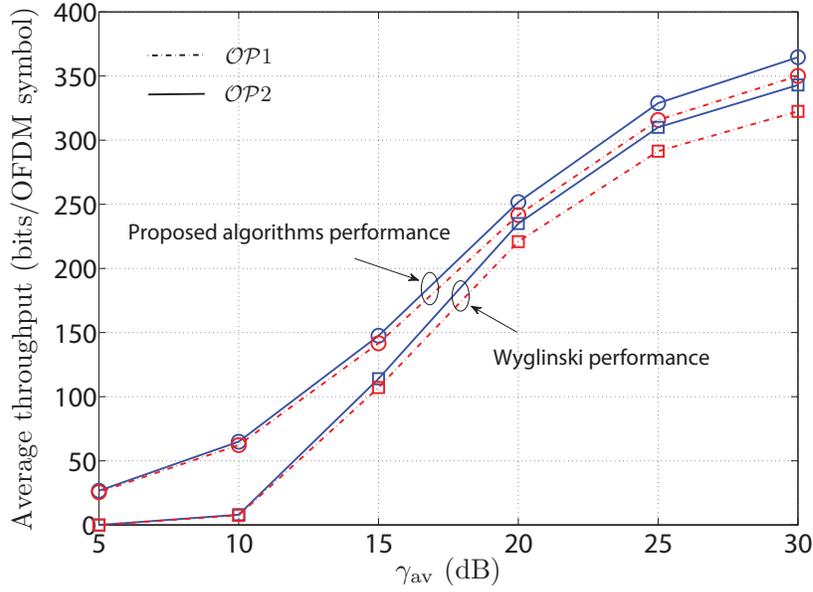}
	\caption{Average throughput as a function of $\gamma_{\rm{av}}$ for the proposed algorithms and Wyglinski's algorithm \cite{wyglinski2005bit}, at $P_{\rm{th}} = 100$ mW and $b_{i,\rm{max}} = 6$.}
	\label{fig:throughput}
\end{figure}
In Fig. \ref{fig:throughput}, the throughput achieved by the proposed algorithms to solve $\mathcal{OP}1$ and $\mathcal{OP}2$ is compared to that obtained by Wyglinski's algorithm \cite{wyglinski2005bit} for the same operating conditions. To make a fair comparison, the uniform power loading used by the loading scheme in \cite{wyglinski2005bit} is computed by dividing the average transmit power allocated by the proposed algorithms to the total number of subcarriers. As can be seen from Fig.~\ref{fig:throughput}, the proposed algorithms provide a higher throughput than the scheme in \cite{wyglinski2005bit} within the low to average SNR range. This result demonstrates that optimal loading of transmit power is crucial for low power budgets.

\begin{figure}[!t]
	\centering
		\includegraphics[width=0.75\textwidth]{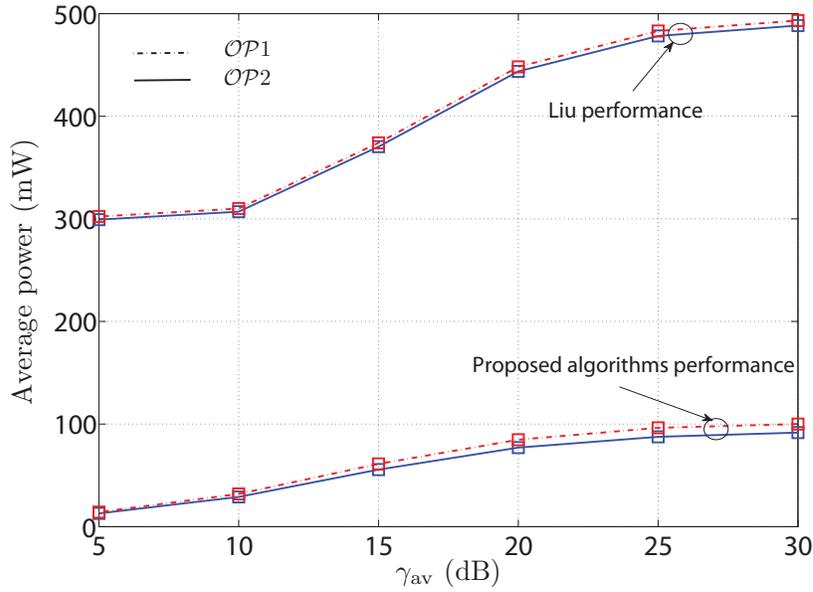}
	\caption{Average power as a function of $\gamma_{\rm{av}}$ for the proposed algorithms and Liu's algorithm \cite{liu2009adaptive}, at $b_{i,\rm{max}} = \infty$.}
	\label{fig:power}
\end{figure}

Fig. \ref{fig:power} compares the average transmit power obtained by the proposed algorithms to solve $\mathcal{OP}1$ and $\mathcal{OP}2$ with the optimum power allocation of Liu \emph{et al.} \cite{liu2009adaptive}. To ensure that the same operating conditions are considered, the fixed bit allocation per subcarrier for Liu's algorithm is set by dividing the average throughput of the proposed algorithms to the total number of subcarriers. As can be seen from Fig.~\ref{fig:power}, the proposed algorithms assign significantly less average power than both schemes in \cite{liu2009adaptive} to achieve the same average BER and average throughput.

Based on the algorithm description in Section \ref{sec_Ch_1:opt}, the computational complexity of the proposed algorithm to solve $\mathcal{OP}2$ is of  $\mathcal{O}(N)$ for inactive power constraint (which is similar to that of Liu's algorithm) and of $\mathcal{O}(N^2)$ for active power constraint (which is similar to that of Wyglinski's algorithm).

\section{Conclusion} \label{sec_Ch_1:conc}
In this paper, we proposed a new formulation for the bit and power allocation problem for OFDM systems. This is a MOOP formulation that simultaneously maximizes the throughput and minimizes the transmit power subject to QoS, total transmit power, and maximum allocated bits per subcarrier constraints. The formulated MOOP was non-convex and solved by using a GA. An approximate convex optimization problem is additionally introduced, with the global optimality guaranteed for the Pareto optimal set. Simulation results showed that the proposed algorithms outperform various allocation schemes in the literature, that separately maximize the throughput or minimize the transmit power, with similar computational effort.

\section*{Appendix A\\Proof of the Optimality of ($\mathbf{b}^*,\mathbf{p}^*$) of $\mathcal{OP}2$}
The KKT conditions are written as \cite{Boyd2004convex}
\setlength{\arraycolsep}{0.0em}
\begin{subequations}
\label{eq:KKT_D}
\begin{IEEEeqnarray}{rCl}
\frac{\partial \mathcal{F}_{\mathcal{OP}2}}{\partial b_{i}} +  \sum_{\varrho = 1}^{N+1} \lambda_{\varrho} \: \frac{\partial G_{\varrho}}{\partial b_{i}} &{} = {}& 0, \label{eq:KH2_5}\\
G_{\varrho} \lambda_{\varrho} &{}={}& 0, \label{eq:KH4_5} \\
G_{\varrho} &{} \leq {}& 0, \label{eq:KH5_5} \\
\lambda_{\varrho} &{} \geq {}& 0, \label{eq:KH3_5}
\end{IEEEeqnarray}
\end{subequations}
$i = 1, ..., N$ and $\varrho = 1, ..., N+1$. One can show that these conditions are satisfied, as sketched in the proof below.
\begin{itemize}
\item \textit{Proof of}  (\ref{eq:KH2_5}): one can find that (\ref{eq:KH2_5}) are satisfied from (\ref{eq:eq2_op5}) directly.
\item \textit{Proof of}  (\ref{eq:KH4_5}):
\begin{enumerate}
  \item Either $\lambda_{\varrho} = 0$, $\varrho = 1, ..., N+1$ (as in case \emph{\MakeUppercase{\romannumeral 1}}); hence, $G_{\varrho} \lambda_{\varrho} = 0$.
  \item Either $y_{\varrho} = 0$, $\varrho = 1, ..., N+1$, so $\mathcal{G}_{\varrho} = G_{\varrho} = 0$ from (\ref{eq:eq_const_op5}) (as in case \emph{\MakeUppercase{\romannumeral 4}}); hence, $G_{\varrho} \lambda_{\varrho} = 0$.
  \item Either: $\lambda_{\varrho_{x}} = 0$, $\varrho_{x} \in \{\varrho = 1, ..., N+1\}$ and $y_{\varrho_{y}} = 0$, $\varrho_{y} \in \{\varrho = 1, ..., N+1\}$, $\varrho_{x} \neq \varrho_{y}$ (as in cases \emph{\MakeUppercase{\romannumeral 2}}---\emph{\MakeUppercase{\romannumeral 3}}), hence, $\mathcal{G}_{\varrho_{y}} = G_{\varrho_{y}}$ from (\ref{eq:eq_const_op5})). Thus, $G_{\varrho} \lambda_{\varrho} = 0$, $\varrho = 1, ..., N+1$.
\end{enumerate}
\item \textit{Proof of}  (\ref{eq:KH5_5}): adding non-negative slack variables in \eqref{eq:eq_const_op5} guarantees that $G_\varrho \leq 0$; hence, \eqref{eq:KH5_5} is always satisfied.
\item \textit{Proof of} (\ref{eq:KH3_5}):
\begin{enumerate}
  \item In case \emph{\MakeUppercase{\romannumeral 2}}, from (\ref{eq:eq33_op5}) $b_i = b_{i,\rm{max}}$ and from (\ref{eq:eq2_op5}) $\lambda_i = \frac{1-\alpha}{u_{\rm{b}}} - \frac{\alpha}{u_{\rm{p}}} \ln(2) \frac{- \ln(5 {\rm{BER}}_{{\rm{th}},i})}{1.6 \: \gamma_i} 2^{b_{i,\rm{max}}}$. To satisfy (\ref{eq:KH3_5}), $\lambda_i$ should be greater than or equal to 0, i.e., $\frac{1-\alpha}{u_{\rm{b}}} - \frac{\alpha}{u_{\rm{p}}} \ln(2)$ $\frac{- \ln(5 {\rm{BER}}_{{\rm{th}},i})}{1.6 \: \gamma_i} 2^{b_{i,\rm{max}}}$ $\geq 0$ which leads to $\gamma_i \geq \gamma_{{\rm{th}},i}^{\rm{max}} = \frac{1}{1.6} \frac{\frac{\alpha}{u_{\rm{p}}} \ln(2)}{\frac{1-\alpha}{u_{\rm{b}}}} (- \ln(5{\rm{BER}}_{{\rm{th}},i})) \: 2^{b_{i,\rm{max}}}$, $i = 1, ..., N$. It is worthy to note that $\lambda_{N+1} = 0$ by definition of case \emph{\MakeUppercase{\romannumeral 2}}, and, hence, (\ref{eq:KH3_5}) is always satisfied.
  \item In case \emph{\MakeUppercase{\romannumeral 4}}, from (\ref{eq:eq33_op5}) $b_i = b_{i,\rm{max}}$ and  $\lambda_i = \frac{1-\alpha}{u_{\rm{b}}} - (\frac{\alpha}{u_{\rm{p}}} + \lambda_{N+1}) \ln(2) \frac{- \ln(5 {\rm{BER}}_{{\rm{th}},i})}{1.6 \: \gamma_i} 2^{b_{i,\rm{max}}}$ from (\ref{eq:eq2_op5}). In order to satisfy (\ref{eq:KH3_5}), $\lambda_i$ should be greater than or equal to 0, i.e., $\frac{1-\alpha}{u_{\rm{b}}} - (\frac{\alpha}{u_{\rm{p}}} + \lambda_{N+1}) \ln(2) \frac{- \ln(5 {\rm{BER}}_{{\rm{th}},i})}{1.6 \: \gamma_i} 2^{b_{i,\rm{max}}} \geq 0$. Hence, $\lambda_{N+1}$ can be found as $\lambda_{N+1} \leq \frac{\frac{1-\alpha}{u_{\rm{b}}}}{\ln(2)} \frac{1.6 \: \gamma_i}{(-\ln(5 {\rm{BER}}_{{\rm{th}},i}))}$  $2^{- b_{i,\rm{max}}}  - \frac{\alpha}{u_{\rm{p}}}$. From (\ref{eq:KH3_5}) $\lambda_{N+1} \geq 0$, and, thus, $\frac{\frac{1-\alpha}{u_{\rm{b}}}}{\ln(2)} \frac{1.6 \: \gamma_i}{(-\ln(5 {\rm{BER}}_{{\rm{th}},i}))}  2^{- b_{i,\rm{max}}}   - \frac{\alpha}{u_{\rm{p}}} \geq \lambda_{N+1} \geq 0$ which leads to $\gamma_i \geq \gamma_{{\rm{th}},i}^{\rm{max}} =  \frac{1}{1.6} \frac{(\frac{\alpha}{u_{\rm{p}}} ) \ln(2)}{\frac{1-\alpha}{u_{\rm{b}}}} (- \ln(5{\rm{BER}}_{{\rm{th}},i})) \: 2^{b_{i,\rm{max}}},$ $i = 1, ..., N$, and, hence, (\ref{eq:KH3_5}) is always satisfied.
\end{enumerate}
\end{itemize}

As can be seen, the KKT conditions are satisfied; thus, based on this result and the convexity of $\mathcal{OP}2$, the solution ($\mathbf{b}^*,\mathbf{p}^*$) represents a global optimum point. \hfill$\blacksquare$ 
\bibliographystyle{IEEEtran}
\bibliography{IEEEabrv,mybib_file} 

\chapter{} \label{ch:WCL}
\section{Abstract}
In this letter, a novel low complexity bit and power loading algorithm is formulated for multicarrier communication systems. The proposed algorithm jointly maximizes the throughput and minimizes the transmit power through a weighting coefficient $\alpha$, while meeting constraints on the target bit error rate (BER) per subcarrier and on the total transmit power. The optimization problem is solved by the Lagrangian multiplier method if the initial $\alpha$ causes the transmit power not to violate the power constraint; otherwise, a bisection search is used to find the appropriate $\alpha$. Closed-form expressions are derived for the close-to-optimal bit and power allocations per subcarrier, average throughput, and average transmit power.  Simulation results illustrate the performance of the proposed algorithm and demonstrate its superiority with respect to existing allocation algorithms. Furthermore, the results show that the performance of the proposed algorithm approaches that of the exhaustive search for the discrete optimal allocations. 
\section{Introduction}

Multicarrier modulation is recognized as a robust and efficient transmission technique, as evidenced by its consideration for diverse communication systems and adoption by several wireless standards \cite{fazel2008multi}. The performance of multicarrier communication systems can be significantly improved by dynamically adapting the transmission parameters, such as power, constellation size, symbol rate, coding rate/scheme, or any combination of these, according to the channel quality or the wireless standard specifications \cite{wyglinski2005bit,  liu2009adaptive, mahmood2010efficient, willink1997optimization, bedeer2011partial, bedeer2012jointVTC, bedeer2012EBERGC, bedeer2012UBERGC, bedeer2012novelICC, bedeer2012adaptiveRWS, bedeer2013resource, bedeer2013adaptive, bedeer2013novel, bedeer2014rateCONF, bedeer2014multiobjective, bedeer2013joint, 
bedeer2014energy, bedeer2015systematic, bedeer2015rate}. 

To date, most of the research literature has focused on the single objective of either maximizing the throughput or minimizing the transmit power separately (see, e.g., \cite{wyglinski2005bit,  liu2009adaptive, willink1997optimization, mahmood2010efficient} and references therein). In \cite{wyglinski2005bit}, Wyglinski \textit{et al.}  proposed an incremental bit loading algorithm with uniform power in order to maximize the throughput while guaranteeing a target BER. Liu \textit{et al.} \cite{liu2009adaptive} proposed a power loading algorithm with uniform bit loading that aims to minimize the transmit power while guaranteeing a target~BER. In \cite{mahmood2010efficient}, Mahmood and Belfiore proposed an efficient greedy bit allocation algorithm that minimizes the transmit power subject to fixed throughput and BER per subcarrier constraints.

In emerging wireless communication systems, various requirements are needed. For example, maximizing the throughput is favoured if sufficient guard bands exist to separate users, while minimizing the transmit power is prioritized when operating in interference-prone shared spectrum environments, to prolong the battery life time of battery-operated nodes, as well as to support environmentally-friendly transmission behaviors.
This motivates us to formulate a multiobjective optimization (MOOP) problem that optimizes the conflicting and incommensurable throughput and power objectives. According to the MOOP principle, there is no solution that improves one of the objectives without deteriorating others. Therefore, MOOP produces a set of optimal solutions and it is the responsibility of the resource allocation entity to choose the most preferred optimal solution depending on its preference \cite{miettinen1999nonlinear}. A well known approach to solve MOOP problems is to linearly combine the competing objective functions into a single objective function, through weighting coefficients that reflect the required preferences \cite{miettinen1999nonlinear}. These preferences can be prescribed and fixed during the solution process (as in posteriori and priori methods) or can be changed during the solution process (interactive methods) \cite{miettinen1999nonlinear}. In this paper, we adopt an interactive approach in order to obtain a low complexity solution.

We propose a low complexity algorithm that jointly maximizes the throughput and minimizes the total transmit power, subject to constraints on the BER per subcarrier and the total transmit power. Limiting the total transmit power is crucial for a variety of reasons, e.g.,  to reflect the transmitter's power amplifier limitations, to satisfy regulatory maximum power limits, and to limit interference/ encourage frequency reuse. Moreover, including the total subcarrier power in the objective function is especially desirable, as it minimizes the transmit power when the power constraint is inactive.
 Closed-form expressions are derived for the close-to-optimal bit and power allocations, average throughput, and average transmit power. Simulation results show that the proposed algorithm outperforms existing bit and power loading schemes in the literature, while requiring similar or reduced computational effort. The results also indicate that the proposed algorithm's performance approaches that of the exhaustive search for the optimal discrete allocations, with significantly reduced computational effort.




\section{Proposed Link Adaptation Scheme} \label{sec_Ch_2:opt}
\subsection{Optimization Problem Formulation}
A multicarrier communication system decomposes the signal bandwidth into a set of $N$ orthogonal narrowband subcarriers of equal bandwidth. Each subcarrier $i$ transmits $b_i$ bits using power $p_i$, $i = 1, ..., N$. Following the common practice in the literature, a delay- and error-free feedback channel is assumed to exist between the transmitter and receiver for reporting the channel state information \cite{liu2009adaptive, willink1997optimization, mahmood2010efficient}.

In order to maximize the throughput and minimize the transmit power  subject to BER and total transmit power constraints, the optimization problem is  formulated as
\begin{IEEEeqnarray}{c}
\underset{b_i}{\textup{Maximize}} \quad \sum_{i = 1}^{N}b_i \quad \textup{and} \quad \underset{p_i}{\textup{Minimize}} \quad \sum_{i = 1}^{N}p_i, \nonumber \\
\textup{subject to} \quad  \textup{BER}_i \leq \textup{BER}_{{\rm{th}},i}, \nonumber \\
  \quad \sum_{i = 1}^{N} p_i \leq P_{\rm{th}}, \qquad i = 1, ..., N, \label{eq_Ch_2:eq_first}
\end{IEEEeqnarray}
where $\textup{BER}_i$ and $\textup{BER}_{{\rm{th}},i}$ are the BER and threshold value of BER per subcarrier\footnote{The constraint on the BER per subcarrier is a suitable formulation that results in similar BER characteristics compared to an average BER constraint, especially at high signal-to-noise ratios (SNRs)~\cite{willink1997optimization}. Further, it significantly reduces the computational complexity by yielding closed-form expressions.} $i$, $i$ = 1, ..., $N$, respectively, and $P_{\rm{th}}$ is the total transmit power threshold. An approximate expression for the BER per subcarrier $i$ for $M$-ary QAM is given by \cite{liu2009adaptive}
\begin{IEEEeqnarray}{rCl}
\textup{BER}_i &{} \approx  {}& 0.2 \: \textup{exp}\left (-1.6 \: \frac{p_i \: \gamma_i}{2^{b_i} - 1} \right ), \label{eq_Ch_2:BER}
\end{IEEEeqnarray}
where $\gamma_i = \: \frac{\left | \mathcal{H}_i \right |^2}{\sigma^2_n}$ is the channel-to-noise ratio for subcarrier $i$, $\mathcal{H}_i$ is the channel gain of subcarrier $i$, and $\sigma^2_n$ is the variance of the additive white Gaussian noise (AWGN).
The multi-objective optimization function in (\ref{eq_Ch_2:eq_first}) can be rewritten as a linear combination of multiple objective functions as follows
\begin{IEEEeqnarray}{c}
\underset{p_i , b_i}{\textup{Minimize}}  \quad  \mathbf{f}(\mathbf{p},\mathbf{b})  = \: \alpha \sum_{i = 1}^{N}p_i - (1-\alpha)\sum_{i = 1}^{N}b_i, \label{eq_Ch_2:p1} \nonumber \\
\textup{subject to} \quad  g_{\varrho}(p_i,b_i)  = \left\{\begin{matrix}
 0.2 \: \textup{exp}\left ( \frac{- 1.6 \: \gamma_i p_i}{2^{b_i} - 1} \right ) - \textup{BER}_{{\rm{th}},i} \leq 0, \quad  \varrho = 1, ..., N,  \\
\sum_{i = 1}^{N} p_i \leq P_{\rm{th}}, \hfill \varrho = N+1,
\end{matrix}\right.
 \label{eq_Ch_2:ineq_const}
\end{IEEEeqnarray}
where $\alpha$ ($0 < \alpha < 1$) is a weighting coefficient which indicates the rate at which the multicarrier system is willing to trade off the values of the objective functions in order to obtain a low complexity solution \cite{miettinen1999nonlinear} (i.e., a higher value of $\alpha$ favors minimizing the transmit power, whereas a lower value of $\alpha$ favors maximizing the throughput).
$\mathbf{p} = [p_1, ..., p_N]^T$ and $\mathbf{b} = [b_1, ..., b_N]^T$ are the \textit{N}-dimensional power and bit distribution vectors, respectively, with $[.]^T$ denoting the transpose operation.

\subsection{Bit and Power Allocations}
The optimization problem in (\ref{eq_Ch_2:ineq_const}) can be solved numerically
; however, this is computationally complex. A low complexity solution can be obtained by relaxing the power constraint in (\ref{eq_Ch_2:ineq_const}), i.e., $\varrho \neq N+1$, and then applying the method of Lagrange multipliers. Accordingly,  the inequality constraints are transformed to equality constraints by adding non-negative slack variables, $y_{i}^2$, $ \varrho = i = 1, ..., N$ \cite{rao2009engineering}. Hence, the constraints are given as
\setlength{\arraycolsep}{0.0em}
\begin{IEEEeqnarray}{RCL}
\mathcal{G}_{i}(\mathbf{p},\mathbf{b},\mathbf{y}) &{} = {}& g_{i}(\mathbf{p},\mathbf{b}) + y_{i}^2 = 0, \quad i = 1, ..., N,
\label{eq_Ch_2:eq_const}
\end{IEEEeqnarray}
where $\mathbf{y} = [y_1^2, ..., y_N^2]^T$ is the vector of slack variables, and the Lagrange function $\mathcal{L}$ is expressed as
\begin{IEEEeqnarray}{RCL}
\mathcal{L}(\mathbf{p},\mathbf{b},\mathbf{y},\boldsymbol\lambda) & = & \mathbf{f}(\mathbf{p},\mathbf{b}) + \sum_{i = 1}^{N} \lambda_{i} \: \mathcal{G}_{i}(\mathbf{p},\mathbf{b},\mathbf{y}), \nonumber \\
 & = &  \alpha \sum_{i = 1}^{N}p_i - (1-\alpha)\sum_{i = 1}^{N}b_i \hspace*{0.7cm} \nonumber \\
 &  & +  \sum_{i = 1}^{N} \lambda_{i}\:\Bigg[ 0.2 \: \textup{exp}\left ( \frac{- 1.6 \: \gamma_i p_i}{2^{b_i} - 1} \right ) - \textup{BER}_{{\rm{th}},i} + y_{i}^2\Bigg],
\end{IEEEeqnarray}
where $\boldsymbol\lambda = [\lambda_1, ..., \lambda_N]^T$ is the vector of Lagrange multipliers. A stationary point is found when $\nabla \mathcal{L}(\mathbf{p},\mathbf{b},\mathbf{y},\boldsymbol\lambda) = 0$ ($\nabla$ denotes the gradient), which yields
\begin{IEEEeqnarray}{RCL}
\frac{\partial \mathcal{L}}{\partial p_i} &{} = {}& \alpha - 0.2 \: \lambda_i \frac{1.6 \: \gamma_i}{2^{b_i}-1} \: \textup{exp}\left ( \frac{- 1.6 \: \gamma_i p_i}{2^{b_i} - 1} \right ) = 0,\label{eq_Ch_2:eq1} \\
\frac{\partial \mathcal{L}}{\partial b_i} &{} = {}& -(1 - \alpha) + 0.2 \ln (2) \: \lambda_i \frac{1.6 \: \gamma_i p_i 2^{b_i}}{(2^{b_i}-1)^2} \: \textup{exp}\left ( \frac{- 1.6 \: \gamma_i p_i}{2^{b_i} - 1} \right ) = 0,\label{eq_Ch_2:eq2}\\
\frac{\partial \mathcal{L}}{\partial \lambda_i} & = & 0.2 \: \textup{exp}\left ( \frac{- 1.6 \: \gamma_i p_i}{2^{b_i} - 1} \right ) - \textup{BER}_{{\rm{th}},i} + y_i^2= 0, \label{eq_Ch_2:eq3} \\
\frac{\partial \mathcal{L}}{\partial y_i} & = & 2\lambda_i y_i = 0. \label{eq_Ch_2:eq4}
\end{IEEEeqnarray}
It can be seen that (\ref{eq_Ch_2:eq1}) to (\ref{eq_Ch_2:eq4}) represent $4N$ equations in the $4N$ unknown components of the vectors $\mathbf{p}, \mathbf{b}, \mathbf{y}$, and $\boldsymbol\lambda$. By solving (\ref{eq_Ch_2:eq1}) to (\ref{eq_Ch_2:eq4}), one obtains the solution $\mathbf{p}^*, \mathbf{b}^*$. Equation (\ref{eq_Ch_2:eq4}) implies that either $\lambda_i$ = 0 or $y_i$ = 0; hence, two possible cases exist and we are going to investigate each case independently.

--- \textit{Case 1}: Setting $\lambda_i = 0$ in (\ref{eq_Ch_2:eq1}) to (\ref{eq_Ch_2:eq4}) results in an underdetermined system of $N$ equations in $3N$ unknowns, and, hence, no unique solution can be reached.

--- \textit{Case 2}: Setting $y_i = 0$ in (\ref{eq_Ch_2:eq1}) to (\ref{eq_Ch_2:eq4}), we can relate $p_i$ and $b_i$ from (\ref{eq_Ch_2:eq1}) and (\ref{eq_Ch_2:eq2}) as follows
\setlength{\arraycolsep}{0.0em}
\begin{IEEEeqnarray}{RCL}
p_i &{} = {}& \frac{1- \alpha}{\alpha \ln(2)}(1 - 2^{-b_i}), \label{eq_Ch_2:eq8}
\end{IEEEeqnarray}
with $p_i \geq 0$ if and only if $b_i \geq 0$. By substituting (\ref{eq_Ch_2:eq8}) into (\ref{eq_Ch_2:eq3}), one obtains the solution
\setlength{\arraycolsep}{0.0em}
\begin{IEEEeqnarray}{c}
b_i^* = \frac{1}{\log(2)}\log\Bigg[- \frac{1-\alpha }{\alpha \ln(2)} \frac{1.6 \: \gamma_i}{\ln(5 \: \textup{BER}_{{\rm{th}},i})}\Bigg]. \label{eq_Ch_2:eq10}
\end{IEEEeqnarray}
Consequently, from (\ref{eq_Ch_2:eq8}) one gets
\setlength{\arraycolsep}{0.0em}
\begin{IEEEeqnarray}{c}
p_i^* = \frac{1-\alpha }{\alpha \ln(2)}\Bigg[ 1 - \Big(- \frac{1-\alpha }{\alpha \ln(2)} \frac{1.6 \: \gamma_i}{\ln(5 \: \textup{BER}_{{\rm{th}},i})} \Big)^{-1} \Bigg]. \label{eq_Ch_2:eq12}
\end{IEEEeqnarray}
Since we consider $M$-ary QAM, $b_i$ should be greater than 2. From (\ref{eq_Ch_2:eq10}), to have $b_i \geq 2$, $\gamma_i$ must satisfy the condition
\setlength{\arraycolsep}{0.0em}
\begin{IEEEeqnarray}{c}
\gamma_i \geq \gamma_{th,i}^{{\rm{min}}} = \frac{4}{1.6} \frac{\alpha \ln(2)}{1-\alpha} (-\ln(5\textup{BER}_{{\rm{th}},i})), i = 1, ..., N. \label{eq_Ch_2:condition} \IEEEeqnarraynumspace
\end{IEEEeqnarray}

The relaxed optimization problem is not convex and, hence, the Karush-Kuhn-Tucker (KKT) conditions do not guarantee that ($\mathbf{p}^*,\mathbf{b}^*$) represents a global optimum \cite{rao2009engineering}; the proof of the KKT conditions is not provided due to the space limitations. To characterize the gap to the global optimum solution, we compare the obtained local optimum results to the global optimum results obtained through the exhaustive search in the next section.


If the total transmit power $\sum_{i = 1}^{N} p_i$ is below $P_{\rm{th}}$, then the final bit and power allocations are reached. On the other hand, if the transmit power exceeds $P_{\rm{th}}$, the algorithm adopts the interactive approach and overrides the initial value of $\alpha$ to meet the power constraint.  This is achieved by giving more weight to the transmit power minimization in (\ref{eq_Ch_2:ineq_const}), i.e., by increasing $\alpha$. The lowest $\alpha^*$ that satisfies the constraint, i.e., $\alpha^*$ that results in the highest total power which is lower than $P_{\rm{th}}$, is found through the bisection search\footnote{This is true as the total transmit power calculated from (\ref{eq_Ch_2:eq12}) is a decreasing function of $\alpha$. The proof is not provided due to the space limitations.} (please note that lower values of $\alpha$ produce lower values of the objective function in (\ref{eq_Ch_2:ineq_const})). The proposed algorithm can be formally stated as follows:

\floatname{algorithm}{}
\begin{algorithm}
\renewcommand{\thealgorithm}{}
\caption{\textbf{Proposed Algorithm}}
\begin{algorithmic}[1]
\small
\State \textbf{INPUT} The AWGN variance ($\sigma^2_n$), channel gain per subcarrier $i$ ($\mathcal{H}_i$), target BER per subcarrier $i$ ($\textup{BER}_{{\rm{th}},i}$), initial weighting parameter $\alpha$, and tolerance $\epsilon$.
\For{$i$ = 1, ..., $N$}
%
\If{$\gamma_i \geq \gamma_{th,i}^{{\rm{min}}} =  - \frac{4}{1.6} \: \frac{\alpha \ln(2)}{1-\alpha} \: \ln(5\:\textup{BER}_{{\rm{th}},i})$}
\State  - $b_i^*$ and $p_i^*$ are given by (\ref{eq_Ch_2:eq10}) and (\ref{eq_Ch_2:eq12}), respectively.
\State - $b^*_{i,{\rm{final}}}$ $\leftarrow$ Round $b_i^*$ to the nearest integer.
\State - $p^*_{i,{\rm{final}}}$ $\leftarrow$ Recalculate $p_i^*$ according to (\ref{eq_Ch_2:BER}).
\Else
\State Null the corresponding subcarrier $i$.
\EndIf
\EndFor
\While{$\sum_{i = 1}^{N}p^*_{i,{\rm{final}}} - P_{\rm{th}} > \epsilon$}
\State - Set $\alpha_L$ = $\alpha$ and $\alpha_U$ = 1.
\State - Set $\alpha^* = (\alpha_L + \alpha_U)/2$.
\State  - Repeat steps: 2 to 10.
\If{$\sum_{i = 1}^{N}p^*_{i,{\rm{final}}} < P_{\rm{th}}$}
\State - Set $\alpha_U$ = $\alpha^*$, then $\alpha^* = (\alpha_L + \alpha_U)/2$.
\State  - Repeat steps: 2 to 10.
\Else
\State - Set $\alpha_L$ = $\alpha^*$, then $\alpha^* = (\alpha_L + \alpha_U)/2$.
\State  - Repeat steps: 2 to 10.
\EndIf
\EndWhile
\State \textbf{OUTPUT} $b^*_{i,{\rm{final}}}$ and $p^*_{i,{\rm{final}}}$, $i$ = 1, ..., $N$.
\end{algorithmic}
\end{algorithm}

\subsection{Analytical Expressions of Average Throughput and Transmit Power}
When the initial value of $\alpha$ results in an inactive power constraint, the closed-form expressions for the average throughput and transmit power can be found by averaging the bit and power allocations given by (\ref{eq_Ch_2:eq10}) and (\ref{eq_Ch_2:eq12}), respectively, over $\gamma_i$. In such a case, the average throughput is expressed as
\begin{IEEEeqnarray}{RCL}
\textup{Throughput}_\textup{av}&{} = {}& \sum_{i = 1}^{N} \mathbb{E}\{b_i(\gamma_i)\} \nonumber \\
 &{} = {}& \sum_{i = 1}^{N} \int_{\gamma_{th,i}^{{\rm{min}}}}^{\infty} b_i(\gamma_i)\Big[\nu \textup{exp}(-\nu \gamma_i)\Big] d\gamma_i, \label{eq_Ch_2:av_b} \IEEEeqnarraynumspace
\end{IEEEeqnarray}
where $\nu \textup{exp}(-\nu \gamma_i)$ is the exponential distribution of~$\gamma_i$ with mean $\frac{1}{\nu}$, given that the channel gain $\mathcal{H}_i$ has a Rayleigh distribution. The integration in (\ref{eq_Ch_2:av_b}) is solved by parts yielding
\begin{IEEEeqnarray}{c}
\textup{Throughput}_\textup{av} = \sum_{i = 1}^{N} \frac{1}{\textup{log}(2)} \Bigg[\textup{log}(4) \: \textup{exp}(-\nu \gamma_{th,i}^{{\rm{min}}})  - \frac{\textup{Ei}(-\nu \gamma_{th,i}^{{\rm{min}}})}{\textup{ln}(10)} \Bigg], \label{eq_Ch_2:av_th}
\end{IEEEeqnarray}
where $\textup{Ei}(-z) = - \int_{z}^{\infty}\frac{\textup{e}^{-t}}{t} \: dt, \: z > 0$ is the exponential integral function. Similarly, the average transmit power is given by
\begin{IEEEeqnarray}{c}
\textup{Power}_\textup{av} = \sum_{i = 1}^{N} \frac{1-\alpha}{\alpha \: \textup{ln}(2)} \Bigg[ \textup{exp}(-\nu \gamma_{th,i}^{{\rm{min}}}) + \frac{\nu \: \gamma_{th,i}^{{\rm{min}}}}{4} \textup{Ei}(-\nu \gamma_{th,i}^{{\rm{min}}}) \Bigg]. \label{eq_Ch_2:av_power}
\end{IEEEeqnarray}



\section{Simulation Results} \label{sec_Ch_2:sim}
This section investigates the performance of the proposed algorithm, and compares its performance with bit and power loading algorithms presented in the literature, as well as with the exhaustive search for the discrete global optimal allocations. The computational complexity of the proposed algorithm is also compared to the other schemes.

\subsection{Simulation Setup}
As an example of a multicarrier system, we consider orthogonal frequency division multiplexing (OFDM) with $N$ = 128 subcarriers. Without loss of generality, the BER constraint per subcarrier, $\textup{BER}_{{\rm{th}},i}$, is assumed to be the same for all subcarriers and set to $10^{-4}$. A Rayleigh fading environment with average channel power gain $\mathbb{E}\{\left | \mathcal{H}_i \right |^2\}$ = 1 is considered.
Representative results are presented, which were obtained through Monte Carlo trials for $10^{4}$ channel realizations with
$\epsilon = 10^{-9}$ mW and initial $\alpha = 0.5$. The transmit power objective function is scaled during simulations so that it is approximately within the same range as the throughput \cite{miettinen1999nonlinear}. For convenience, presented numerical results are displayed in the original scales.

\subsection{Performance of the Proposed Algorithm}

\begin{figure}[t]
	\centering
		\includegraphics[width=0.75\textwidth]{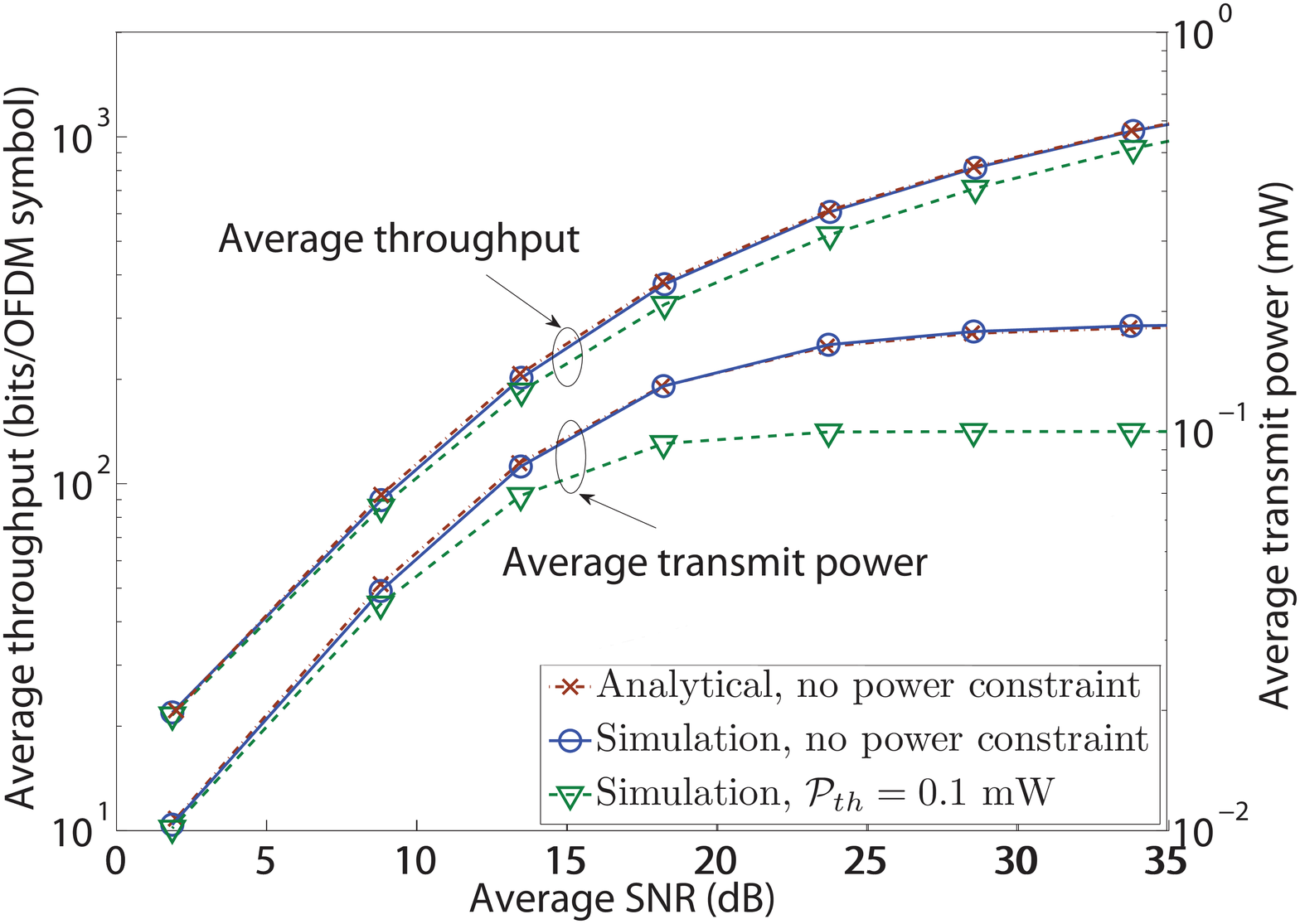}
	\caption{Average throughput and average transmit power as a function of average SNR, with and without a power constraint.}
	\label{fig_Ch_2:proposed}
\end{figure}

Fig. \ref{fig_Ch_2:proposed} depicts the average throughput and transmit power as a function of the average SNR\footnote{The average SNR is calculated by averaging the instantaneous SNR values
per subcarrier over the total number of subcarriers and the total number of
channel realizations, respectively.}, with and without considering the total power constraint. In the latter case, the average throughput and transmit power, obtained by averaging (\ref{eq_Ch_2:eq10}) and  (\ref{eq_Ch_2:eq12}), respectively, over the total number of channel realizations through Monte Carlo simulations, show an excellent match to their counterparts in (\ref{eq_Ch_2:av_th}) and (\ref{eq_Ch_2:av_power}), respectively.  Further, for an average SNR $\leq$ 24 dB, one finds that both the average throughput and transmit power increase as the SNR increases, whereas for an average SNR $\geq$ 24 dB, the transmit power saturates while the throughput continues to increase. This observation can be explained as follows. The relation between $b_i$ and $p_i$ in (\ref{eq_Ch_2:eq8}) implies that increasing the number of bits at the low range of $b_i$ (that exists at low average SNR values) occurs at the expense of additional transmit power, while increasing the number of bits at the high range of $b_i$ (that exists at high average SNR values) occurs at negligible increase in the transmit power. Accordingly, for lower values of the average SNR, increasing the average throughput is accompanied by a corresponding increase in the transmit power. On the other hand, for higher values of the average SNR, the average transmit power saturates and the average throughput is increased.
By considering a total power constraint, $\mathcal{P}_{th}$ = 0.1 mW, at lower SNRs, when the total transmit power is below the threshold, the average transmit power and throughput are similar to their respective values for the no power constraint case.
As the SNR increases, the transmit power reaches the power threshold and the average throughput is reduced accordingly.

\begin{figure}[t]
	\centering
\includegraphics[width=0.75\textwidth]{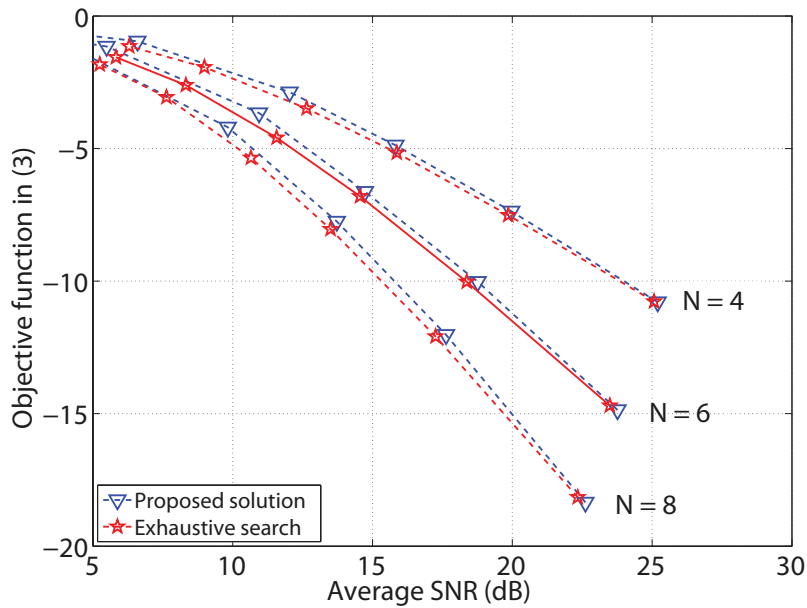}
\caption{Objective function for the proposed algorithm and the exhaustive search when $N$ = 4, 6, and 8.}
\label{fig_Ch_2:ex}
\end{figure}

Fig. \ref{fig_Ch_2:ex} compares the objective function achieved with the proposed algorithm and the exhaustive search that finds the discretized global optimal allocation for the problem in (\ref{eq_Ch_2:ineq_const}). Results are presented for $P_{\rm{th}}$ = 5 $\mu$W and $N$ = 4, 6, and 8; a small number of subcarriers is chosen, such that the exhaustive search is feasible. As can be seen, the proposed algorithm approaches the optimal results of the exhaustive search, and, hence, provides a close-to-optimal solution.

\subsection{Performance Comparison with Algorithms in the Literature}
\begin{figure}[t]
	\centering
\includegraphics[width=0.75\textwidth]{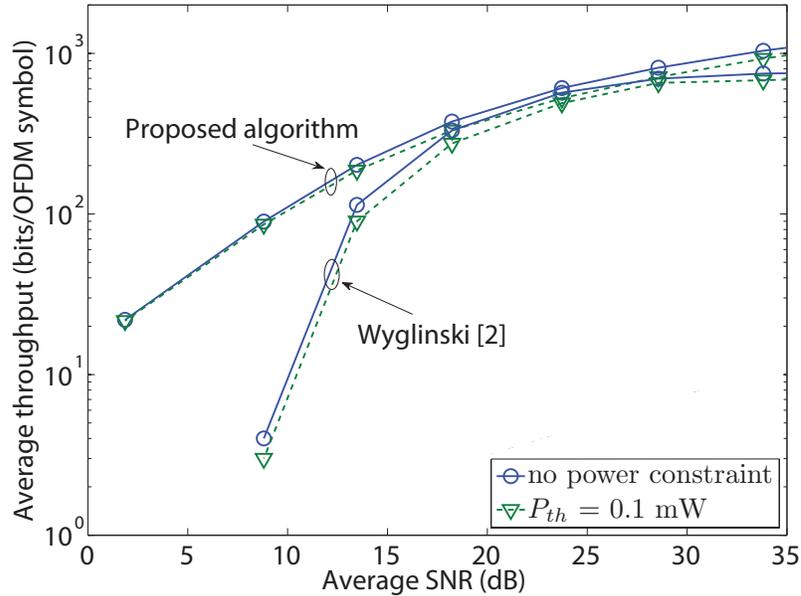}
\caption{Average throughput as a function of average SNR for the proposed algorithm and Wyglinski's algorithm in \cite{wyglinski2005bit}.}
\label{fig_Ch_2:Throughput_comp}
\end{figure}

In Fig. \ref{fig_Ch_2:Throughput_comp}, the throughput achieved by the proposed algorithm is compared to that obtained by Wyglinski's algorithm \cite{wyglinski2005bit} for the same operating conditions, with and without considering the total power constraint. For a fair comparison, the uniform power allocation used by the allocation scheme in \cite{wyglinski2005bit} is computed by dividing the average transmit power allocated by our algorithm by the total number of subcarriers. As shown in Fig.~\ref{fig_Ch_2:Throughput_comp}, the proposed algorithm provides a significantly higher throughput than the scheme in \cite{wyglinski2005bit} for low average SNRs. This result demonstrates that optimal allocation of transmit power is crucial for low power budgets.

\begin{figure}
	\centering
\includegraphics[width=0.75\textwidth]{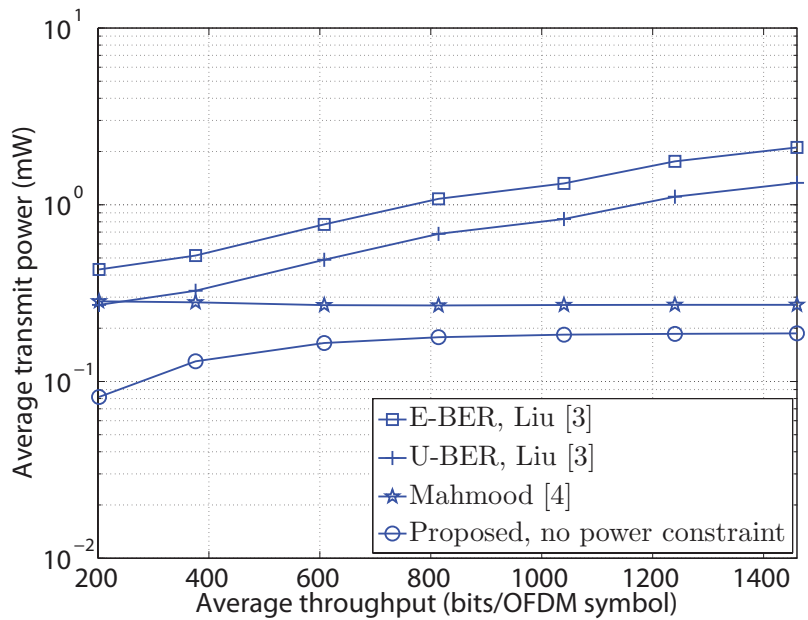}
\caption{Average transmit power as a function of average throughput for the proposed algorithm and the algorithms in \cite{liu2009adaptive} and \cite{mahmood2010efficient}.}
\label{fig_Ch_2:power}
\end{figure}

Fig. \ref{fig_Ch_2:power} compares the average transmit power obtained by the proposed algorithm, in the case of no power constraint, with the optimum power allocation of Liu \textit{et al.} \cite{liu2009adaptive} that assumes unequal BER (U-BER) per subcarrier, a variation called E-BER \cite{liu2009adaptive} that assumes an equal BER per subcarrier, and the algorithm of Mahmood and Belfiore \cite{mahmood2010efficient}. 
After matching the operating conditions, one can see that the proposed allocation scheme assigns less average power than the schemes in \cite{liu2009adaptive} and \cite{mahmood2010efficient} to achieve the same average BER and throughput. The different results between \cite{liu2009adaptive} and \cite{mahmood2010efficient} (while both guarantee the same fixed throughput) are mainly because the algorithms in \cite{liu2009adaptive} allocate the same number of bits per subcarrier, while the algorithm in \cite{mahmood2010efficient} allocates a different number of bits per subcarrier, which is intuitively more efficient.

The improved performance of the proposed joint bit and power allocation algorithm does not come at the cost of additional complexity. Its computational complexity is of $\mathcal{O}(N)$ when the initial value of $\alpha$ results in an inactive power constraint, which is similar to that of Liu's algorithm. Otherwise, it is of  $\mathcal{O}(N \textup{log} (N))$, which is lower than that of Wyglinski's $\mathcal{O}(N^2)$ algorithm and significantly lower than $\mathcal{O}(N!)$ of the exhaustive search.


\section{Conclusion} \label{sec_Ch_2:conc}
In this letter, we proposed a novel algorithm that jointly maximizes the throughput and minimizes the transmit power given constraints on the BER per subcarrier and the total transmit power. Closed-form expressions were derived for the close-to-optimal bit and power allocations per subcarrier, average throughput, and average transmit power. Simulation results demonstrated that the proposed algorithm outperforms different allocation schemes that separately maximizes the throughput or minimizes the transmit power, under the same operating conditions, while requiring similar or reduced computational effort. Additionally, it was shown that its performance approaches that of the exhaustive search with significantly lower complexity.

\bibliographystyle{IEEEtran}
\bibliography{IEEEabrv,mybib_file} 

\chapter{} \label{ch:TW}
\section{Abstract}
This paper adopts a multiobjective optimization (MOOP) approach to investigate the optimal link adaptation problem of orthogonal frequency division multiplexing (OFDM)-based cognitive radio (CR) systems, where secondary users (SUs) can opportunistically access the spectrum of primary users (PUs). For such a scenario, we solve the problem of jointly maximizing the CR system throughput and minimizing its transmit power, subject to constraints on both SU and PUs. The optimization problem imposes predefined interference thresholds for the PUs, guarantees the SU quality of service in terms of a maximum bit-error-rate (BER), and satisfies a transmit power budget and a maximum number of allocated bits per subcarrier. Unlike most of the work in the literature that considers perfect SU spectrum sensing capabilities, the problem formulation takes into account errors due to imperfect sensing of 	the PUs bands.
Closed-form expressions are obtained for the optimal bit and power allocations per SU subcarrier. Simulation results illustrate the performance of the proposed algorithm and demonstrate the superiority of the MOOP approach when compared to single optimization approaches presented in the literature, without additional complexity. Furthermore, results show that the interference thresholds at the PUs receivers can be severely exceeded due to the perfect spectrum sensing assumption or due to partial channel information on links between the SU and the PUs receivers.
Additionally, the results show that the performance of the proposed algorithm approaches that of an exhaustive search for the discrete optimal allocations with a significantly reduced computational effort. 
\section{Introduction}

The wireless radio spectrum has become a scarce resource due to the ceaseless demands for spectrum by new applications and services. However, this spectrum scarcity happens while most of the allocated spectrum is under-utilized, as reported by many jurisdictions \cite{fcc2002spectrum}. This paradox occurs due to the inefficiency of the traditional static spectrum allocation policies. Cognitive radio (CR) \cite{hossain2007cognitive} provides a solution to the spectrum utilization inefficiency by allowing unlicensed/secondary users (SUs)  to opportunistically access spectrum holes in licensed/primary users (PUs) frequency bands/time slots under the condition that no harmful interference occurs to PUs. Orthogonal frequency division multiplexing (OFDM) is recognized as an attractive modulation technique for CR due to its spectrum shaping flexibility, adaptivity in allocating vacant radio resources, and capability in monitoring the spectral activities of PUs \cite{joshi2012joint,wang2011new,mahmoud2009ofdm}. Link adaptation for OFDM-based CR systems is the terminology used to describe techniques that improve the system performance by dynamically changing various transmission parameters, e.g., the number of allocated bits and power per subcarrier,  based on the quality of the wireless link and the imposed PU interference constraints \cite{kang2009optimal, zhang2010efficient, attar2008interference, bansal2008optimal, hasan2009energy, zhao2010power, bansal2011adaptive, kaligineedi2012power, kang2009sensing, wang2009joint, zhao2007decentralized, srinivasa2008much, almalfouh2011interference, bedeer2011partial, bedeer2012jointVTC, bedeer2012EBERGC, bedeer2012UBERGC, bedeer2012novelICC, bedeer2012adaptiveRWS, bedeer2013resource, bedeer2013adaptive, bedeer2013novel, bedeer2014rateCONF, bedeer2014multiobjective, bedeer2013joint, 
bedeer2014energy, bedeer2015systematic, bedeer2015rate}.

Generally speaking, the interference introduced to the PUs bands in OFDM-based CR networks can be classified as: 1) mutual interference (co-channel interference (CCI) and adjacent channel interference (ACI)) between the SU and PUs due to non-orthogonality of their respective transmissions \cite{bansal2008optimal, zhang2010efficient, kang2009optimal, attar2008interference, zhao2010power,  hasan2009energy, bansal2011adaptive, kaligineedi2012power} and 2) interference due to the SU's imperfect spectrum sensing capabilities \cite{kaligineedi2012power, kang2009sensing, wang2009joint, srinivasa2008much, zhao2007decentralized, almalfouh2011interference}. Spectrum sensing is not fully reliable due to the SU hardware limitations and the variable channel conditions. Therefore, the SU may identify certain PUs bands as occupied when they are truly vacant. This results in the sensing error known as a \textit{false-alarm}.  On the other hand, if the SU identifies certain PUs bands to be vacant while they are truly occupied, this leads to the sensing error known as a \textit{mis-detection}. The probability of mis-detection increases the interference to the undetected PUs, while the probability of false-alarm reduces the transmission opportunities of SUs.

To date, most of the research literature has focused on the \emph{single} objective function of maximizing the SU capacity/throughput with constraints on the SU total transmit power and the interference introduced to existing PUs, while less attention has been given to the effects of the SU's imperfect sensing 	capabilities  or to guarantee a certain SU bit error rate (BER) \cite{kang2009optimal, zhang2010efficient, attar2008interference, bansal2008optimal, hasan2009energy, zhao2010power, bansal2011adaptive, kaligineedi2012power, kang2009sensing, wang2009joint, srinivasa2008much, zhao2007decentralized, almalfouh2011interference}. For example, Kang \textit{et al.}, in \cite{kang2009optimal}, studied the problem of optimal power allocation to achieve the ergodic, delay-sensitive, and outage capacities of a SU under a constrained average/peak  SU transmit power and interference to the PUs, with no interference from the PUs to the  SU taken into consideration. In \cite{zhang2010efficient}, Zhang and Leung proposed a low complexity suboptimal algorithm for an OFDM-based CR system in which SUs may access both nonactive and active PU frequency bands, as long as the total CCI and ACI are within acceptable limits. Attar \textit{et al.} \cite{attar2008interference} proposed an algorithm that maximizes the throughput of both SUs and PUs under constraints on the experienced interference by each user. Bansal \textit{et al.} \cite{bansal2008optimal} investigated the optimal power allocation problem in CR networks to maximize the  SU downlink transmission capacity under a constraint on the instantaneous interference to PUs. The proposed algorithm was complex and several suboptimal algorithms were developed to reduce the computational complexity. In \cite{hasan2009energy}, Hasan \textit{et al.}  presented a solution to maximize the SU capacity while taking into account the availability of subcarriers, i.e., the activity of PUs in the licensed bands, and the interference leakage to PUs. Zhao and Kwak \cite{zhao2010power} maximized the throughput of the  SU while keeping the interference to PUs below a certain threshold. A low-complexity iterative power loading algorithm and a suboptimal iterative bit loading algorithm were proposed to solve the optimization problem. Almalfouh and St{\"u}ber \cite {almalfouh2011interference} maximized the overall rate of  the SU OFDMA-based CR network subject to maximum power constraint and average interference constraints to the PUs due to the mis-detection and false-alarm probabilities. The resource allocation problem was classified as a  mixed-integer nonlinear programming that is NP-hard to obtain the optimal solution. An iterative algorithm based on the multiple-choice knapsack problem was proposed to find a sub-optimal solution. 


CR systems will have different requirements than those listed above.
For example, if only partial channel information is known on the links between the SU and the PUs receivers or the sensing is not fully reliable, then minimizing the transmit power is prioritized in order not to violate the interference constraints.
On the other hand, maximizing the CR system throughput is of interest to improve the overall network performance. This motivates us to adopt a multiobjective optimization (MOOP) approach that optimizes the conflicting and incommensurable throughput and power objectives.
For most of the MOOP problems, it is not possible to find a single solution that optimizes all the conflicting objectives simultaneously, i.e., there is no solution that improves one of the objective functions without deteriorating other objectives. However, a set of non-dominated, \emph{weak} Pareto optimal solutions exists and it is the decision maker's (the SU in our case) responsibility to choose its preferred optimal solution \cite{miettinen1999nonlinear}. Various methods for solving MOOP problems exist and are classified according to the level of preferences of the competing objective functions as \emph{posteriori} methods and \emph{priori} methods \cite{miettinen1999nonlinear}. For the former, the (whole, if possible) set of the Pareto optimal solutions are generated and presented to the decision maker who selects the preferred one. On the other hand, for the latter, the decision maker must specify the preferences before  the optimization process starts. In this paper, we adopt the \emph{priori} method, with the SU linearly combining the competing throughput and power objectives  into a single objective function. For that, positive weighting coefficients are used \cite{miettinen1999nonlinear}, which reflects the SU preferences  according to the surrounding environment, the application, and/or the target performance.
Recently, MOOP has attracted researchers' attention due to its flexible and superior performance over single objective optimization approaches \cite{bedeer2013joint, elmusrati2008applications, devarajan2012energy, sun2009modified, elmusrati2007multiobjective, yang2012robust}.
In a non-CR environment, jointly maximizing the throughput and minimizing the transmit power provides significant performance improvements when compared with other works in the literature that separately maximize the throughput (with a constraint on  the transmit power) or minimize the transmit power (with a constraint on the throughput), respectively \cite{bedeer2013joint}.

In this paper, we formulate a multiobjective optimization problem $\mathcal{OP}1$ that jointly maximizes the OFDM SU throughput and minimizes its total transmit power subject to constraints on the BER, the total transmit power, the CCI and ACI to existing PUs, and the maximum allocated bits per subcarrier for the SU. Furthermore, $\mathcal{OP}1$  considers the spectrum sensing errors; this is achieved by formulating the CCI and ACI constraints as a function of the mis-detection and false-alarm probabilities.
We transform the non-convex $\mathcal{OP}1$ to an equivalent convex problem $\mathcal{OP}2$ where closed-form expressions are derived for the bit and power allocations per each SU subcarrier. Unlike the works in  \cite{bansal2008optimal, kang2009optimal, zhang2010efficient, attar2008interference, kang2009sensing, wang2009joint, zhao2007decentralized} that assume full channel state information (CSI), we adopt the more practical assumption of only knowing the path loss on the links between the SU transmitter and the PUs receivers \cite{zhao2010power, hasan2009energy}. Additionally, we run simulations to quantify the violation of both the CCI and ACI constraints that results at the PUs receivers due to the incomplete link information between the SU transmitter and the PUs receivers. The effect of adding a fading margin to compensate for this violation is studied. Also, simulation results show that the interference constraints are violated in practice at the PUs receivers if perfect spectrum sensing is assumed.
The results illustrate the performance of the proposed algorithm and show its closeness to the global optimal allocations obtained by an exhaustive search for the equivalent discrete problem. Furthermore, the results show the performance improvements of the proposed algorithm when compared to other works in the literature at the cost of no additional complexity.

The remainder of the paper is organized as follows. Section \ref{sec_Ch_3:model} introduces the system model and Section \ref{sec_Ch_3:opt} formulates and analyzes the optimization problems. Section \ref{sec_Ch_3:proposed} summarizes the proposed algorithm and provides a complexity analysis. Simulation results are presented in Section \ref{sec_Ch_3:sim}, while conclusions are drawn in Section \ref{sec_Ch_3:conc}.

Throughout the paper we use bold-faced upper case letters to denote matrices, e.g., $\mathbf{X}$,
bold-faced lower case letters for vectors, e.g., $\mathbf{x}$, and light-faced letters for scalar quantities, e.g., $x$. $[.]^T$ denotes the transpose operation, $\nabla$ represents the gradient, $\lfloor x \rfloor$ is the largest integer not greater than $x$, $\lfloor x \rceil$ is the nearest integer to $x$, $[x,y]^-$ represents $\textup{min}(x,y)$, $\mathbf{I}_{X}$ is the $X \times X$ identity matrix, and $\bar{\bar{\mathbb{X}}}$ is the cardinality of the set $\mathbb{X}$. 
\section{System Model} \label{sec_Ch_3:model}

\begin{figure}[t]
\centering
\includegraphics[width=0.750\textwidth]{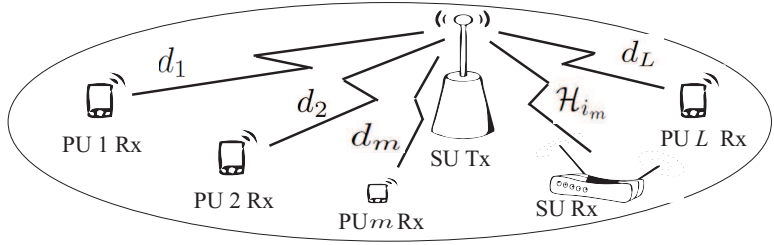}
\caption{Cognitive radio system model.}
\label{fig_Ch_3:sys}
\end{figure}

Fig.~\ref{fig_Ch_3:sys} depicts the CR system model under consideration, where the available spectrum is assumed to be divided into $L$ subchannels that are licensed to $L$ PUs. 
We assume that the SU periodically senses the PUs spectrum in order to identify vacant bands for its transmission. Without loss of generality, we consider that the SU senses that subchannel $m$, of bandwidth $B$, is vacant and decides to access it with $N$ subcarriers and $i$ denotes the $i$th  subcarrier  in the subchannel $m$, $i = 1, ..., N$. However, due to the varying channel conditions between the SU and PUs, the $m$th PU signal may drop below the SU sensing threshold. This means that the SU identifies the $m$th PU band as vacant when it is truly occupied. This is referred to as a mis-detection error and it occurs with probability $\rho^{(m)}_{{\rm{md}}}$. On the other hand, the SU may identify the $\ell$th PU band as occupied when it is truly vacant. This is referred to as a false-alarm error and it occurs with probability $\rho^{(\ell)}_{fa}$.  Mis-detection errors lead to severe co-channel interference to the $m$th PU, while false-alarm errors result in the SU losing transmission opportunities. Using the Bayes' theorem and the law of total probability, the probability that subchannel $m$ is truly occupied under the condition that the SU identified it to be vacant can be defined as \cite{almalfouh2011interference}
\begin{IEEEeqnarray}{c}
\beta_{{\rm{ov}}}^{(m)} = \frac{\rho^{(m)}_{{\rm{md}}}  \rho^{(m)}}{\rho^{(m)}_{{\rm{md}}}  \rho^{(m)} + (1 - \rho^{(m)}_{{\rm{fa}}}) (1 - \rho^{(m)})}, \label{eq_Ch_3:b_ov}
\end{IEEEeqnarray}
where $\rho^{(m)}$ is the probability that the PU transmits on subchannel $m$. Furthermore, the probability that subchannel $\ell$ is truly occupied by the PU under the condition that the SU identified it to be occupied can be written as
\begin{IEEEeqnarray}{c}
\beta_{{\rm{oo}}}^{(\ell)} = \frac{ (1 - \rho^{(\ell)}_{{\rm{md}}})  \rho^{(\ell)}}{ (1 - \rho^{(\ell)}_{{\rm{md}}})  \rho^{(\ell)} + \rho^{(\ell)}_{fa} (1 - \rho^{(\ell)})}. \label{eq_Ch_3:b_oo}
\end{IEEEeqnarray}
The conditional probability $\beta_{{\rm{ov}}}^{(m)}$ represents the probability that the interference due to mis-detection errors will be present in suchannel $m$, which is determined to be vacant by the SU, and, hence, $1 - \beta_{{\rm{ov}}}^{(m)}$ represents the confidence level of the SU that subchannel $m$ is truly vacant \cite{wang2009joint}. It is worthy to mention that for perfect sensing 
$\beta_{{\rm{ov}}}^{(m)} = 0$ and $\beta_{{\rm{oo}}}^{(\ell)} = 1$.

While it is possible to estimate the instantaneous channel gains between the SU transmitter and receiver pairs, it is more challenging to estimate the instantaneous channel gains from the SU transmitter to the PUs receivers without the PUs cooperation. That being said, we assume perfect CSI between the SU transmitter and receiver pairs, while only the path loss is assumed to be known between the SU transmitter and PUs receivers.
In practical scenarios, using only partial information on the links between the SU transmitter and PU receivers may result in the violation of the CCI and ACI constraints at the PU receivers. This problem is discussed in Section \ref{sec_Ch_3:sim}.

The CCI to subchannel $m$ that the SU decides to be vacant, but may or may not be truly vacant needs to be less than a certain threshold $P_{\textup{th}}^{(m)}$ as
\begin{eqnarray}
\beta_{{\rm{ov}}}^{(m)} 10^{-0.1PL(d_m)} 10^{-0.1 \: \textup{FM}} \sum_{i=1}^{N} p_{i}  \leq  P_{\textup{th}}^{(m)},
\end{eqnarray}
where $PL(d_m)$ is the path loss in dB at distance $d_m$ from the SU, $\textup{FM}$ is the fading margin\footnote{The fading margin is added to compensate for the possible violation of the interference constraints at the PUs receivers due to the imperfect CSI on the links between the SU transmitter and the PUs receivers.} in dB, and $p_{i}$ is the allocated power per subcarrier $i$, $i = 1, ..., N$. On the other hand, if spectrum sensing is perfect (i.e., the $m$th PU does not truly exist), then the SU transmit power in the $m$th subchannel should reflect the SU transmitter's power amplifier limitations or/and satisfy regulatory maximum power limits as
\begin{eqnarray}
\sum_{i=1}^{N} p_{i}\leq P_{\rm{th}}.
\end{eqnarray}
Hence, for either perfect or imperfect spectrum sensing, the condition on the CCI/total transmit power is generalized as\footnote{For the perfect sensing assumption, $\beta_{{\rm{ov}}}^{(m)} = 0$ and $\sum_{i=1}^{N} p_{i}\leq \Big[P_{\rm{th}}, \infty \Big]^- = P_{\rm{th}}$; otherwise, for the case of imperfect sensing, we consider $\frac{1}{\beta_{{\rm{ov}}}^{(m)}} 10^{0.1PL(d_m)} 10^{0.1 \: \textup{FM}} \: P_{\textup{th}}^{(m)} < P_{\rm{th}}$, and, hence, $\sum_{i=1}^{N} p_{i}\leq \frac{1}{\beta_{{\rm{ov}}}^{(m)}} 10^{0.1PL(d_m)} 10^{0.1 \: \textup{FM}} \: P_{\textup{th}}^{(m)}$. This is a reasonable assumption as $P_{\rm{th}}$ represents the maximum power the SU can transmit.}
\begin{eqnarray}
\sum_{i=1}^{N} p_{i}\leq \Big[P_{\rm{th}}, \frac{1}{\beta_{{\rm{ov}}}^{(m)}} 10^{0.1PL(d_m)} 10^{0.1 \: \textup{FM}} \: P_{\textup{th}}^{(m)}\Big]^-. \label{eq_Ch_3:CCI_2}
\end{eqnarray}
The ACI to subchannel $\ell$ that the SU decides to be occupied, but may or may not be truly occupied should be kept below a certain threshold $P_{\textup{th}}^{(\ell)}$ as follows \cite{kang2009optimal, zhang2010efficient, attar2008interference, bansal2008optimal, hasan2009energy, zhao2010power, bansal2011adaptive, kaligineedi2012power, weiss2004mutual}
\begin{IEEEeqnarray}{c}
\beta_{{\rm{oo}}}^{(\ell)} 10^{-0.1 PL(d_{\ell})} 10^{-0.1 \: \textup{FM}} \sum_{i = 1}^{N} p_{i} \varpi_{i}^{(\ell)} \leq P_{\textup{th}}^{(\ell)}, \quad \ell = 1, ..., L,
\end{IEEEeqnarray}
where $\varpi_{i}^{(\ell)} = T_{s,m} \: \int_{f_{i,\ell}-\frac{B_\ell}{2}}^{f_{i,\ell}+\frac{B_\ell}{2}} \textup{sinc}^2(T_{s,m} f) \: df$,  $T_{s,m}$ is the duration of the OFDM symbol of the SU, $f_{i,\ell}$ is the spectral distance between the SU subcarrier $i$ and the $\ell$th PU frequency band,  $B_\ell$ is the bandwidth of the $\ell$th PU subchannel, and $\textup{sinc}(x) = \frac{\textup{sin}(\pi x)}{\pi x}$.

\section{Optimization Problems: Formulation and Analysis} \label{sec_Ch_3:opt}
In MOOP principle, if the objective functions and constraints are convex, then the obtained Pareto optimal solution is referred to as a \emph{global} Pareto optimal solution; otherwise, it is refereed to as a \emph{local} Pareto optimal solution \cite{miettinen1999nonlinear}. Furthermore, the obtained solution is a \emph{weak} Pareto optimal solution if there is no other solution that causes every objective to improve;
otherwise, it is refereed to as a \emph{strong} Pareto optimal solution \cite{miettinen1999nonlinear}.

Our target is to jointly maximize the SU throughput and minimize its transmit power while satisfying target quality-of-service QoS (in terms of BER),  certain levels of CCI/total transmit power and ACI to the PUs receivers, and a maximum number of bits per each subcarrier while considering the errors due to imperfect sensing.
We assume that the SU accesses the spectrum if the QoS is achievable. For an average BER constraint, this corresponds to a non-convex optimization problem where the obtained numerical solution is not guaranteed to be a global Pareto optimal solution. According to the results in \cite{willink1997optimization}, the constraint on the average BER can be relaxed to a constraint on the BER per subcarrier, especially for high SNRs. The benefit of this relaxation is that the resultant optimization problem can be convex after some mathematical manipulations, in which case the global optimality of the Pareto set of solutions is guaranteed. Also, such relaxation allows us to obtain closed-form expressions for the optimal bit and power allocations,
and, hence, the obtained solution will be of significantly lower complexity when compared to the solution of the problem with the constraint on the average BER. Therefore, the obtained solution of the problem in hand will be a globally (as the MOOP problem is convex) weak (as the objective functions are conflicting) Pareto optimal solution.

The multiobjective optimization problem is formulated as\footnote{The optimization problem with discrete constraints for the number of the allocated bits per subcarrier is a mixed integer nonlinear programming problem that can be solved by the branch and bound algorithm \cite{floudas1995nonlinear}. However, this will be significantly complex and not tractable for large number of subcarriers. In the rest of the paper and according to  the common practice in the literature, we assume continuous values for the number of bits per subcarrier in order to obtain a low complexity solution, and then discretize the number of allocated bits per subcarrier. To address the gap to the discrete global optimal solution, the obtained results are compared with an exhaustive search in Section \ref{sec_Ch_3:sim}. It is worthy to mention that the exhaustive search is based on continues (and not discrete) values of the power that is calculated from the discrete values of the bit allocation according to (\ref{eq_Ch_3:sub_P_b}).}

\begin{IEEEeqnarray}{rcl}
 &\qquad &  \underset{b_{i}}{\textup{Maximize}} \: \sum_{{i} = 1}^{N}b_{i}  \quad \textup{and} \quad \underset{p_{i}}{\textup{Minimize}} \: \sum_{{i} = 1}^{N}p_{i}, \nonumber \\
\textup{s.t.} \quad \textup{C}1: &{}\qquad {}& b_i \leq b_{i,\textup{max}},  \qquad i = 1, ..., N, \nonumber \vspace{5pt}\\
\textup{C}2: & & \textup{BER}_{i}  \leq  \textup{BER}_{{\rm{th}},i},  \qquad i = 1, ..., N, \label{eq_Ch_3:eq_first} \nonumber\\
\textup{C}3: & & \sum_{i=1}^{N} p_{i}  \leq  \Big[P_{\rm{th}}, \frac{1}{\beta_{{\rm{ov}}}^{(m)}} 10^{0.1PL(d_m)} 10^{0.1 \: \textup{FM}} \: P_{\textup{th}}^{(m)}\Big]^-,  \label{eq_Ch_3:eq_1} \nonumber \\
\textup{C}4: & & \sum_{i = 1}^{N} p_{i} \varpi_{i}^{(\ell)}  \leq  \frac{1}{\beta_{{\rm{oo}}}^{(\ell)}} 10^{0.1 PL(d_{\ell})} 10^{0.1 \: \textup{FM}} P_{\textup{th}}^{(\ell)}, \qquad  \ell = 1, ..., L, \label{eq_Ch_3:eq_1_2} \IEEEeqnarraynumspace
\end{IEEEeqnarray}
where $b_i$ and $b_{i,\textup{max}}$ are the number of bits and the maximum number of bits per subcarrier $i$, $i = 1, ..., N$, respectively, and $\textup{BER}_i$ and $\textup{BER}_{{\rm{th}},i}$  are the BER and the threshold value of the BER per subcarrier $i$, $i$ = 1, ..., $N$, respectively. An approximate expression for the BER per subcarrier $i$ in case of $M$-ary QAM \cite{chung2001degrees}, while taking the interference from the PUs into account, is given by\footnote{This expression is tight within 1 dB for BER $\leq$ $10^{-3}$ \cite{chung2001degrees}.}
\setlength{\arraycolsep}{0.0em}
\begin{IEEEeqnarray}{c}
\textup{BER}_{i}  \approx   0.2 \: \textup{exp}\left ( -1.6 \: \frac{\gamma_i \: p_i}{2^{b_{i}} - 1} \right ), \label{eq_Ch_3:BER}
\end{IEEEeqnarray}
where $\gamma_i = \frac{\left | \mathcal{H}_{i} \right |^2}{(\sigma^2_n + \mathcal{J}_{i})}$ is the channel-to-noise-plus-interference ratio for subcarrier $i$, $\mathcal{H}_{i}$ is the channel gain of subcarrier $i$ between the SU transmitter and receiver pair, $\sigma^2_n$ is the variance of the additive white Gaussian noise (AWGN), and $\mathcal{J}_{i}$ is the interference from the PUs to subcarrier $i$ of the SU. We solve the MOOP problem in (\ref{eq_Ch_3:eq_1_2}) by linearly combining the normalized competing throughput and power objectives into a single objective function (note that in the solution, the throughput and power objectives are normalized to their maximum values $N b_{i,\mathrm{max}}$ and $P_{\rm{th}}$, respectively, so they are approximately within the same range [0,1]; for convenience of notation, the normalization factors are not presented in the problem formulation/solution). For that, positive weighting coefficients are used \cite{miettinen1999nonlinear}, which reflects the SU preferences according to the surrounding environment, the application, and/or the target performance. As the power and throughput objectives are conflicting, the obtained solution represents a weak Pareto optimal solution.
The MOOP problem in (\ref{eq_Ch_3:eq_1_2}) can be rewritten as
\begin{IEEEeqnarray}{c}
\mathcal{OP}1: \qquad  \underset{b_{i},p_i}{\textup{Minimize}} \qquad \alpha \sum_{{i} = 1}^{N}p_{i} - (1-\alpha) \sum_{{i} = 1}^{N}b_{i}, \nonumber \\
\textup{subject to} \quad \textup{C}1\textup{---}\textup{C}4, \label{eq_Ch_3:eq_1_21}
\end{IEEEeqnarray}
where $\alpha$ ($0 \leq \alpha \leq 1$) is the weighting coefficient that represents the relative importance of the competing objectives, i.e., higher values of $\alpha$ favors minimizing the transmit power, while lower values of $\alpha$ favors maximizing the throughput. We assume that the SU chooses the proper value of $\alpha$ depending on the mode of operation. For example, if the transmission rate, and, hence, the transmission time is crucial, then the SU chooses lower values of $\alpha$. On the other hand, if minimizing the transmit power/protecting the environment, and, hence, the energy efficiency is more important, then higher values of $\alpha$ are selected.

$\mathcal{OP}1$ is not convex as the constraint on the BER is not convex in both $p_i$ and $b_i$, and, hence, the global optimality of the Pareto set of solutions is not guaranteed. An important remark that helps to resolve the non convexity issue is that the constraint on the BER per subcarrier, i.e., C2 in $\mathcal{OP}1$ which is the source of the non convexity, is always active\footnote{A constraint on the form $G(x) \leq 0$ is said to be active if it holds with equality sign, i.e., $G(x) = 0$; otherwise, it is inactive, i.e., $G(x) < 0$ \cite{Boyd2004convex}.} and it can be relaxed in order to obtain a convex problem equivalent to $\mathcal{OP}1$. We can prove that C2 in $\mathcal{OP}1$ is always active by contradiction, as follows. Let us assume that the optimal bit and power allocations ($b_i^*,p_i^*$) exist at a value for the BER per subcarrier that is not at the boundary, i.e., at $\textup{BER}_{{\rm{th}},i}^* < \textup{BER}_{{\rm{th}},i}$. In this case, a new solution could be obtained at $\textup{BER}_{{\rm{th}},i}^{\textup{new}}$, $\textup{BER}_{{\rm{th}},i}^*  < \textup{BER}_{{\rm{th}},i}^{\textup{new}} \leq   \textup{BER}_{{\rm{th}},i}$,  where the power could be decreased, i.e., $p_{i,\textup{new}} < p_i^*$ or the rate can be increased, i.e., $b_{i,\textup{new}} > b_i^*$ without violating the BER constraint. Clearly, this results in a lower objective function value in (\ref{eq_Ch_3:eq_1_21}), and, hence, the allocation of the bit and power $(b_{i}^*, p_{i}^*)$ that is  at $\textup{BER}_{{\rm{th}},i}^* < \textup{BER}_{{\rm{th}},i}$ cannot be an optimal solution. This can be mathematically proved by applying the Karush-Khun-Tucker (KKT) conditions to $\mathcal{OP}1$; the proof is not provided due to space limitations.

As such, the power per subcarrier $i$ can be related to the number of bits per subcarrier $i$ through the active BER constraint as
\begin{IEEEeqnarray}{c}
p_i =  \frac{- \ln(5 \textup{BER}_{{\rm{th}},i})}{1.6 \: \gamma_i} (2^{b_i} - 1),
\label{eq_Ch_3:sub_P_b}
\end{IEEEeqnarray}
and $\mathcal{OP}1$ can be reformulated as
\begin{IEEEeqnarray}{c}
\mathcal{OP}2: \underset{b_{i}}{\textup{Minimize}}  \:\:  \mathbf{f}_{\mathcal{OP}2}(\mathbf{b}) = \: \alpha \sum_{i = 1}^{N}  \frac{- \ln(5 \textup{BER}_{{\rm{th}},i})}{1.6 \: \gamma_i} (2^{b_{i}} - 1) - (1-\alpha)\sum_{i = 1}^{N}b_{i}, \nonumber
\end{IEEEeqnarray}
\begin{IEEEeqnarray}{c}
\textup{s.t.} \quad g_{\varrho_{\mathcal{OP}2}}(\mathbf{b}) =
\left\{\begin{array}{ll}
b_{i} - b_{i,\textup{max}} \leq 0, \hfill \varrho_{\mathcal{OP}2} = i = 1, ..., N, \vspace{5pt}\\
\sum_{i = 1}^{N}  \frac{-\ln(5 \textup{BER}_{{\rm{th}},i})}{1.6 \: \gamma_i} (2^{b_{i}} - 1) -   \Big[P_{\rm{th}}, \frac{1}{\beta_{{\rm{ov}}}^{(m)}} 10^{0.1PL(d_m)} 10^{0.1 \: \textup{FM}} \: P_{\textup{th}}^{(m)}\Big]^-
 \leq 0,  \\ \hfill \varrho_{\mathcal{OP}2} = N+1,  \\
\sum_{i = 1}^{N}  \frac{- \ln(5 \textup{BER}_{{\rm{th}},i})}{1.6 \: \gamma_i} \varpi_{i}^{(\ell)} (2^{b_{i}} - 1) -  \frac{1}{\beta_{{\rm{oo}}}^{(\ell)}} 10^{0.1 PL(d_{\ell})} 10^{0.1 \: \textup{FM}} P_{\textup{th}}^{(\ell)} \leq 0, \\ \hfill \varrho_{\mathcal{OP}2} = N + 2, ..., N + L + 2,
\end{array}\right. \label{eq_Ch_3:ineq_const_op5} \IEEEeqnarraynumspace
\end{IEEEeqnarray}
where $\ell = 1, ..., L$. $\mathcal{OP}2$ is a convex optimization problem, and, hence, the global optimality of the Pareto set of solutions is guaranteed (the proof is provided in Appendix A). $\mathcal{OP}2$  can be solved by applying the KKT conditions (i.e., transforming the inequalities constraints to equality constraints by adding non-negative slack variables, $y^2_{\varrho_{\mathcal{OP}2}}$, $\varrho_{\mathcal{OP}2} = 1, ..., N+L + 1$) \cite{Boyd2004convex}. Hence, the constraints are rewritten as
\begin{IEEEeqnarray}{rCl}
\mathcal{G}_{\varrho_{\mathcal{OP}2}}(\mathbf{b}, \mathbf{y}_{{\mathcal{OP}2}}) & = & g_{\varrho_{\mathcal{OP}2}}(\mathbf{b}) + y_{\varrho_{\mathcal{OP}2}}^2 = 0,  \qquad \varrho_{\mathcal{OP}2} = 1, ..., N+L+1,
\label{eq_Ch_3:eq_const_op5}
\end{IEEEeqnarray}
where $\mathbf{y}_{{\mathcal{OP}2}} = [y_1^2, ..., y_{N+1}^2, y_{N+2}^{2,(\ell)}]^T$ is the vector of slack variables. The Lagrange function $\mathcal{L}_{\mathcal{OP}2}$ is expressed as
\begin{IEEEeqnarray}{rcl}
\mathcal{\mathcal{L}}_{\mathcal{OP}2}(\mathbf{b}, \mathbf{y}_{{\mathcal{OP}2}},\boldsymbol\lambda_{{\mathcal{OP}2}}) & = &  \mathbf{f}_{\mathcal{OP}2}(\mathbf{b}) + \sum_{\varrho_{\mathcal{OP}2} = 1}^{N+L+1} \lambda_{\varrho_{\mathcal{OP}2}} \: \mathcal{G}_{\varrho_{\mathcal{OP}2}}(\mathbf{b}, \mathbf{y}_{{\mathcal{OP}2}}) \hspace{2cm} \nonumber \\
\hspace{0.7cm} &{} = {}& \alpha \sum_{i = 1}^{N} \frac{- \ln(5 \textup{BER}_{{\rm{th}},i})}{1.6 \: \gamma_i} (2^{b_{i}} - 1) - (1-\alpha)\sum_{i = 1}^{N}b_{i} \nonumber \\
\hspace{0.7cm} &  &+ \sum_{i = 1}^{N} \lambda_{i} \: \Bigg[ b_{i} - b_{i,\textup{max}} + y^2_{i} \Bigg] \label{eq_Ch_3:L_op5} +  \lambda_{N+1} \: \Bigg[ \sum_{i = 1}^{N} \frac{- \ln(5 \textup{BER}_{{\rm{th}},i})}{1.6 \: \gamma_i} (2^{b_{i}} - 1) \nonumber \\& & \hspace{5pt}- \Big[P_{\rm{th}}, \frac{1}{\beta_{{\rm{ov}}}^{(m)}} 10^{0.1PL(d_m)} 10^{0.1 \: \textup{FM}} \: P_{\textup{th}}^{(m)}\Big]^- + y_{N+1}^2  \Bigg] \nonumber \\
\hspace{0.7cm} & &+  \sum_{\ell = 1}^{L} \lambda_{N+2}^{(\ell)} \: \Bigg[ \sum_{i = 1}^{N}  \frac{- \ln(5 \textup{BER}_{{\rm{th}},i})}{1.6 \: \gamma_i} \varpi_{i}^{(\ell)} (2^{b_{i}} - 1) \nonumber \\ & & \hspace{5pt}- \frac{1}{\beta_{{\rm{oo}}}^{(\ell)}} 10^{0.1 PL(d_{\ell})} 10^{0.1 \: \textup{FM}} P_{\textup{th}}^{(\ell)} + y_{N+2}^{2,(\ell)} \Bigg], \nonumber \\
\end{IEEEeqnarray}
where $\boldsymbol\lambda_{{\mathcal{OP}2}} = [\lambda_1, ..., \lambda_{N+1}, \lambda_{N+2}^{(\ell)}]^T$  is the vector of Lagrange multipliers. A stationary point can be found when $\nabla \mathcal{\mathcal{L}}_{\mathcal{OP}2}(\mathbf{b}, \mathbf{y}_{{\mathcal{OP}2}},\boldsymbol\lambda_{{\mathcal{OP}2}}) = 0$, which yields

\begin{subequations}
\label{eq_Ch_3:lag_op5}
\begin{IEEEeqnarray}{RCL}
\frac{\partial \mathcal{L}_{\mathcal{OP}2}}{\partial b_{i}} &{}  = {}& \alpha \ln(2) \frac{- \ln(5 \textup{BER}_{{\rm{th}},i})}{1.6 \: \gamma_i} 2^{b_{i}}  - (1-\alpha) + \lambda_{i}  + \ln(2) \lambda_{N+1} \frac{ - \ln(5 \textup{BER}_{{\rm{th}},i})}{1.6 \: \gamma_i} 2^{b_{i}} \nonumber \\ & &  + \ln(2)  \sum_{\ell = 1}^{L} \lambda_{N+2}^{(\ell)} \frac{ - \ln(5 \textup{BER}_{{\rm{th}},i})}{1.6 \: \gamma_i} \varpi_{i}^{(\ell)} 2^{b_{i}} = 0, \label{eq_Ch_3:eq2_op5} \IEEEeqnarraynumspace\\
\frac{\partial \mathcal{L}_{\mathcal{OP}2}}{\partial \lambda_{i}} &{}  = {}& b_{i} - b_{i,\textup{max}} +  y_{i}^2 = 0, \IEEEeqnarraynumspace \label{eq_Ch_3:eq33_op5}\\
\frac{\partial \mathcal{L}_{\mathcal{OP}2}}{\partial \lambda_{N+1}} &{}  = {}& \sum_{i = 1}^{N} \frac{- \ln(5 \textup{BER}_{{\rm{th}},i})}{1.6 \: \gamma_i} (2^{b_{i}} - 1)   - \Big[P_{\rm{th}}, \frac{1}{\beta_{{\rm{ov}}}^{(m)}} 10^{0.1PL(d_m)} 10^{0.1 \: \textup{FM}} \: P_{\textup{th}}^{(m)}\Big]^-  \nonumber \\ & & + y_{N+1}^2 = 0, \IEEEeqnarraynumspace \label{eq_Ch_3:eq3_op5}\\
\frac{\partial \mathcal{L}_{\mathcal{OP}2}}{\partial \lambda_{N+2}^{(\ell)}} &{}  = {}& \sum_{i = 1}^{N} \frac{- \ln(5 \textup{BER}_{{\rm{th}},i})}{1.6 \: \gamma_i} \varpi_{i}^{(\ell)} (2^{b_{i}} - 1)  - \frac{1}{\beta_{{\rm{oo}}}^{(\ell)}} 10^{0.1 PL(d_{\ell})} 10^{0.1 \: \textup{FM}} P_{\textup{th}}^{(\ell)} \nonumber \\ & & + y_{N+2}^{2,(\ell)} = 0, \label{eq_Ch_3:aci_op5}\\
\frac{\partial \mathcal{L}_{\mathcal{OP}2}}{\partial y_{i}} &{}  = {}& \quad 2\lambda_{i} y_{i} = 0, \label{eq_Ch_3:eq4_op5} \\
\frac{\partial \mathcal{L}_{\mathcal{OP}2}}{\partial y_{N+1}} &{}  = {}& \quad 2\lambda_{N+1} y_{N+1} = 0, \label{eq_Ch_3:eq44_op5} \\
\frac{\partial \mathcal{L}_{\mathcal{OP}2}}{\partial y_{N+2}^{(\ell)}} &{}  = {}& \quad 2\lambda_{N+2}^{(\ell)} y_{N+2}^{(\ell)} = 0. \label{eq_Ch_3:eq444_op5}
\end{IEEEeqnarray}
\end{subequations}
It can be seen that (\ref{eq_Ch_3:eq2_op5})-(\ref{eq_Ch_3:eq444_op5}) represent $3N+2L+2$ equations in the $3N+2L+2$ unknown components of the vectors $\mathbf{b}, \mathbf{y}_{{\mathcal{OP}2}}$,  and $\boldsymbol\lambda_{{\mathcal{OP}2}}$. Equation (\ref{eq_Ch_3:eq4_op5}) implies that either $\lambda_{i}$ = 0 or $y_{i}$ = 0, (\ref{eq_Ch_3:eq44_op5}) implies that either $\lambda_{N+1}$ = 0 or $y_{N+1}$ = 0, while (\ref{eq_Ch_3:eq444_op5}) implies that either $\lambda_{N+2}^{(\ell)}$ = 0 or $y_{N+2}^{(\ell)}$ = 0. Accordingly, eight possible cases exist, as follows:

--- \textit{Case 1}: Setting $\lambda_{i} = 0$ ($y_{i} \neq 0$, i.e., inactive maximum allocated bits per subcarrier constraint),  $\lambda_{N + 1} = 0$ ($y_{N + 1} \neq 0$, i.e., inactive CCI/total transmit power constraint), and  $\lambda_{N + 2}^{(\ell)} = 0$ ($y_{N + 2}^{(\ell)} \neq 0$, i.e., inactive ACI constraint) results in the  bit allocation given by
\begin{IEEEeqnarray}{c}
b_i^* = \Bigg\lfloor \log_2\Big[\frac{1-\alpha }{\alpha \ln(2)} \frac{1.6 \: \gamma_i}{( - \ln(5 \: \textup{BER}_{{\rm{th}},i}))}\Big] \Bigg\rceil, \label{eq_Ch_3:eq10}
\end{IEEEeqnarray}
and the power allocation as in (\ref{eq_Ch_3:sub_P_b}). Since $M$-ary QAM is considered, $b_{i}^*$ should be greater than 2. From (\ref{eq_Ch_3:eq10}), to have $b_{i}^* \geq 2$, $\gamma_{i}$ must satisfy the condition
\begin{IEEEeqnarray}{C}
\gamma_{i} \geq \frac{1}{1.6} \frac{\alpha \ln(2)}{1-\alpha} (- \ln(5\textup{BER}_{th,i_m})) \: 2^2, \hfill i = 1, ..., N,  \label{eq_Ch_3:condition}
\end{IEEEeqnarray}
otherwise $b_i^* = p_i^* = 0$.

--- \textit{Case 2}: Setting $y_{i} = 0$ (active maximum allocated bits per subcarrier constraint),  $\lambda_{N + 1} = 0$ ($y_{N + 1} \neq 0$, i.e., inactive CCI/total transmit power constraint), and  $\lambda_{N + 2}^{(\ell)} = 0$ ($y_{N + 2}^{(\ell)} \neq 0$, i.e., inactive ACI constraint) results in the  bit allocation $b_{i}^* = b_{i,\textup{max}}$ if and only if $\gamma_i \geq \frac{1}{1.6} \frac{\alpha \ln(2)}{1-\alpha} (-\ln(5\textup{BER}_{{\rm{th}},i})) 2^{b_{i,\textup{max}}}, i = 1, ..., N$. This is proved as follows. From (\ref{eq_Ch_3:eq33_op5}) $b_{i} = b_{i,\textup{max}}$ and from (\ref{eq_Ch_3:eq2_op5}) $\lambda_{i} = (1-\alpha) - \alpha \ln(2) \frac{- \ln(5 \textup{BER}_{{\rm{th}},i})}{1.6 \: \gamma_{i}}$ $2^{b_{i,\textup{max}}}$. In order to have a non-negative Lagrange multipliers, $\lambda_{i}$ should be greater than or equal to 0, i.e., $(1-\alpha) - \alpha \ln(2)$ $\frac{- \ln(5 \textup{BER}_{{\rm{th}},i})}{1.6 \: \gamma_{i}} \: 2^{b_{i,\textup{max}}} \geq 0$ which leads to $\gamma_{i} \geq \frac{1}{1.6} \frac{\alpha \ln(2)}{1-\alpha} (- \ln(5\textup{BER}_{{\rm{th}},i})) \: 2^{b_{i,\textup{max}}},$ $i = 1, ..., N$.

--- \textit{Case 3}: Setting $\lambda_{i} = 0$ ($y_{i} \neq 0$, i.e., inactive maximum allocated bits per subcarrier constraint),  $y_{N + 1} = 0$ (i.e., active CCI/total transmit power constraint), and  $\lambda_{N + 2}^{(\ell)} = 0$ ($y_{N + 2}^{(\ell)} \neq 0$, i.e., inactive ACI constraint) results in the  bit allocation is given by
\begin{IEEEeqnarray}{c}
b_i^* = \Bigg\lfloor \log_2\Big[\frac{1-\alpha }{\ln(2)(\alpha + \lambda_{N+1})} \frac{1.6 \: \gamma_i}{( - \ln(5 \: \textup{BER}_{{\rm{th}},i}))}\Big] \Bigg\rceil, \IEEEeqnarraynumspace \label{eq_Ch_3:eq10new}
\end{IEEEeqnarray}
and $p_i^*$ is obtained from (\ref{eq_Ch_3:sub_P_b}). $\lambda_{N+1}$ is calculated to satisfy the active CCI/total transmit power constraint in (\ref{eq_Ch_3:eq3_op5}); if non-negative then the optimal solution is reached, otherwise, $b_i^* = p_i^* = 0$. The value of $\lambda_{N+1}$ is found to be
\begin{IEEEeqnarray}{l}
\lambda_{N+1} = \bar{\bar{\mathbb{N}}}_a \frac{1 - \alpha}{\ln 2} \frac{1}{\Big[P_{\rm{th}}, \frac{1}{\beta_{{\rm{ov}}}^{(m)}} 10^{0.1\: PL(d_m)} 10^{0.1 \: \textup{FM}} \: P_{\textup{th}}^{(m)}\Big]^- + \sum_{i \in \mathbb{N}_a}^{}\frac{- \ln(5 \: \textup{BER}_{{\rm{th}},i})}{1.6 \: \gamma_i}}   - \alpha, \label{eq_Ch_3:lambda} \IEEEeqnarraynumspace
\end{IEEEeqnarray}
where $\bar{\bar{\mathbb{N}}}_a$ is the cardinality of the set of active subcarriers $\mathbb{N}_a$.

--- \textit{Case 4}: Setting $y_{i} = 0$ (i.e., active maximum allocated bits per subcarrier constraint),  $y_{N + 1} = 0$ (i.e., active CCI/total transmit power constraint), and  $\lambda_{N + 2}^{(\ell)} = 0$ ($y_{N + 2}^{(\ell)} \neq 0$, i.e., inactive ACI constraint) results in the  bit allocation $b_{i}^* = b_{i,\textup{max}}$ if and only if $\gamma_i \geq \frac{1}{1.6} \frac{\ln(2)(\alpha + \lambda_{N+1})}{1-\alpha}$ $(-\ln(5\textup{BER}_{{\rm{th}},i})) 2^{b_{i,\textup{max}}}, \quad i = 1, ..., N$ and $\lambda_{N+1}$ is non-negative.

--- \textit{Case 5}: Setting $\lambda_{i} = 0$ ($y_{i} \neq 0$, i.e., inactive maximum allocated bits per subcarrier constraint),  $\lambda_{N + 1} = 0$ ($y_{N + 1} \neq 0$, i.e., inactive CCI/total transmit power constraint), and  $y_{N + 2}^{(\ell)} = 0$ (i.e., active ACI constraint) results in the  bit allocation given by
\begin{IEEEeqnarray}{c}
b_i^* = \Bigg\lfloor \log_2\Big[\frac{1-\alpha }{\ln(2)(\alpha + \sum_{\ell = 1}^{L} \varpi_i^{(\ell)} \lambda_{N+2}^{(\ell)})} \frac{1.6 \: \gamma_i}{(-\ln(5 \textup{BER}_{{\rm{th}},i}))}\Big] \Bigg\rceil, \label{eq_Ch_3:eq10newACI}
\end{IEEEeqnarray}
and $p_i^*$ is obtained from (\ref{eq_Ch_3:sub_P_b}). $\lambda_{N+2}^{(\ell)}$ is calculated numerically using the Newton's method \cite{burden2010numerical} to satisfy the active ACI constraint in (\ref{eq_Ch_3:aci_op5}); if non-negative then the optimal solution is reached, otherwise, $b_i^* = p_i^* = 0$.

--- \textit{Case 6}: Setting $y_{i} = 0$ (i.e., active maximum allocated bits per subcarrier constraint),  $\lambda_{N + 1} = 0$ ($y_{N + 1} \neq 0$, i.e., inactive CCI/total transmit power constraint), and  $y_{N + 2}^{(\ell)} = 0$ (i.e., active ACI constraint) results in the  bit allocation $b_{i}^* = b_{i,\textup{max}}$ if and only if $\gamma_i \geq \frac{1}{1.6} \frac{\ln(2)(\alpha + \sum_{\ell = 1}^{L} \varpi_{i}^{(\ell)} \lambda_2^{(\ell)})}{1-\alpha}$ $(-\ln(5\textup{BER}_{{\rm{th}},i})) 2^{b_{i,\textup{max}}}, \quad i = 1, ..., N$ and $\lambda_{N+2}^{(\ell)}$ is non-negative, $\ell = 1, ..., L$.

--- \textit{Case 7}: Setting $\lambda_{i} = 0$ ($y_{i} \neq 0$, i.e., inactive maximum allocated bits per subcarrier constraint),  $y_{N + 1} = 0$ (i.e., active CCI/total transmit power constraint), and  $y_{N + 2}^{(\ell)} = 0$ (active ACI constraint) results in the  bit allocation given by
\begin{IEEEeqnarray}{c}
b_i^* = \Bigg\lfloor \log_2\Big[\frac{1-\alpha }{\ln(2)(\alpha + \lambda_{N+1} + \sum_{\ell = 1}^{L}\varpi_i^{(\ell)} \lambda_{N+2}^{(\ell)})}  \frac{1.6 \: \gamma_i}{(-\ln(5 \textup{BER}_{{\rm{th}},i}))}\Big] \Bigg\rceil, \label{eq_Ch_3:eq10newACI_CCI}
\end{IEEEeqnarray}
and $p_i^*$ is obtained from (\ref{eq_Ch_3:sub_P_b}). $\lambda_{N+1}$ and $\lambda_{N+2}^{(\ell)}$ are calculated numerically to satisfy the active CCI/total transmit power and ACI constraints in (\ref{eq_Ch_3:eq3_op5}) and (\ref{eq_Ch_3:aci_op5}), respectively; if non-negative then the optimal solution is reached, otherwise, $b_i^* = p_i^* = 0$.

--- \textit{Case 8}: Setting $y_{i} = 0$ (i.e., active maximum allocated bits per subcarrier constraint),  $y_{N + 1} = 0$ (i.e., active CCI/total transmit power constraint), and  $y_{N + 2}^{(\ell)} = 0$ (active ACI constraint) results in the  bit allocation $b_{i}^* = b_{i,\textup{max}}$ if and only if $\gamma_i \geq \frac{1}{1.6} \frac{\ln(2)(\alpha + \lambda_{N+1} + \sum_{\ell = 1}^{L} \varpi_{i}^{(\ell)} \lambda_{N+2}^{(\ell)})}{1-\alpha}$ $(-\ln(5\textup{BER}_{{\rm{th}},i})) 2^{b_{i,\textup{max}}}, \quad i = 1, ..., N$ and $\lambda_{N+1}$ and $\lambda_{N+2}^{(\ell)}$ are non-negative, $\ell = 1, ..., L$.

The  solution ($\mathbf{p}^*,\mathbf{b}^*$) represents a global minimum of $\mathbf{f}_{\mathcal{OP}2}(\mathbf{p},\mathbf{b})$ as the problem is convex and the KKT conditions \cite{Boyd2004convex} are satisfied, as given in Appendix A and Appendix B, respectively.

\section{Proposed Algorithm and Complexity Analysis} \label{sec_Ch_3:proposed}

\subsection{Proposed Algorithm}

The proposed algorithms to solve $\mathcal{OP}2$ can be formally stated as follows:

\floatname{algorithm}{}
\begin{algorithm}
\renewcommand{\thealgorithm}{}
\caption{\textbf{Proposed Algorithm}}
\begin{algorithmic}[1]
\small
\State \textbf{INPUT} $\gamma_i$, $\textup{BER}_{{\rm{th}},i}$, $\alpha$, $P_{\rm{th}}$, $P_{\textup{th}}^{(m)}$, $P_{\textup{th}}^{(\ell)}$, $d_m$, $d_{\ell}$ $\beta_{{\rm{ov}}}^{(m)}$, $\beta_{{\rm{oo}}}^{(\ell)}$, and $b_{i,\textup{max}}$.
\For{$i$ = 1, ..., $N$}
\If{$\gamma_{i} <  \frac{1}{1.6} \: \frac{\alpha \ln(2)}{1-\alpha} \: (- \ln(5\:\textup{BER}_{{\rm{th}},i})) 2^2$}
\State Null subcarrier $i$.
\ElsIf {$\frac{1}{1.6} \: \frac{\alpha \ln(2)}{1-\alpha} \: (- \ln(5\:\textup{BER}_{{\rm{th}},i})) 2^2 \leq \gamma_i < \frac{1}{1.6} \: \frac{\alpha \ln(2)}{1-\alpha} \: (- \ln(5\:\textup{BER}_{{\rm{th}},i})) 2^{b_{i,\textup{max}}}$}
\State  $b_{i}^*$ and $p_{i}^*$ are given by (\ref{eq_Ch_3:eq10}) and (\ref{eq_Ch_3:sub_P_b}), respectively.
\Else
\State $b_i^* = b_{i,\textup{max}}$ and $p_{i}^*$ is given by (\ref{eq_Ch_3:sub_P_b}).
\EndIf
\EndFor
\If {$\sum_{i=1}^{N} p_{i}^* \geq \Big[P_{\rm{th}}, \frac{1}{\beta_{{\rm{ov}}}^{(m)}} 10^{0.1PL(d_m)} 10^{0.1 \: \textup{FM}} P_{\textup{th}}^{(m)}\Big]^-$ and $\sum_{i = 1}^{N} p_{i}^* \varpi_{i}^{(\ell)} < \frac{1}{\beta_{{\rm{oo}}}^{(\ell)}} 10^{0.1 PL(d_{\ell})} 10^{0.1 \: \textup{FM}} P_{\textup{th}}^{(\ell)}$}
\State  $b_{i}^*$ and $p_{i}^*$ are given by (\ref{eq_Ch_3:eq10new}) and (\ref{eq_Ch_3:sub_P_b}), respectively.
\State $\lambda_{N+1}$ is non-negative value given by (\ref{eq_Ch_3:lambda}) (otherwise, $b_i^* = p_i^* = 0$) and $\lambda_{N+2}^{(\ell)} = 0$.
        \algstore{myalg}
  \end{algorithmic}
\end{algorithm}

\floatname{algorithm}{}
\begin{algorithm}
 \renewcommand{\thealgorithm}{}
  \caption{\textbf{Proposed Algorithm} (continued)}
  \begin{algorithmic}
      \algrestore{myalg}
     \small
\If {$\gamma_{i} \geq \frac{1}{1.6} \frac{\ln(2)(\alpha + \lambda_{N+1})}{1-\alpha} (-\ln(5\textup{BER}_{{\rm{th}},i})) 2^{b_{i,\textup{max}}}$}
\State $b_i^* = b_{i,\textup{max}}$ and $p_{i}^*$ is given by (\ref{eq_Ch_3:sub_P_b}) for non-negative values of $\lambda_{N+1}$; otherwise, $b_i^* = p_i^* = 0$.
\EndIf
\ElsIf {$\sum_{i=1}^{N} p_{i}^* < \Big[P_{\rm{th}}, \frac{1}{\beta_{{\rm{ov}}}^{(m)}} 10^{0.1PL(d_m)} 10^{0.1 \: \textup{FM}} P_{\textup{th}}^{(m)}\Big]^-$ and $\sum_{i = 1}^{N} p_{i}^* \varpi_{i}^{(\ell)} \geq \frac{1}{\beta_{{\rm{oo}}}^{(\ell)}} 10^{0.1 PL(d_{\ell})} 10^{0.1 \: \textup{FM}} P_{\textup{th}}^{(\ell)}$}
\State  $b_{i}^*$ and $p_{i}^*$ are given by (\ref{eq_Ch_3:eq10newACI}) and (\ref{eq_Ch_3:sub_P_b}), respectively.
\State $\lambda_{N + 1} = 0$ and $\lambda_{N + 2}^{(\ell)}$ is non-negative value calculated numerically to satisfy $\sum_{i = 1}^{N} p_{i}^* \varpi_{i}^{(\ell)} = \frac{1}{\beta_{{\rm{oo}}}^{(\ell)}} 10^{0.1 PL(d_{\ell})} 10^{0.1 \: \textup{FM}} P_{\textup{th}}^{(\ell)}$ (otherwise, $b_i^* = p_i^* = 0$).
\If {$\gamma_{i} \geq \frac{1}{1.6} \frac{\ln(2)(\alpha + \sum_{\ell = 1}^L \varpi_{i}^{(\ell)} \lambda_{N+2}^{(\ell)})}{1-\alpha} (-\ln(5\textup{BER}_{{\rm{th}},i})) 2^{b_{i,\textup{max}}}$}
\State $b_i^* = b_{i,\textup{max}}$ and $p_{i}^*$ is given by (\ref{eq_Ch_3:sub_P_b}) for non-negative values of $\lambda_{N+2}^{(\ell)}$; otherwise, $b_i^* = p_i^* = 0$.
\EndIf
\Else \:  \textbf{if} {\: $\sum_{i=1}^{N} p_{i}^* \geq \Big[P_{\rm{th}}, \frac{1}{\beta_{{\rm{ov}}}^{(m)}} 10^{0.1PL(d)} 10^{0.1 \: \textup{FM}} P_{\textup{th}}^{(m)}\Big]^-$ and $\sum_{i = 1}^{N} p_{i}^* \varpi_{i}^{(\ell)} \geq \frac{1}{\beta_{{\rm{oo}}}^{(\ell)}} 10^{0.1 PL(d_{\ell})} 10^{0.1 \: \textup{FM}} P_{\textup{th}}^{(\ell)}$} \textbf{then}
\State  $b_{i}^*$ and $p_{i}^*$ are given by (\ref{eq_Ch_3:eq10newACI_CCI}) and (\ref{eq_Ch_3:sub_P_b}), respectively.
\State $\lambda_{N + 1}$ and $\lambda_{N + 2}^{(\ell)}$ are non-negative values calculated numerically to satisfy $\sum_{i=1}^{N} p_{i}^* = \Big[P_{\rm{th}}, \frac{1}{\beta_{{\rm{ov}}}^{(m)}} 10^{0.1PL(d_m)} 10^{0.1 \: \textup{FM}} \: P_{\textup{th}}^{(m)}\Big]^-$ and $\sum_{i = 1}^{N} p_{i}^* \varpi_{i}^{(\ell)} = \frac{1}{\beta_{{\rm{oo}}}^{(\ell)}} 10^{0.1 PL(d_{\ell})} 10^{0.1 \: \textup{FM}} P_{\textup{th}}^{(\ell)}$, respectively (otherwise, $b_i^* = p_i^* = 0$).
\If {$\gamma_{i} \geq \frac{1}{1.6} \frac{\ln(2)(\alpha + \lambda_{N+1} + \sum_{\ell = 1}^L \varpi_{i}^{(\ell)} \lambda_{N+2}^{(\ell)})}{1-\alpha}$ $(-\ln(5\textup{BER}_{{\rm{th}},i})) 2^{b_{i,\textup{max}}}$}
\State $b_i^* = b_{i,\textup{max}}$ and $p_{i}^*$ is given by (\ref{eq_Ch_3:sub_P_b}) for non-negative values of $\lambda_{N+1}$ and $\lambda_{N+2}^{(\ell)}$; otherwise, $b_i^* = p_i^* = 0$.
\EndIf
\EndIf
\State If the conditions on the CCI/total transmit power and the ACI are violated due to rounding, decrement the number of bits on the subcarrier that has the largest $\Delta p_{i}(b_{i}) = p_{i}(b_{i}) - p_{i}(b_{i} - 1)$ until satisfied.
\State \textbf{OUTPUT} $b^*_{i}$ and $p^*_{i}$, $i$ = 1, ..., $N$.
\end{algorithmic}
\end{algorithm}

According to the MOOP problem analysis in Section \ref{sec_Ch_3:opt}, the optimal solution belongs to one of the following four scenarios: 1) both the CCI/total transmit power and ACI constraints are inactive, 2) the CCI/total transmit power constraint is active and the ACI constraint is inactive, 3) the CCI/total transmit power constraint is inactive and the ACI constraint is active, and 4) both the CCI/total transmit power and ACI constraints are active. For each of the four scenarios, the constraint on the maximum allocated bits per subcarrier can be either inactive or active.
\begin{itemize}
\item \textbf{Steps 2 to 10}: the proposed algorithm starts by assuming that both the CCI/total transmit power and ACI constraints are inactive. Then, based on the the value of $\gamma_i$, the proposed algorithm finds the optimal solution per subcarrier for inactive/active maximum allocated bit constraint or nulls the corresponding subcarrier if $\gamma_i$ is below a certain threshold. If both the CCI/total transmit power and the ACI constraints are inactive, then the optimal solution is reached.
\item \textbf{Steps 11 to 16}: based on the assumption that the optimal solution belongs to scenario 1 (i.e., inactive CCI/total transmit power and ACI constraints), the CCI/total transmit power constraint may be not inactive while the ACI is inactive. This means that the initial solution (from steps 2 to 10)
is infeasible and the proposed algorithm finds the Lagrangian multipliers that enforce the
solution to be in the feasible region. More specifically, the proposed algorithm finds the Lagrangian multiplier $\lambda_{N+1}$ that makes the CCI/total transmit power active (i.e., satisfied with equal sign)--scenario 2; if $\lambda_{N+1}$ is non-negative then the optimal solution is reached, otherwise $b_i^* = p_i^* = 0$ (for inactive/active maximum allocated bit constraint).
\item \textbf{Steps 17 to 22}: based on the assumption that the optimal solution belongs to scenario 1 (i.e., inactive CCI/total transmit power and ACI constraints), the ACI constraint may be not inactive while the CCI/total transmit power is inactive. This means that the initial solution (from steps 2 to 10)
is infeasible and the proposed algorithm finds the Lagrangian multipliers that enforce the
solution to be in the feasible region. More specifically, the proposed algorithm finds the Lagrangian multiplier $\lambda_{N+2}^{(\ell)}$ that makes the ACI constraint active--scenario 3; if $\lambda_{N+2}^{(\ell)}$ is non-negative then the optimal solution is reached, otherwise $b_i^* = p_i^* = 0$ (for inactive/active maximum allocated bit constraint).
\item \textbf{Steps 23 to 28}: based on the assumption that the optimal solution belongs to scenario 1 (i.e., inactive CCI/total transmit power and ACI constraints), the CCI/total transmit power may be not inactive and the ACI constraint may be not inactive. This means that the initial solution (from steps 2 to 10)
is infeasible and the proposed algorithm finds the Lagrangian multipliers that enforce the
solution to be in the feasible region. More specifically, the proposed algorithm finds the Lagrangian multipliers $\lambda_{N+1}$ and $\lambda_{N+2}^{(\ell)}$ that make the CCI/total transmit power and ACI constraints, respectively, active--scenario 4; if $\lambda_{N+1}$ or $\lambda_{N+2}^{(\ell)}$ are non-negative then the optimal solution is reached, otherwise $b_i^* = p_i^* = 0$ (for inactive/active maximum allocated bit constraint).
\item \textbf{Step 30}: The purpose of step 30 is to guarantee that neither the CCI/total transmit power nor the ACI constraints are violated due to rounding the continuous allocated bits to the nearest integer. As the common practice in the literature, the MOOP problem in (\ref{eq_Ch_3:eq_1_21}) assumes continuous number of allocated bits per each SU subcarrier. This is to avoid the significantly complex and intractable solution of the equivalent problem with discrete constraints on the number of bits per subcarrier \cite{floudas1995nonlinear}. Therefore, the CCI/transmit power and ACI constraints are checked. If violated, the subcarrier corresponding to the largest power reduction when the number of bits is decremented by 1 bit is chosen, and the number of bits is decreased by 1 bit on that subcarrier. The process repeats until the CCI/transmit power and/or the ACI constraints are satisfied.
The obtained solution is shown in Section \ref{sec_Ch_3:sim} to be near  the discrete optimal solution obtained through an exhaustive search.

\end{itemize}

\subsection{Complexity Analysis}

The worst case computational complexity of the proposed algorithms to solve $\mathcal{OP}2$ can be analyzed as follows. Steps 2 to 10 require a complexity of  $\mathcal{O}(N)$; steps 11 to 16 require a complexity of $\mathcal{O}(N)$; steps 17 to 22 require a complexity of $\mathcal{O}(\mathcal{U}N)$, where $\mathcal{O}(\mathcal{U})$ is the complexity of finding $\lambda_{N+2}^{(\ell)}$; and steps 23 to 28 require a complexity of $\mathcal{O}(\mathcal{V}N)$, where $\mathcal{O}(\mathcal{V})$ is the complexity of finding $\lambda_{N+1}$ and $\lambda_{N+2}^{(\ell)}$. As the computational requirement of the Newton's method to solve a system of $\mathcal{N}$ equations in $\mathcal{N}$ unknowns is $\mathcal{O}(\mathcal{N} \mathcal{K})$, where $\mathcal{K}$ is the number of required iterations \cite{more1979numerical}, $\mathcal{O}(\mathcal{U})$ and $\mathcal{O}(\mathcal{V})$ equal $\mathcal{O}(L\mathcal{K}_1)$ and $\mathcal{O}((L+1)\mathcal{K}_2)$, respectively.
Step 30 requires a computational complexity of $\mathcal{O}(N^2)$. This can be explained as follows: First, step 30 finds the subcarrier $i'$ with the maximum $\Delta p_{i'}(b_{i'}) = p_{i'}(b_{i'}) - p_{i'}(b_{i'} - 1)$ due to rounding, this is of complexity of  $\mathcal{O}(N)$. Then, step 30 decrements the allocated bits on $i'$ until the CCI/total transmit power and ACI constraints are satisfied. In the worst case, this process will be repeated $N$ times and, hence, the computational complexity is of $\mathcal{O}(N)$ if all the allocated bits are rounded up to the nearest integer. Thus, the worst case computational complexity  of the proposed algorithms to solve $\mathcal{OP}2$ is calculated as $\mathcal{O}(N) + \mathcal{O}(N) + \mathcal{O}(L \mathcal{K}_1 N) + \mathcal{O}((L+1)\mathcal{K}_2 N) + \mathcal{O}(N) \: \mathcal{O}(N) = \mathcal{O}(N^2)$. Note that the asymptotic complexity of $\mathcal{O}(N) \: \mathcal{O}(N)$ dominates the complexities of $\mathcal{O}(L \mathcal{K}_1 N)$ and $\mathcal{O}((L+1)\mathcal{K}_2 N)$ given that the number of iterations  $\mathcal{K}_1$ and $\mathcal{K}_2$ are found to be around 6--7 iterations, which is significantly less than $N$, and the number of PUs $L$ is assumed to be less than $N$.


\section{Numerical Results} \label{sec_Ch_3:sim}
This section investigates the performance of the proposed algorithm and compares it with other techniques in the literature, as well as with an exhaustive search for the discrete global optimal allocations. The computational complexity of the proposed algorithm is also compared to that of other schemes.

\subsection{Simulation Setup}
Without loss of generality, we assume that the OFDM SU coexists with one adjacent channel PU $\ell$ and one co-channel PU $m$. The SU parameters are as follows: number of subcarriers $N = 128$ and subcarrier spacing $\Delta f = 9.7656$ kHz. The propagation path loss parameters are: exponent $ = 4$, wavelength $ = \frac{3 \times 10^8}{900 \times 10^6} = 0.33\:\textup{meters}$, distance to the $\ell$th PU $d_{\ell} = 1.5$ km, distance to the $m$th PU  $d_m = 1$ km, and reference distance $d_0 = 100$ m. The BER constraint per subcarrier, $\textup{BER}_{{\rm{th}},i}$, is set to $10^{-4}$.  Unless otherwise mentioned, the fading margin $\textup{FM}$ is set to 0 dB. A Rayleigh fading environment
is considered and
representative results are presented in this section, which were obtained through Monte Carlo trials for $10^{4}$ channel realizations.
The value of the AWGN noise variance $\sigma_n^2$ is assumed to be $10^{-16}$ W and the PU signal is assumed to be an elliptically-filtered white random process \cite{bansal2008optimal, zhao2010power, hasan2009energy, bansal2011adaptive, weiss2004mutual}.
Unless otherwise mentioned, imperfect spectrum sensing is assumed, with the mis-detection probability $\rho_{md}^{(m)}$  uniformly distributed over the interval [0.01,  0.05], the false-alarm probability $\rho_{fa}^{(m)}$ uniformly distributed over the interval [0.01,  0.2], and the probability of the PU activity $\rho^{(m)}$ and $\rho^{(\ell)}$ uniformly distributed between [0, 1].
According to the common practice in the MOOP problem solving techniques, the throughput and transmit power objective functions are scaled during simulations so that they are approximately within the same range \cite{miettinen1999nonlinear}. For convenience, presented numerical results are displayed in the original scales.

\subsection{Performance of the Proposed Algorithm}

\begin{figure}[!t]
\centering
\includegraphics[width=0.750\textwidth]{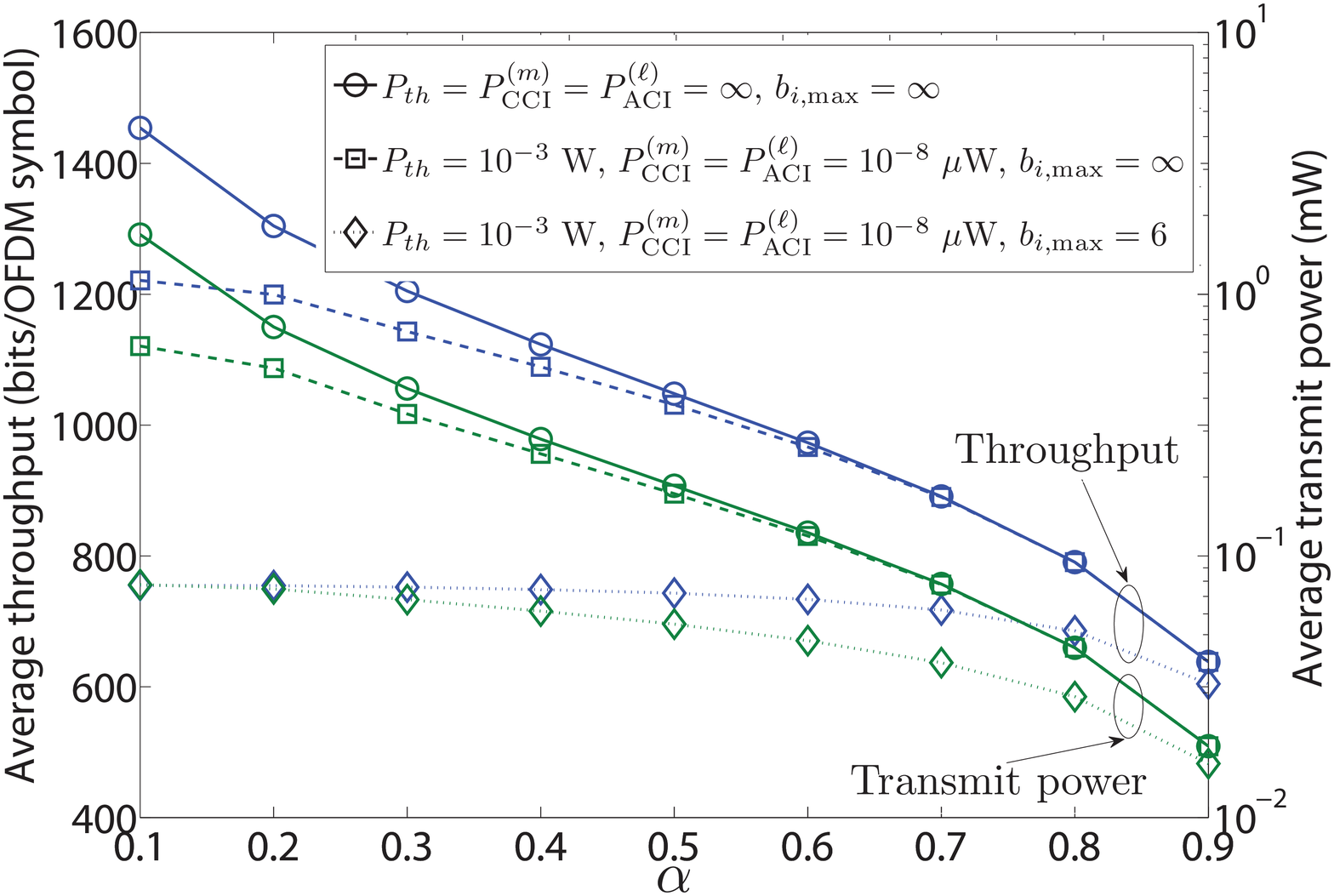}
\caption{Effect of $\alpha$ on the SU performance for different values of $P_{\rm{th}}$, $P_{\textup{CCI}}$,  $P_{\textup{ACI}}$, and $b_{i,\textup{max}}$.}
\label{fig_Ch_3:Performance_alpha}
\end{figure}

Fig. \ref{fig_Ch_3:Performance_alpha} shows the average throughput and average transmit power as a function of the weighting coefficient $\alpha$, for different values of $P_{\rm{th}}$, $P_{\textup{th}}^{(m)}$, $P_{\textup{th}}^{(\ell)}$, and $b_{i,\textup{max}}$. In order to understand the effect of the weighting coefficient $\alpha$ on the MOOP problem formulation, we set $P_{\rm{th}} = P_{\textup{th}}^{(m)} = P_{\textup{th}}^{(\ell)} = \infty$ and $b_{i,\textup{max}} = \infty$; in this scenario, one can notice that an increase of the weighting coefficient $\alpha$ yields a decrease of both the average throughput and average transmit power. This can be explained as follows. By increasing $\alpha$, more weight is given to the transmit power  minimization (the minimum transmit power is further reduced), whereas less weight is given to the throughput maximization (the maximum throughput is reduced), according to the MOOP problem formulations in (\ref{eq_Ch_3:eq_1_21}).  For another scenario we set $P_{\rm{th}} = 10^{-3}$ W  and $P_{\textup{th}}^{(m)} = P_{\textup{th}}^{(\ell)} = 10^{-8}$ $\mu$W, the average transmit power and throughput are similar to their respective values if the total transmit power is less than $\big[P_{\rm{th}}, \frac{1}{\beta_{ov}} 10^{0.1PL(d_m)} 10^{0.1 \: \textup{FM}} \: P_{\textup{th}}^{(m)}\big]^-$ and $\sum_{i = 1}^{N} p_i \varpi_i^{(\ell)} \leq P_{\textup{th}}^{(\ell)}$, while the average throughput and power decrease if the total transmit power and $\sum_{i = 1}^{N} p_i \varpi_i^{(\ell)}$ exceed $\big[P_{\rm{th}}, \frac{1}{\beta_{ov}} 10^{0.1PL(d_m)} 10^{0.1 \: \textup{FM}} \: P_{\textup{th}}^{(m)}\big]^-$ and $P_{\textup{th}}^{(\ell)}$, respectively. If we have a further constraint on $b_{i,\textup{max}} = 6$, then the average throughput and transmit power are reduced accordingly.
Fig. \ref{fig_Ch_3:Performance_alpha} illustrates the flexibility of the proposed algorithm to achieve different levels of the average throughput and transmit power by changing the weighting coefficient $\alpha$.

\begin{figure}[!t]
\centering
\includegraphics[width=0.750\textwidth]{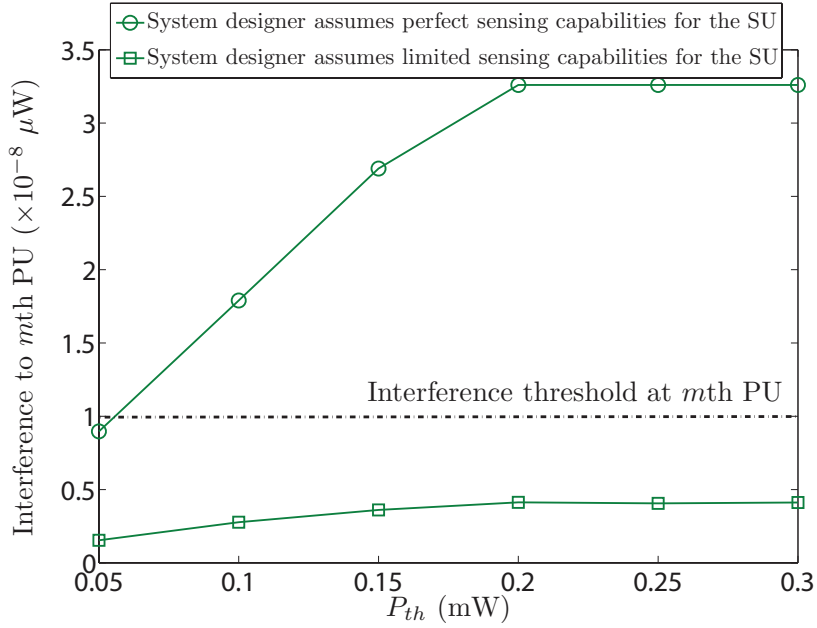}
\caption{Effect of perfect and imperfect sensing on the SU performance for $P_{\textup{CCI}} = P_{\textup{ACI}} = 10^{-8}$ $\mu$W, $\alpha$ = 0.5, and $b_{i,\textup{max}} = 6$.}
\label{fig_Ch_3:Performance_P_th}
\end{figure}

In Fig. \ref{fig_Ch_3:Performance_P_th}, the interference introduced into the $m$th PU band is depicted as a function of $P_{\rm{th}}$ for $\alpha = 0.5$ and the cases of perfect and imperfect spectrum sensing. For the case of perfect sensing, the system designer assumes that the SU has perfect sensing capabilities (which is not true in practice) and the values of $\beta_{{\rm{ov}}}^{(m)}$ and $\beta_{{\rm{oo}}}^{(\ell)}$ are 0 and 1, respectively. On the other hand, for the case of imperfect sensing, the system designer considers the limited sensing capabilities of the SU and  the values of $\beta_{{\rm{ov}}}^{(m)}$ and $\beta_{{\rm{oo}}}^{(\ell)}$ are given as in (\ref{eq_Ch_3:b_ov}) and (\ref{eq_Ch_3:b_oo}), respectively. As can be seen, if the sensing errors are not taken into account, then the interference leaked in the $m$th PU band exceeds the threshold. On the other hand, if the sensing errors are considered, the interference to the $m$th PU band is below the threshold. In other words, if perfect sensing is assumed, then the SU transmits higher power that leads to higher interference levels at the $m$th PU.

\begin{figure}[!t]
\centering
\includegraphics[width=0.750\textwidth]{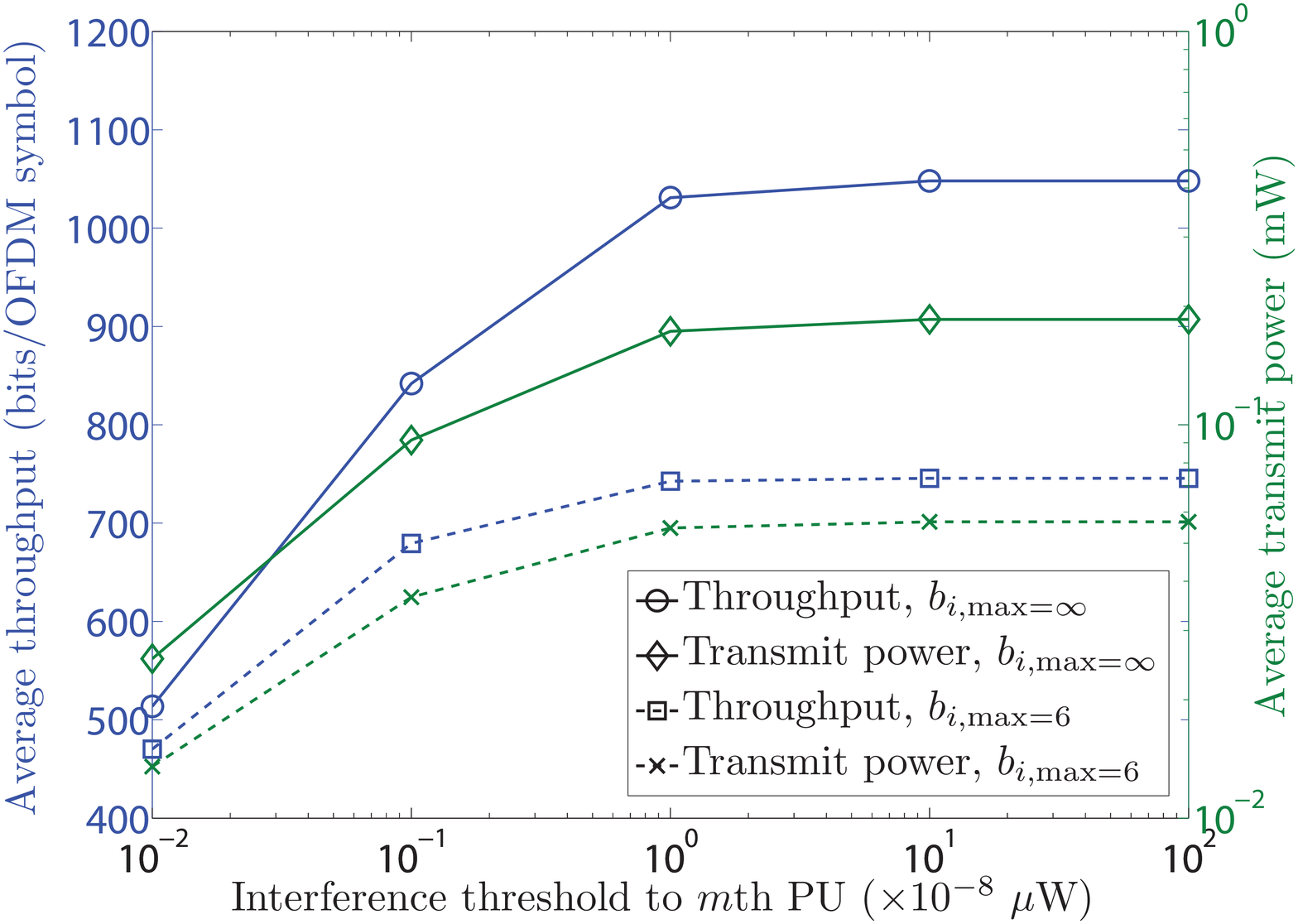}
\caption{Effect of $P_{\textup{th}}^{(m)}$ on the SU performance for $P_{\rm{th}} = 10^{-3}$ W, $P_{\textup{th}}^{(\ell)} = 10^{-8}$ $\mu$W, $\alpha$ = 0.5, and $b_{i,\textup{max}} = \infty$ and $6$.}
\label{fig_Ch_3:performance_CCI}
\end{figure}

Fig. \ref{fig_Ch_3:performance_CCI} depicts the average throughput and average transmit power as a function of the CCI threshold $P_{\textup{th}}^{(m)}$ for $P_{\rm{th}} = 10^{-3}$ W, $P_{\textup{th}}^{(\ell)} = 10^{-8}$ $\mu$W, $\alpha = 0.5$, and  $b_{i,\textup{max}} = \infty$ and $6$. As can be seen for $b_{i,\textup{max}} = \infty$, both the average throughput and average transmit power increase as $P_{\textup{th}}^{(m)}$ increases, and saturates for higher values of $P_{\textup{th}}^{(m)}$. This can be explained, as for lower values of $P_{\textup{th}}^{(m)}$ the CCI/total transmit power constraint is active and, hence, the total transmit power is limited by this constraint. Increasing $P_{\textup{th}}^{(m)}$ results in a corresponding increase in both the average throughput and total transmit power. For higher values of $P_{\textup{th}}^{(m)}$ (the CCI/total transmit power constraint is inactive), the proposed algorithm minimize the transmit power by keeping it constant, and, hence, the average throughput saturates. For $b_{i,\textup{max}} = 6$, as expected, the average throughput and transmit power are reduced due to the constraint on the maximum allocated bits per subcarrier.

\begin{figure}[!t]
\centering
\includegraphics[width=0.750\textwidth]{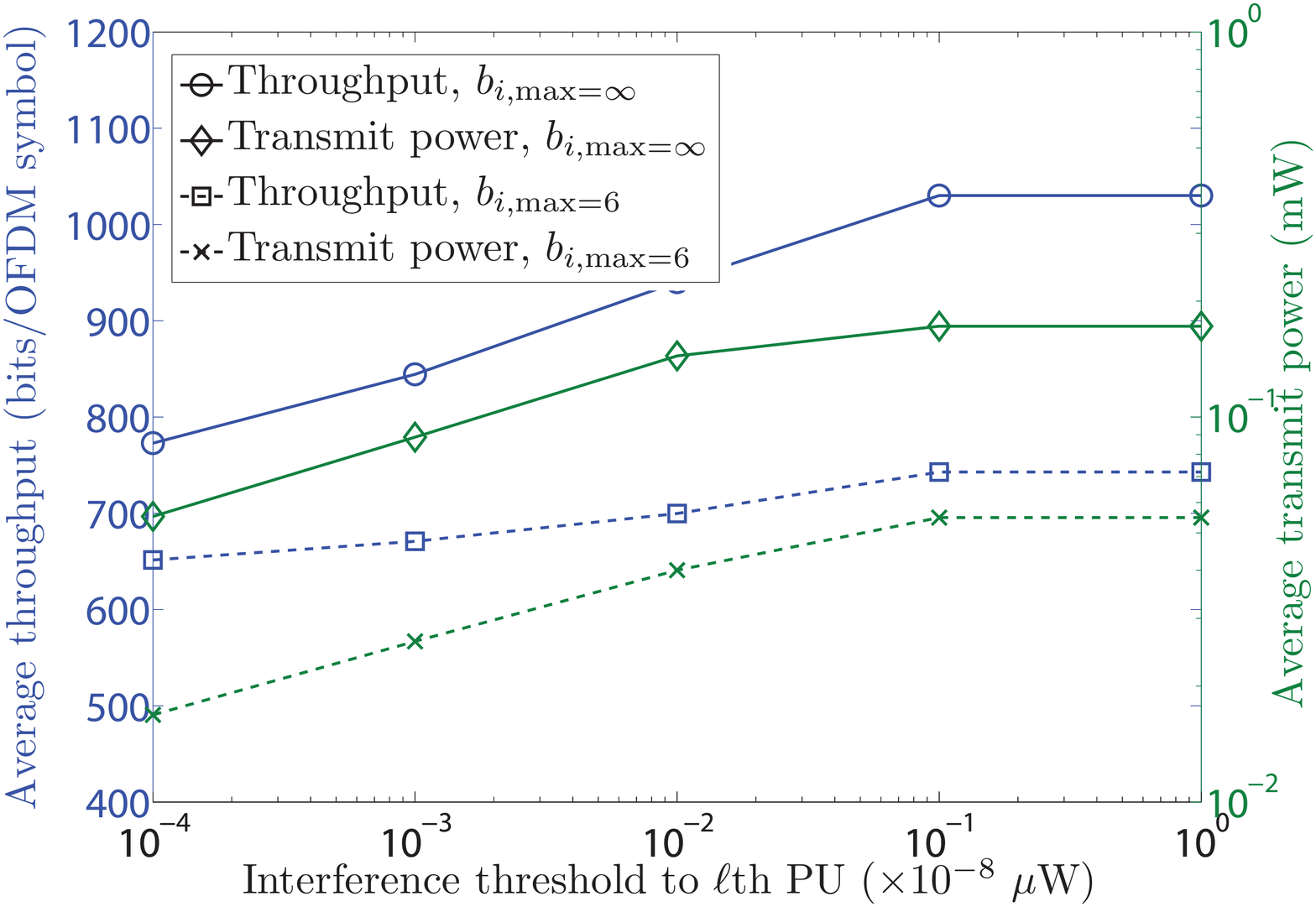}
\caption{Effect of $P_{\textup{th}}^{(\ell)}$ on the SU performance for $P_{\rm{th}} = 10^{-3}$ W, $P_{\textup{th}}^{(m)} = 10^{-8}$ $\mu$W, $\alpha$ = 0.5, and $b_{i,\textup{max}} = \infty$ and $6$.}
\label{fig_Ch_3:performance_ACI}
\end{figure}
Fig. \ref{fig_Ch_3:performance_ACI} depicts the average throughput and average transmit power as a function of the ACI threshold $P_{\textup{ACI}}$ for $P_{\rm{th}} = 10^{-3}$ W, $P_{\textup{th}}^{(m)} = 10^{-8}$ $\mu$W, $\alpha = 0.5$, and  $b_{i,\textup{max}} = \infty$ and $6$. Similar to the discussion  on Fig. \ref{fig_Ch_3:performance_CCI}, for $b_{i,\textup{max}} = \infty$, both the average throughput and average transmit power increase as $P_{\textup{th}}^{(\ell)}$ increases, and saturates for higher values of $P_{\textup{th}}^{(\ell)}$.
For $b_{i,\textup{max}} = 6$, as expected the average throughput and transmit power are reduced due to the constraint on the maximum allocated bits per subcarrier.

\begin{figure}[!t]
\centering
\includegraphics[width=0.750\textwidth]{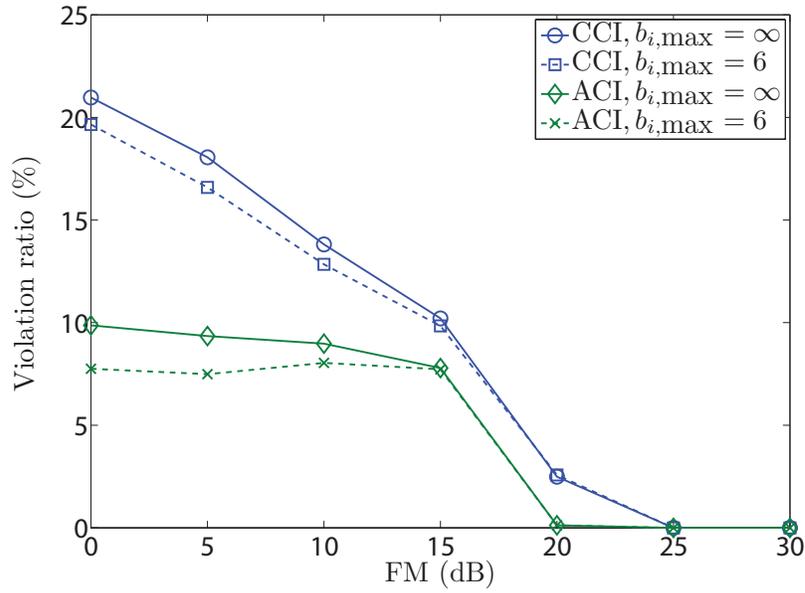}
\caption{Effect of FM on the violation ratio of the CCI and ACI constraints for $P_{\textup{th}}^{(m)} = P_{\textup{th}}^{(\ell)} = 10^{-10}$ $\mu$W, and $b_{i,\textup{max}} = \infty$ and $6$.}
\label{fig_Ch_3:VR}
\end{figure}

In Fig. \ref{fig_Ch_3:VR}, the violation ratios of the CCI and ACI constraints at the $m$th and $\ell$th PUs receivers, respectively, are plotted as a function of FM for $\alpha = 0.5$ and $P_{\textup{th}}^{(m)} = P_{\textup{th}}^{(\ell)} = 10^{-10}$ $\mu$W. We choose small values for the CCI and ACI thresholds so that their constraints are always active.  The violation ratios represent the percentage of simulation trials in which the CCI and ACI constraints are respectively violated at the $m$th and $\ell$th PUs receivers due to the partial channel information.  As can be seen, increasing the $\textup{FM}$ value reduces the violation ratio values as expected.
As it is difficult to estimate the channel between the SU transmitter and PU receivers, the CCI and ACI constraints are violated in practice when only knowledge of the path loss is available, and, hence, adding a fading margin becomes crucial to protect the PUs receivers.

\begin{figure}[!t]
\centering
\includegraphics[width=0.750\textwidth]{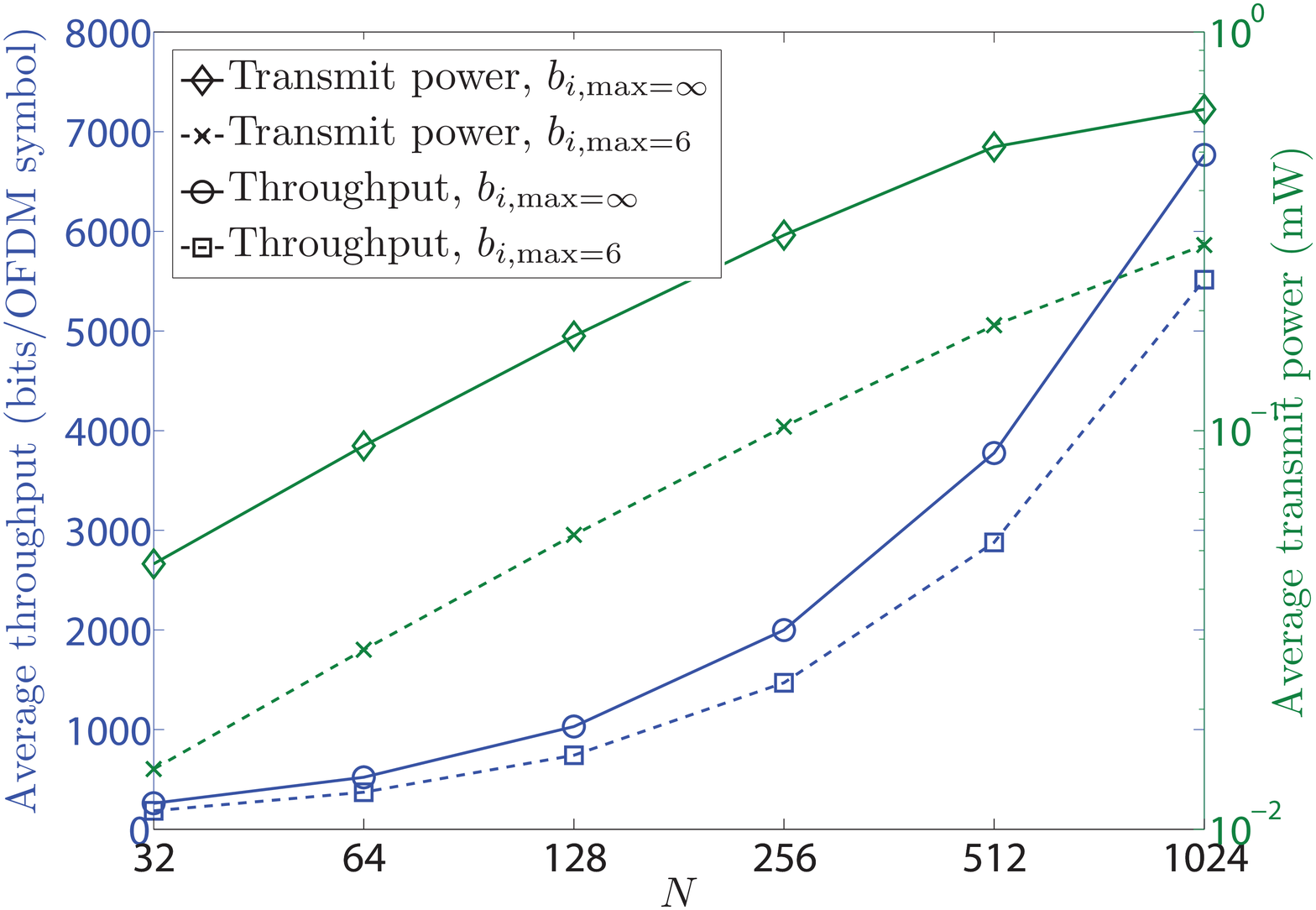}
\caption{Effect of $N$ on the SU performance for $P_{\rm{th}} = 10^{-3}$ W, $P_{\textup{th}}^{(m)} = P_{\textup{th}}^{(\ell)} = 10^{-8}$ $\mu$W, $\alpha = 0.5$, and $b_{i,\textup{max}} = \infty$ and $6$.}
\label{fig_Ch_3:performance_Nm}
\end{figure}

The effect of the number of subcarriers $N$ on the SU performance is depicted in Fig. \ref{fig_Ch_3:performance_Nm} for $\alpha = 0.5$, $P_{\rm{th}} = 10^{-3}$ W, $P_{\textup{th}}^{(m)} = P_{\textup{th}}^{(\ell)} = 10^{-8}$ $\mu$W, and $b_{i,\textup{max}} = \infty$ and $6$. For $b_{i,\textup{max}} = \infty$, increasing $N$ increases the average throughput and also increases the transmit power as long as neither the CCI nor the ACI constraints are violated. Such behaviour occurs as increasing the number of OFDM subcarriers reduces the out-of-band spectral leakage, and, hence, contributes lower interference levels to adjacent PUs. Accordingly, this increases the SUs chances to transmit more bits/power per subcarrier. For $b_{i,\textup{max}} = 6$, as expected, the average throughput and transmit power reduces due to the constraint on the maximum allocated bits per subcarrier.

\subsection{Performance Comparison with some Works in the literature and the Exhaustive Search}

\begin{figure}[!t]
\centering
		\includegraphics[width=0.75\textwidth]{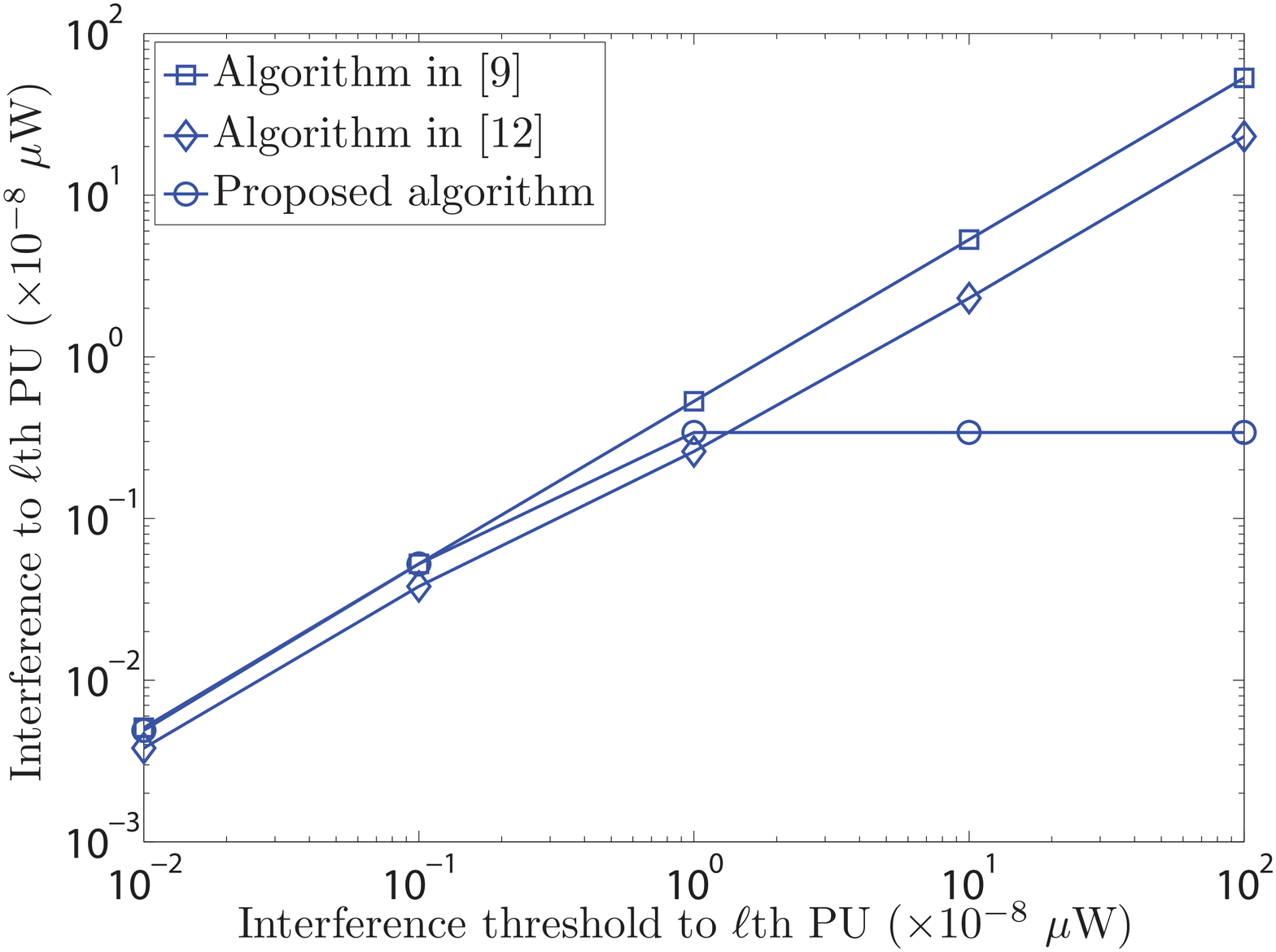}
	\caption{Comparison between the interference leaked to the $\ell$th PU for the proposed algorithm and the algorithm in~\cite{bansal2008optimal}. }
	\label{fig_Ch_3:Comp_SC_1_interference}
\end{figure}

In Fig. \ref{fig_Ch_3:Comp_SC_1_interference}, we compare the leaked interference to the $\ell$th PU receiver  for the proposed algorithm at $\alpha = 0.5$ and the works in \cite{bansal2008optimal, bansal2011adaptive} that assume perfect spectrum sensing. While the work in \cite{bansal2008optimal} assumes full CSI knowledge and maximizes the SU transmission rate with constraints on ACI and with no constraints on the CCI/total transmit power, the work in \cite{bansal2011adaptive} maximizes the SU transmission rate and satisfies the CCI/total transmit power and the ACI constraints in a probabilistic manner (i.e., meets the constraints with  a predefined probability).
For this and in order to match the operating conditions, we set $P_{\rm{th}} = P_{\textup{th}}^{(m)} = \infty$ and $b_{i,\textup{max}} = \infty$ in the proposed algorithm and consider knowledge of the path loss for the work in \cite{bansal2008optimal}. Furthermore, we set the signal-to-noise ratio (SNR) gap $\frac{- \ln(5 \: \textup{BER}_{{\rm{th}},i})}{1.6}$ to $10 \log_{10}(4.7506) = 6.77$ dB in \cite{bansal2008optimal, bansal2011adaptive}, and the predefined probability to meet the ACI constraints in \cite{bansal2011adaptive} to 90\%.
As can be observed, the work in \cite{bansal2008optimal, bansal2011adaptive} produces higher interference levels to the $\ell$th PU receiver (as well as, higher SU transmission rate) when compared to the proposed algorithm. This is expected as increasing the value of $\alpha$ in (\ref{eq_Ch_3:eq_1_21}) gives more weight to minimizing the SU transmit power and, hence, it is reduced. It is worthy to mention that the work in \cite{bansal2011adaptive} produces lower interference levels when compared to the work in \cite{bansal2008optimal} as it imposes a certain probability on violating the ACI constraint which does not exist in \cite{bansal2008optimal}. So, it is expected that reducing the value of the predefined probability in \cite{bansal2011adaptive} allows the SU to transmit higher power levels and to produce higher interference levels to existing PUs.

\begin{figure}[!t]
\centering
		\includegraphics[width=0.75\textwidth]{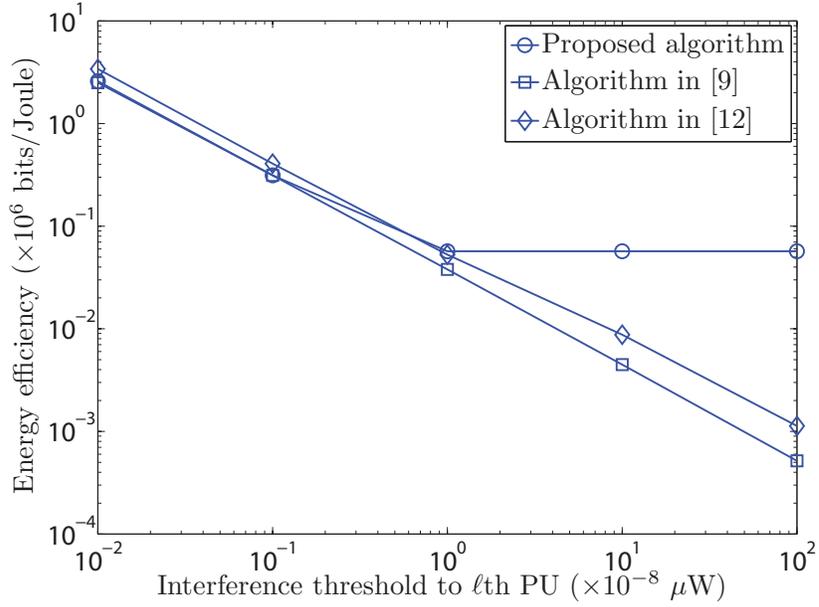}
	\caption{Comparison between the SU energy efficiency for the proposed algorithm and the algorithm in \cite{bansal2008optimal}.}
	\label{fig_Ch_3:efficiency_sc_1}
\end{figure}

Fig. \ref{fig_Ch_3:efficiency_sc_1} compares the energy efficiency (in bits/Joule) for the work in  \cite{bansal2008optimal, bansal2011adaptive} and the proposed algorithm at $\alpha = 0.5$ for the same operating conditions. As can be seen, the energy efficiency of the proposed algorithm is higher than its counterpart in \cite{bansal2008optimal, bansal2011adaptive} that decreases with increasing $P_{\textup{th}}^{(\ell)}$. This decrease is due to the logarithmic expression of the rate, i.e., $\log_2(1 + \gamma_i p_i)$, where increasing $P_{\textup{th}}^{(\ell)}$ (that corresponds to increasing the value of $p_i$) at the low range of the power results in a notable increase in the rate, while increasing $P_{\textup{th}}^{(\ell)}$ at the high range of the power results in a negligible increase in the rate. On the other hand, the energy efficiency of the proposed algorithm saturates as both the transmit power and the throughput saturate for the latter range of $P_{\textup{th}}^{(\ell)}$. The computational complexity of the works in \cite{bansal2008optimal, bansal2011adaptive} is $\mathcal{O}(N^3)$ when compared with $\mathcal{O}(N^2)$ of the proposed algorithm; hence, the improved energy efficiency of the proposed algorithm does not come at the cost of additional complexity.

\begin{figure}[!t]
	\centering
		\includegraphics[width=0.75\textwidth]{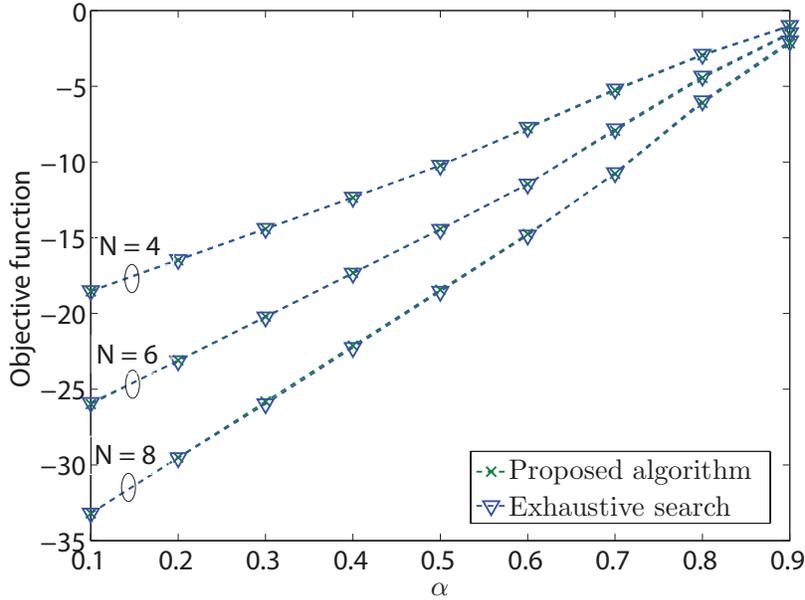}
	\caption{Comparison between the proposed algorithm and the exhaustive search for $P_{\rm{th}} = 5 \times 10^{-6}$ W, $P_{\textup{th}}^{(m)} = P_{\textup{th}}^{(\ell)} = 10^{-10}$ $\mu$W, and $b_{i,\textup{max}} = 6$.}
	\label{fig_Ch_3:ex}
\end{figure}
To characterize the gap between the proposed algorithm that finds the solution of the MOOP problem in (\ref{eq_Ch_3:eq_1_21}) and the discrete optimal solution, Fig. \ref{fig_Ch_3:ex} compares the values of the objective function achieved with the proposed algorithm and the optimal exhaustive search.  Note that the latter finds the discretized optimal allocation for the problems in (\ref{eq_Ch_3:eq_1_21}) by testing all  possible combinations of the bit and power allocations (the power per subcarrier is calculated from the discrete value of the bit allocation according to (\ref{eq_Ch_3:sub_P_b})) and selecting the pair with the least objective function value. Results are presented for $P_{\rm{th}} = 5 \times 10^{-6}$ W, $P_{\textup{th}}^{(m)} = P_{\textup{th}}^{(\ell)} = 10^{-10}$ $\mu$W, and $N$ = 4, 6, and 8; a small number of subcarriers is chosen, such that the exhaustive search is feasible. As can be seen in Fig. \ref{fig_Ch_3:ex}, the proposed algorithm approaches the discrete optimal results of the exhaustive search. Note that the complexity of the proposed algorithms is of $\mathcal{O}(N^2)$, which is significantly lower than $\mathcal{O}(N!)$ of the exhaustive search.


\section{Conclusions} \label{sec_Ch_3:conc}

Unlike prior work in the literature, this paper proposed a multiobjective optimization approach for the optimal link adaptation of OFDM-based CR systems. We jointly maximized the SU throughput and minimized its transmit power subject to total transmit power threshold and predefined CCI and ACI constraints to existing PUs. Additionally, we guaranteed a minimum BER and a maximum allocated bits per subcarrier for the SU, and considered the effect of imperfect spectrum sensing.
Closed-form expressions were derived for the close-to-optimal bit and power distributions per subcarrier.
Simulation results demonstrated the flexibility of the proposed algorithm to support different operating modes of the SU (i.e., to tune for various levels of throughput and transmit power as needed by the CR system) while meeting the constraints. For example, the SU may choose to maximize its throughput/transmission rate, and, hence, to reduce the transmission time by choosing lower values of $\alpha$. On the other hand, the SU may choose to reduce its transmit power, and hence, the interference to existing PUs, if the channel conditions to the PUs are not completely known or the spectrum sensing is not fully reliable by selecting higher values of $\alpha$ (interestingly, the SU transmission in this case is more energy-efficient when compared to the other case and/or to the work in the literature).
Moreover, the results show that the violation of the interference constraints can be due to 1) partial channel information of the links between the SU and the PUs receivers, where a fading margin becomes crucial to protect the PUs receivers, and 2) assuming perfect spectrum sensing. When compared to the single objective solutions, the multiobjective optimization approach tends to be more energy efficient at the cost of no additional complexity. Additionally,  the results indicated that the performance of the proposed algorithm approaches the discrete optimal results obtained by an exhaustive search, with significantly reduced computational effort.


\section*{Appendix A\\Proof of the Convexity of $\mathcal{OP}2$}

The Hessian of the objective function $\mathbf{f}_{\mathcal{OP}2}(\mathbf{b})$ can be written as
\begin{IEEEeqnarray}{RCL}
\nabla^2 \mathbf{f}_{\mathcal{OP}2}(\mathbf{b}) & = & \frac{\partial^2 \mathbf{f}_{\mathcal{OP}2}(\mathbf{b})}{\partial b_{i}^2} \nonumber \\
& = & \alpha (\ln(2)^2) \left( \frac{ - \ln(5 \textup{BER}_{{\rm{th}},i})}{1.6 \: \gamma_{i}}\right) 2^{b_{i}} \: \mathbf{I}_N. \IEEEeqnarraynumspace
\end{IEEEeqnarray}
For an arbitrary vector $\textbf{x}$, the value of $\textbf{x}^T \: \nabla^2 \mathbf{f}_{\mathcal{OP}2}(\mathbf{b}) \: \textbf{x}$ can be thus expressed as
\begin{IEEEeqnarray}{RCL}
\textbf{x}^T \: \nabla^2 \mathbf{f}_{\mathcal{OP}2}(\mathbf{b}) \: \textbf{x} &= & \alpha (\ln(2)^2) \left( \frac{ - \ln(5 \textup{BER}_{{\rm{th}},i})}{1.6 \: \gamma_{i}}\right) \: 2^{b_{i}} \: \textbf{x}^T \: \mathbf{I}_N \: \textbf{x}, \nonumber \\
\end{IEEEeqnarray}
which is positive semi-definite for any arbitrary vector $\textbf{x}$ and $b_{i}$; hence, the objective function $\mathbf{f}_{\mathcal{OP}2}(\mathbf{b})$ is convex. Note that the term  $\left(\frac{ - \ln(5 \textup{BER}_{{\rm{th}},i})}{1.6 \: \gamma_{i}}\right)$ is positive given that the value of the BER threshold per subcarrier $\textup{BER}_{{\rm{th}},i}$ is always less than $\frac{1}{5}$ for practical scenarios.

Similarly, the Hessian of the CCI/total transmit power and ACI constraints ($g_{\varrho_{\mathcal{OP}2}}(\mathbf{b}), \: \varrho_{\mathcal{OP}2} = 1, ..., L+1$) is positive semi-definite, and, hence, $\mathcal{OP}2$ is convex.
\hfill$\blacksquare$ 

\section*{Appendix B\\Proof of the Optimality of the Solution ($\mathbf{b}^*,\mathbf{p}^*$) of $\mathcal{OP}2$}

The KKT conditions are written as \cite{Boyd2004convex}
\begin{subequations}
\label{eq_Ch_3:KKT}
\begin{IEEEeqnarray}{rCl}
\frac{\partial \mathbf{f}_{\mathcal{OP}2}}{\partial b_{i}} +  \sum_{\varrho_{\mathcal{OP}2 = 1}}^{N+L+1} \lambda_{\varrho_{\mathcal{OP}2}} \: \frac{\partial g_{\varrho_{\mathcal{OP}2}}}{\partial b_{i}} &{} = {}& 0, \label{eq_Ch_3:KH2_5}\\
g_{\varrho_{\mathcal{OP}2}} \lambda_{\varrho_{\mathcal{OP}2}} &{}={}& 0, \label{eq_Ch_3:KH4_5} \\
g_{\varrho_{\mathcal{OP}2}} &{} \leq {}& 0, \label{eq_Ch_3:KH5_5} \\
\lambda_{\varrho_{\mathcal{OP}2}} &{} \geq {}& 0, \label{eq_Ch_3:KH3_52}
\end{IEEEeqnarray}
\end{subequations}
$i = 1, ..., N$ and $\varrho_{\mathcal{OP}2} = 1, ..., N+L+1$. One can show that (\ref{eq_Ch_3:KH2_5})-(\ref{eq_Ch_3:KH3_52}) are satisfied, as sketched in the proof below.

\begin{itemize}
  \item \textit{Proof of} (\ref{eq_Ch_3:KH2_5}): one can find that (\ref{eq_Ch_3:KH2_5}) is satisfied from (\ref{eq_Ch_3:eq2_op5}) directly.
  \item \textit{Proof of} (\ref{eq_Ch_3:KH4_5}): one of the following possibilities exist in \emph{cases} 1--8:
\begin{enumerate}
  \item Either $\lambda_{\varrho_{\mathcal{OP}2}} = 0$, $\varrho_{\mathcal{OP}2} = 1, ..., N+L+1$ (as in \emph{case} 1); hence, $g_{\varrho_{\mathcal{OP}2}} \lambda_{\varrho_{\mathcal{OP}2}} = 0$.
  \item Either $y_{\varrho_{\mathcal{OP}2}} = 0$, $\varrho_{\mathcal{OP}2} = 1, ..., N+L+1$, so $\mathcal{G}_{\varrho_{\mathcal{OP}2}} = g_{\varrho_{\mathcal{OP}2}} = 0$ from (\ref{eq_Ch_3:eq_const_op5}) (as in \emph{case} 8); hence, $g_{\varrho_{\mathcal{OP}2}} \lambda_{\varrho_{\mathcal{OP}2}} = 0$.
  \item Either: $\lambda_{\varrho_{x}} = 0$, $\varrho_{x} \in \{\varrho_{\mathcal{OP}2} = 1, ..., N+L+1\}$ and $y_{\varrho_{y}} = 0$, $\varrho_{y} \in \{\varrho_{\mathcal{OP}2} = 1, ..., N+L+1\}$, $\varrho_{x} \neq \varrho_{y}$ (as in \emph{cases} 2--7), hence, $\mathcal{G}_{\varrho_{y}} = g_{\varrho_{y}}$ from (\ref{eq_Ch_3:eq_const_op5})). Thus, $g_{\varrho_{\mathcal{OP}2}} \lambda_{\varrho_{\mathcal{OP}2}} = 0$, $\varrho_{\mathcal{OP}2} = 1, ..., N+L+1$.
\end{enumerate}
    \item \textit{Proof of} (\ref{eq_Ch_3:KH5_5}): adding non-negative slack variables in (\ref{eq_Ch_3:eq_const_op5}) guarantees that $g_{\varrho_{\mathcal{OP}2}} \leq 0$, $\varrho_{\mathcal{OP}2} = 1, ..., N+L+1$; hence, (\ref{eq_Ch_3:KH5_5}) is satisfied.
    \item \textit{Proof of} (\ref{eq_Ch_3:KH3_52}): the Lagrangian multipliers are found to be non-negative in order to obtain the optimal solution.\hfill$\blacksquare$ 
\end{itemize}


\bibliographystyle{IEEEtran}
\bibliography{IEEEabrv,mybib_file} 

\chapter{} \label{ch:TVT_MOOP}
\section{Abstract}
In this paper, we investigate the tradeoff between increasing the secondary users (SUs) transmission rate and reducing the interference levels at the primary users (PUs) for orthogonal frequency division multiplexing based cognitive radio systems. To achieve this target, we formulate a generalized multiobjective optimization (MOOP) problem that jointly maximizes the transmission rate of the SU and minimizes the co-channel interference (CCI) and adjacent channel interference (ACI) to existing PUs.
We additionally constrain the allowed CCI and ACI to the PUs in order to guarantee the PUs protection from harmful interference.
The MOOP problem is solved by linearly combining the normalized competing objective functions---through weighting coefficients---into a single objective function.
Prior work in the literature that maximizes the SU transmission rate can be considered as a special case of the generalized MOOP problem by setting the weighting coefficients associated with interference minimization to zero.
Since estimating the full channel-state information (CSI) of the links between the SU transmitter and the PUs receivers is practically challenging, we assume only partial CSI knowledge of these links. Simulation results illustrate the performance of the proposed algorithm  and quantify the SU performance loss due to incomplete CSI knowledge. Furthermore, the proposed algorithm is compared to state-of-the-art techniques and our performance results show that the proposed algorithm is more energy-aware, yet with reduced complexity.
\section{Introduction}
\vspace{-5pt}
The Federal Communications Commission's report \cite{fcc2002spectrum} reveals that the spectrum underutilization problem faced by the wireless industry is a result of traditional inefficient spectrum allocation policies rather than an actual scarcity of radio spectrum. Therefore, the concept of dynamic spectrum access is proposed to improve the spectrum utilization \cite{cabric2008addressing}. Cognitive radio (CR) promotes this concept by permitting secondary users (SUs) to opportunistically access spectrum holes in primary users (PUs) frequency bands, subject to constrained degradation of the PUs performance \cite{cabric2008addressing}.

Cognitive radio is based on a flexible software-defined-radio (SDR) platform that is capable of adapting its transmission parameters to surrounding environmental conditions, with two target objectives \cite{cabric2008addressing}: 1) improving the spectrum utilization by maximizing the transmission rate of SUs for a given bandwidth  and 2) controlling the amount of co-channel interference (CCI) and adjacent channel interference (ACI) leaked to PUs receivers due to the SUs transmission. Considering both objectives is a challenging task for the SDR platform, as they are conflicting, i.e., increasing the transmission rate of SUs is accompanied by an increase in the SU transmit power and, hence, potentially  excessive interference levels to PUs. Therefore, a tradeoff exists between the two objectives and it should be carefully investigated in order to have a flexible design that improves the overall performance of the CR systems. In prior work, this design flexibility was not fully exploited as all the proposed algorithms focused on maximizing the SUs transmission rate, with predefined thresholds for the leaked interference, and less attention was given to minimizing the leaked interference to PUs \cite{bansal2008optimal, zhang2010efficient, zhao2010power, bansal2011adaptive, hasan2009energy, kang2009optimal}.

Orthogonal frequency division multiplexing (OFDM) is widely recognized as an attractive candidate for SUs transmission due to its capabilities in analyzing the spectral activities of PUs \cite{weiss2004spectrum, bedeer2011partial, bedeer2012jointVTC, bedeer2012EBERGC, bedeer2012UBERGC, bedeer2012novelICC, bedeer2012adaptiveRWS, bedeer2013resource, bedeer2013adaptive, bedeer2013novel, bedeer2014rateCONF, bedeer2014multiobjective, bedeer2013joint, 
bedeer2014energy, bedeer2015systematic, bedeer2015rate}. 
Bansal \textit{et al.} \cite{bansal2008optimal} investigated the optimal power allocation problem in OFDM-based CR systems to maximize the  SU downlink transmission rate under a constraint on the instantaneous interference to PUs.
Zhang and Leung \cite{zhang2010efficient} proposed a low complexity suboptimal algorithm  in which SUs may access both non-active and active PUs frequency bands, as long as the total CCI and ACI are within acceptable limits. Zhao and Kwak \cite{zhao2010power} maximized the throughput of the  SU while keeping the interference to PUs below a certain threshold.
In \cite{bansal2011adaptive}, Bansal \textit{et al.} maximized the transmission rate of an OFDM-based CR network while satisfying probabilistic interference constraints to the PUs.
In \cite{hasan2009energy}, Hasan \textit{et al.}  presented a solution to maximize the SU transmission rate while taking into account the interference leakage to PUs and the availability of subcarriers, i.e., the activity of PUs in the licensed bands.
In general, it is preferable for SUs to generate interference levels that are lower than predefined limits to compensate for spectrum sensing or channel estimation errors, both of which may lead to violation of the CCI and ACI constraints.
Moreover, reducing the transmit power (that results from minimizing the CCI or ACI) is important due to various environmental and technical reasons, e.g.,  reducing global $\rm{CO}_2$ emissions and the power needed to operate future mobile broadband systems.
This motivates us to adopt a multiobjective optimization (MOOP) approach for the resource allocation problem to investigate the rate-interference tradeoff of OFDM-based CR systems. Recently, MOOP has attracted researchers' attention due to its flexible and superior performance over single objective optimization approaches, e.g., having two objectives in the cost function provides significant performance improvements when compared with having a single objective in the cost function and using the other objective as a constraint \cite{miettinen1999nonlinear, bedeer2013joint}.

In this paper, we provide a mathematical framework for the rate-interference tradeoff  of OFDM-based CR systems. This is achieved by formulating a MOOP problem that jointly maximizes the SU transmission rate and minimizes the leaked CCI and ACI interferences  to the PUs receivers.
We additionally set  predefined interference thresholds per each PU as constraints. We consider partial channel-state information (CSI) knowledge on the links between the SU transmitter and the PUs receivers and full CSI knowledge between the SU transmitter and receiver pair.
Simulation results show the performance of the proposed algorithm and illustrate the SU performance degradation due to the partial CSI knowledge. Additionally, the results show the advantages that the MOOP approach provides compared to the classical single optimization approaches proposed in the literature, with no additional complexity.


The remainder of the paper is organized as follows. Section \ref{sec_Ch_4:model} introduces the system model. Section \ref{sec_Ch_4:opt} analyzes the MOOP problem, outlines the proposed algorithm, and provides a complexity analysis. Simulation results are presented in Section \ref{sec_Ch_4:sim}, while conclusions are drawn in Section \ref{sec_Ch_4:conc}.

\section{System Model} \label{sec_Ch_4:model}

\subsection{System Description}

The available spectrum is divided into $L$ subchannels that are licensed to $L$ PUs. PUs do not necessarily fully occupy their licensed spectrum temporally and/or spatially; hence, an SU may access such spectrum holes as long as no harmful interference occurs to frequency-adjacent PUs due to ACI or to other PUs operating in the same frequency band at distant locations due to CCI \cite{zhao2007survey}. Without loss of generality, we assume that the SU decides to access subchannel $m$ of bandwidth $B_m$ using OFDM; this decision can be reached by consulting a database administrated by a government or third party, or by optionally sensing the PUs radio spectrum \cite{gao2012taxonomy}. 


As common practice in the literature, we assume that the instantaneous channel gains between the SU transmitter and receiver pair are available through a delay- and error-free feedback channel \cite{bansal2008optimal, zhang2010efficient, kang2009optimal, zhao2010power,  hasan2009energy, bansal2011adaptive}.
As estimating the instantaneous channel gains between from the SU transmitter to PUs receivers is practically challenging without the PUs cooperation, we assume partial CSI knowledge on the links between the SU transmitter and PUs receivers. More specifically, we assume: 1) knowledge of the path loss, which is practically possible especially in applications with stationary nodes, where the path loss exponent and the node locations can be estimated with high accuracy \cite{salman2012low} and 2) knowledge of the path loss and the channel statistics (i.e., the fading distribution and its parameters), which is a reasonable assumption for certain wireless environments, e.g., in non-line-of-sight urban environments, a Rayleigh distribution is usually assumed for the magnitude of the fading channel coefficients. The case of full CSI knowledge on the links between the SU transmitter and PUs receivers represents an upper bound on the achievable SU performance and is additionally provided in the numerical results section to characterize the performance loss due to partial CSI knowledge.
\vspace{-10pt}
\subsection{Modeling of the CCI and ACI Constraints with Partial CSI Knowledge}


\subsubsection{Case 1---Knowledge of the path loss} The transmit power on subchannel $m$ should be limited to a certain threshold $P_{{\rm{th}}}^{(m)}$ to protect the $m$th distant PU receiver from harmful CCI. This can be expressed as $10^{-0.1 \: PL(d_{m})}     \sum_{i = 1}^{N} p_i \leq P_{{\rm{th}}}^{(m)}$,
where $PL(d_m)$ is the distance-based path loss in dB at distance $d_m$ from the SU and $p_i$ is the allocated power per subcarrier $i$, $i = 1, ..., N$.
To reflect the SU transmitter's power amplifier limitations and/or to satisfy regulatory maximum power limits, the total SU transmit power should be limited to a certain threshold $P_{th}$ as $\sum_{i = 1}^{N} p_i \leq P_{th}.$
Hence, the constraint on the total transmit power is formulated as $\sum_{i = 1}^{N} p_i \leq \left[P_{th}, \frac{P_{{\rm{th}}}^{(m)}}{10^{-0.1 \: PL(d_{m})}    } \right]^-,$ where $[x,y]^-$ represents $\min(x,y)$.
To simplify the notation and without loss of generality, we assume that $\frac{P_{{\rm{th}}}^{(m)}}{10^{-0.1 \: PL(d_{m})}    } < P_{th}$. Hence, the CCI constraint is written as 
\begin{IEEEeqnarray}{c}
\sum_{i=1}^{N} p_i \leq P_{{\rm{th}}}^{(m)} X_{\textup{Case 1}}^{(m)},
\end{IEEEeqnarray}
where $X_{\textup{Case 1}}^{(m)} = \frac{1}{10^{-0.1 \: PL(d_{m})}    }$ represents the channel knowledge coefficient from the SU transmitter to the $m$th PU receiver for the case of only knowing the path loss.

The ACI is mainly due to the power spectral leakage of the SU subcarriers to the PUs receivers. This depends on the power allocated to each SU subcarrier and the spectral distance between the SU subcarriers and the PUs receivers. The ACI to the $\ell$th PU receiver should be limited to a certain threshold $P_{{\rm{th}}}^{(\ell)}$ as $10^{-0.1 \: PL(d_{\ell})}     \sum_{i=1}^{N} p_i \: \varpi_i^{(\ell)} \leq P_{{\rm{th}}}^{(\ell)}, \ell = 1, ..., L,$
where $\varpi_i^{(\ell)} = T_{s} \: \int_{f_{i,\ell}-\frac{B_\ell}{2}}^{f_{i,\ell}+\frac{B_\ell}{2}} \textup{sinc}^2(T_{s} f) \: df$, $T_s$ is the SU OFDM symbol duration, $f_{i,\ell}$ is the spectral distance between the SU subcarrier $i$ and the $\ell$th PU  frequency band, $B_{\ell}$ is the bandwidth of $\ell$th PU, and $\textup{sinc}(x) = \frac{\sin(\pi x)}{\pi x}$. The ACI constraint can be further written as 
\begin{IEEEeqnarray}{c}
\sum_{i = 1}^{N} p_i \: \varpi_i^{(\ell)} \leq P_{{\rm{th}}}^{(\ell)} X_{\textup{Case 1}}^{(\ell)}, \quad \ell = 1, ..., L,
\end{IEEEeqnarray}
where $X_{\textup{Case 1}}^{(\ell)} = \frac{1}{10^{-0.1 \: PL(d_{\ell})}    }$ is the channel knowledge coefficient from the SU transmitter to the $\ell$th PU receiver for the case of only knowing the path loss.


\subsubsection{Case 2---Knowledge of the path loss and channel statistics}
The CCI constraint is written as $|\mathcal{H}_{{\rm{sp}}}^{(m)} |^2 10^{-0.1 \: PL(d_{m})}     \sum_{i = 1}^{N} p_i \leq P_{{\rm{th}}}^{(m)},$
where $\mathcal{H}_{{\rm{sp}}}^{(m)}$ is the channel gain to the distant $m$th PU receiver. Since $\mathcal{H}_{{\rm{sp}}}^{(m)}$ is not perfectly known at the SU transmitter, the CCI constraint is limited below the threshold $P_{{\rm{th}}}^{(m)}$ with at least a probability of $\Psi_{{\rm{th}}}^{(m)}$. This is formulated as $\textup{Pr}\left( | \mathcal{H}_{{\rm{sp}}}^{(m)} |^2 10^{-0.1 \: PL(d_{m})}     \sum_{i = 1}^{N} p_i \leq P_{{\rm{th}}}^{(m)} \right)$ $\geq \Psi_{{\rm{th}}}^{(m)}.$
A non-line-of-sight propagation environment is assumed; therefore, the channel gain $\mathcal{H}_{{\rm{sp}}}^{(m)}$ can be modeled as a zero-mean complex Gaussian random variable, and, hence, $| \mathcal{H}_{{\rm{sp}}}^{(m)} |^2$ follows an exponential distribution \cite{bansal2011adaptive}. After some mathematical manipulations, the CCI statistical constraint can be expressed as 
\begin{IEEEeqnarray}{c}
\sum_{i = 1}^{N} p_i \leq P_{{\rm{th}}}^{(m)} X_{\textup{Case 2}}^{(m)},
\end{IEEEeqnarray}
where $X_{\textup{Case 2}}^{(m)} = \frac{\nu^{(m)}}{\left(-\ln(1 - \Psi_{{\rm{th}}}^{(m)})\right) 10^{- 0.1 \: PL(d_{m})}}$ is the channel knowledge coefficient from the SU transmitter to the $m$th PU receiver for the case of knowing the path loss and the channel statistics and $\frac{1}{\nu^{(m)}}$ is the mean of the exponential distribution. Similarly, the ACI constraint can be written as 
\begin{IEEEeqnarray}{c}
\sum_{i = 1}^{N} p_i \: \varpi_i^{(\ell)} \leq P_{{\rm{th}}}^{(\ell)} X_{\textup{Case 2}}^{(\ell)}, \quad \ell = 1, ..., L,
\end{IEEEeqnarray}
where $X_{\textup{Case 2}}^{(\ell)} = \frac{\nu^{(\ell)}}{\left(-\ln(1 - \Psi_{{\rm{th}}}^{(\ell)})\right) 10^{- 0.1 \: PL(d_{\ell})}}$ is the channel knowledge coefficient to the $\ell$th PU receiver for the case of knowing the path loss and the channel statistics and $\frac{1}{\nu^{(\ell)}}$ is the mean of the exponential distribution.


\section{Joint Rate and Interference Optimization} \label{sec_Ch_4:opt}

\subsection{Problem Formulation and Analysis}
For most of the MOOP problems, due to the contradiction and incommensurability of the competing objective functions it is not possible to find a single solution that optimizes all the objectives simultaneously, i.e., there is no solution that improves one of the objective functions without deteriorating other objectives. However, a set of non-dominated Pareto optimal solutions exists and it is the decision maker's (the SU in our case) responsibility to choose its preferred optimal solution \cite{miettinen1999nonlinear}. If the objective functions and constraints are convex, then the obtained Pareto optimal solution is referred to as a \emph{global} Pareto optimal solution; otherwise, it is refereed to as a \emph{local} Pareto optimal solution \cite{miettinen1999nonlinear}. Furthermore, the obtained solution is a \emph{weak} Pareto optimal solution if there is no other solution that causes every objective to improve;
otherwise, it is refereed to as a \emph{strong} Pareto optimal solution \cite{miettinen1999nonlinear}.

We formulate an MOOP problem that jointly minimizes the CCI to a distant PU (working on the same frequency band as the SU), minimizes the ACI to adjacent PUs, and maximizes the SU transmission rate, while guaranteeing acceptable levels of CCI and ACI to the existing PUs receivers, as
\begin{IEEEeqnarray}{c} \label{eq_Ch_4:MOOP_1}
\underset{p_i}{\textup{min}} \quad \frac{1}{X^{(m)}} \sum_{i = 1}^{N} p_i \quad \textup{and} \quad \underset{p_i}{\textup{min}} \quad \frac{1}{X^{(\ell)}} \sum_{i = 1}^{N} p_i \varpi_i^{(\ell)} \quad \textup{and} \quad \underset{p_i}{\textup{max}} \quad \Delta f \sum_{i = 1}^{N} \log_2(1 + p_i \: \frac{|\mathcal{H}_i|^2}{\sigma_n^2 + \mathcal{J}_i}), \nonumber
\end{IEEEeqnarray}
\begin{IEEEeqnarray}{c}
\textup{subject to} \qquad \textup{C1}: \sum_{i = 1}^{N} p_i  \leq  P_{{\rm{th}}}^{(m)} X^{(m)}, \quad \textup{and} \quad
\textup{C2}: \sum_{i = 1}^{N} p_i \varpi_i^{(\ell)} \leq P_{{\rm{th}}}^{(\ell)} X^{(\ell)},  \ell = 1, ..., L, \nonumber \\
\end{IEEEeqnarray}
where $X^{(m)} \in \left\{ X_{\textup{Case 1}}^{(m)}, X_{\textup{Case 2}}^{(m)} \right\}$ and $X^{(\ell)} \in \left\{ X_{\textup{Case 1}}^{(\ell)}, X_{\textup{Case 2}}^{(\ell)}\right\}$ represent the channel knowledge coefficients from the SU transmitter to the $m$th and $\ell$th PUs receivers, respectively, $\Delta f$ is the subcarrier spacing of the OFDM SU, $\mathcal{H}_i$ is the channel gain of subcarrier $i, i = 1, ..., N$, between the SU transmitter and receiver pair, $\sigma_n^2$ is the variance of the additive while Gaussian noise (AWGN), and $\mathcal{J}_i$ is the average interference power from all the PUs to the SU subcarrier $i, i = 1, ..., N$ where the PU signal is modeled as an elliptical filtered white noise process \cite {bansal2008optimal}. 
We solve the MOOP problem in (\ref{eq_Ch_4:MOOP_1}) by linearly combining the competing CCI, ACI, and rate objectives into a single objective function through weighting coefficients $\alpha_{\mathrm{CCI}}$, $\alpha_{\mathrm{ACI}}^{(\ell)}$, and $\alpha_{\mathrm{rate}}$, respectively. In order for the weighting coefficients to directly reflect the importance of the objectives, the CCI, ACI, and rate objectives are scaled using the normalization factors $u_{\mathrm{CCI}}$, $u_{\mathrm{ACI}}^{(\ell)}$, and $u_{\mathrm{rate}}$, respectively, such that they are approximately within the same range \cite{miettinen1999nonlinear}.  That being said, the normalization factors are set to the maximum of each objective, i.e., $u_{\mathrm{CCI}} = \frac{1}{P_{{\rm{th}}}^{(m)}}$, $u_{\mathrm{ACI}}^{(\ell)} = \frac{1}{P_{{\rm{th}}}^{(\ell)}}$, and $u_{\mathrm{rate}}$ is the inverse of the maximum achievable rate, so that the three objectives are within the range [0,1].
The MOOP in (\ref{eq_Ch_4:MOOP_1}) is written as
\begin{IEEEeqnarray}{c}
\underset{p_i}{\textup{min}} \quad \alpha_{\mathrm{CCI}} \frac{u_{\mathrm{CCI}}}{X^{(m)}} \sum_{i = 1}^{N} p_i +  \sum_{\ell = 1}^{L} \alpha_{\mathrm{ACI}}^{(\ell)} \frac{u_{\mathrm{ACI}}^{(\ell)}}{X^{(\ell)}}  \sum_{i = 1}^{N} p_i \varpi_i^{(\ell)}   - \alpha_{\mathrm{rate}} u_{\mathrm{rate}} \Delta f \sum_{i = 1}^{N} \log_2(1 + \gamma_i p_i), \nonumber \\
\textup{subject to} \qquad \textup{C1---C2}, \label{eq_Ch_4:OP1}
\end{IEEEeqnarray}
where
$\alpha_{\mathrm{CCI}} + \sum_{\ell = 1}^{L} \alpha_{\mathrm{ACI}}^{(\ell)} + \alpha_{\mathrm{rate}} = 1$. We assume that the SU chooses the proper values of $\alpha_{\mathrm{CCI}}, \alpha_{\mathrm{ACI}}^{(\ell)}, \alpha_{\mathrm{rate}}$ depending on the application, the surrounding environment, and/or the target performance \cite{miettinen1999nonlinear, bedeer2013joint}. For example, if the transmission rate, and, hence, the transmission time is crucial, then the SU chooses higher values for $\alpha_{\mathrm{rate}}$. On the other hand, if minimizing the CCI/ACI, and, hence, improving the energy efficiency is more important, then the SU chooses higher values for $\alpha_{\mathrm{CCI}}$/$\alpha_{\mathrm{ACI}}^{(\ell)}$.
The optimization problem in \eqref{eq_Ch_4:OP1} is convex, as the objective function is the sum of convex functions  and the constraints are convex \cite{Boyd2004convex}, and it can be solved by applying the Karush-Khun-Tucker (KKT) conditions (i.e., transforming the inequalities constraints to equality constraints by adding non-negative slack variables) \cite{Boyd2004convex}. The Lagrangian function $\mathcal{L}(\mathbf{p},\mathbf{y},\boldsymbol \lambda)$ is expressed as
\begin{IEEEeqnarray}{rcl}
\mathcal{L}(\mathbf{p},\mathbf{y},\boldsymbol \lambda) &{} = {}& \alpha_{\mathrm{CCI}} \frac{u_{\mathrm{CCI}}}{X^{(m)}} \sum_{i = 1}^{N} p_i +  \sum_{\ell = 1}^{L} \alpha_{\mathrm{ACI}}^{(\ell)} \frac{u_{\mathrm{ACI}}^{(\ell)}}{X^{(\ell)}}  \sum_{i = 1}^{N} p_i \varpi_i^{(\ell)}   - \alpha_{\mathrm{rate}} u_{\mathrm{rate}} \Delta f \sum_{i = 1}^{N} \log_2(1 + \gamma_i p_i) \nonumber \\
&{} + {}& \lambda_{1} \left[\sum_{i = 1}^{N} p_i - P_{{\rm{th}}}^{(m)} X^{(m)} + y_{1}^2 \right]  + \sum_{\ell = 1}^{L} \lambda_{2}^{(\ell)} \left[\sum_{i = 1}^{N} p_i \varpi_i^{(\ell)} -  P_{{\rm{th}}}^{(\ell)} X^{(\ell)} + (y_{2}^{(\ell)})^2 \right],
\end{IEEEeqnarray}
where $\mathbf{y} = \left[y_1^2, (y_{2}^{(\ell)})^2 \right]^T$ and $\boldsymbol \lambda = \left[\lambda_{1}, \lambda_{2}^{(\ell)} \right]^T$, $\ell = 1, ..., L$, are the vectors of the slack variables and Lagrange multipliers of length  $L + 1$, respectively. The optimal solution is found when $\nabla \mathcal{L}(\mathbf{p},\mathbf{y},\boldsymbol \lambda) = 0$, which yields
\begin{IEEEeqnarray}{rcl}
\frac{\partial \mathcal{L}}{\partial p_i} &{}={}& \alpha_{\mathrm{CCI}} \frac{u_{\mathrm{CCI}}}{X^{(m)}} +  \sum_{\ell = 1}^{L} \alpha_{\mathrm{ACI}}^{(\ell)} \frac{u_{\mathrm{ACI}}^{(\ell)}}{X^{(\ell)}} \varpi_i^{(\ell)} - \frac{\alpha_{\mathrm{rate}} u_{\mathrm{rate}} \Delta f}{\ln(2) (p_i + \gamma_i^{-1})} + \lambda_{1} + \sum_{\ell = 1}^{L} \lambda_{2}^{(\ell)} \varpi_i^{(\ell)} = 0, \label{eq_Ch_4:OP_1_first} \\
\frac{\partial \mathcal{L}}{\partial \lambda_{1}} &{}={}& \sum_{i = 1}^{N} p_i - P_{{\rm{th}}}^{(m)} X^{(m)} + y_{1}^2 = 0, \\
\frac{\partial \mathcal{L}}{\partial \lambda_{2}^{(\ell)}} &{}={}& \sum_{i = 1}^{N} p_i \varpi_i^{(\ell)} -  P_{{\rm{th}}}^{(\ell)} X^{(\ell)} + (y_{2}^{(\ell)})^2 = 0, \\
\frac{\partial \mathcal{L}}{\partial y_{1}} &{}={}& 2 \lambda_{1} y_{1} = 0, \label{eq_Ch_4:OP_1_6}\\
\frac{\partial \mathcal{L}}{\partial y_{2}^{(\ell)}} &{}={}& 2 \lambda_{2}^{(\ell)} y_{2}^{(\ell)} = 0. \label{eq_Ch_4:OP_1_last}
\end{IEEEeqnarray}
It can be seen that (\ref{eq_Ch_4:OP_1_first})--(\ref{eq_Ch_4:OP_1_last}) represent $N + 2 L + 2$ equations in the $N + 2 L + 2$ unknown components of the vectors $\mathbf{p}, \mathbf{y}$, and $\boldsymbol \lambda$. From \eqref{eq_Ch_4:OP_1_first}, the optimal power allocation per subcarrier is given as
\begin{IEEEeqnarray}{c} \label{eq_Ch_4:power}
p_i^* = \left[\frac{\alpha_{\mathrm{rate}} u_{\mathrm{rate}} \Delta f/\ln(2)}{\alpha_{\mathrm{CCI}} \frac{u_{\mathrm{CCI}}}{X^{(m)}} + \sum_{\ell = 1}^{L} \alpha_{\mathrm{ACI}}^{(\ell)}\frac{u_{\mathrm{ACI}}^{(\ell)}}{X^{(\ell)}} \varpi_i^{(\ell)}  + \lambda_{1} + \sum_{\ell = 1}^{L} \lambda_{2}^{(\ell)}\varpi_i^{(\ell)}} - \gamma_i^{-1}\right]^+, \quad i = 1, ..., N, \IEEEeqnarraynumspace
\end{IEEEeqnarray}
where $[x]^+$ represents $\max(0,x)$. In \eqref{eq_Ch_4:power}, the value of the Lagrangian multipliers $\lambda_{1}$ and $\lambda_{2}^{(\ell)}$ are determined as explained below depending on whether the CCI and ACI constraints are active or inactive\footnote{A constraint on the form $\Gamma(x) \leq \Gamma_{\rm{th}}$ is said to be inactive if $\Gamma(x) < \Gamma_{\rm{th}}$, while it is active if $\Gamma(x) = \Gamma_{\rm{th}}$.}, respectively. Equation (\ref{eq_Ch_4:OP_1_6}) implies that either $\lambda_{1} = 0$ or $y_{1} = 0$ and (\ref{eq_Ch_4:OP_1_last}) implies that either $\lambda_{2}^{(\ell)} = 0$ or $y_{2}^{(\ell)} = 0$, $\ell = 1, ..., L$. Hence, four possible cases exist, as follows:

---\textit{Case 1}: Setting $\lambda_{1} = 0$ (i.e., $\sum_{i = 1}^{N} p_i^* < P_{{\rm{th}}}^{(m)} X^{(m)}$) and $\lambda_{2}^{(\ell)} = 0$ (i.e., $\sum_{i = 1}^{N} p_i^* \varpi_i^{(\ell)} < P_{{\rm{th}}}^{(\ell)} X^{(\ell)}$) results in the optimal solution for inactive CCI and ACI constraints.

---\textit{Case 2}: Setting $y_{1} = 0$ (i.e., $\sum_{i = 1}^{N} p_i^* = P_{{\rm{th}}}^{(m)} X^{(m)}$) and $\lambda_{2}^{(\ell)} = 0$ (i.e., $\sum_{i = 1}^{N} p_i^* \varpi_i^{(\ell)} < P_{{\rm{th}}}^{(\ell)} X^{(\ell)}$) results in the optimal solution for active CCI and inactive ACI constraints.

---\textit{Case 3}: Setting $\lambda_{1} = 0$ (i.e., $\sum_{i = 1}^{N} p_i^* < P_{{\rm{th}}}^{(m)} X^{(m)}$) and $y_{2}^{(\ell)} = 0$ (i.e., $\sum_{i = 1}^{N} p_i^* \varpi_i^{(\ell)} = P_{{\rm{th}}}^{(\ell)} X^{(\ell)}$) results in the optimal solution for inactive CCI and active ACI constraints.

---\textit{Case 4}: Setting $y_{1} = 0$ (i.e., $\sum_{i = 1}^{N} p_i^* = P_{{\rm{th}}}^{(m)} X^{(m)}$) and $y_{2}^{(\ell)} = 0$ (i.e., $\sum_{i = 1}^{N} p_i^* \varpi_i^{(\ell)} = P_{{\rm{th}}}^{(\ell)} X^{(\ell)}$) results in the optimal solution for active CCI and ACI constraints.

Similar to the discussion in Appendix B of Chapter \ref{ch:TW}, the solution $p_i^*$ can be shown to satisfy the KKT conditions \cite{Boyd2004convex}, and, hence, it is an optimal solution.




\subsection{Proposed Algorithm and Complexity Analysis}

The proposed algorithm can be formally stated as follows:
\floatname{algorithm}{}
\begin{algorithm}
\renewcommand{\thealgorithm}{}
\caption{\textbf{Proposed Algorithm}}
\begin{algorithmic}[1]
\small
\State \textbf{INPUT} $\sigma^2_n$, $\mathcal{H}_{i}$, $\alpha_{\mathrm{CCI}}$, $\alpha_{\mathrm{ACI}}^{(\ell)}$, $\alpha_{\mathrm{rate}}$, $P_{{\rm{th}}}^{(m)}$, $P_{{\rm{th}}}^{(\ell)}$, $X^{(m)}$, and $X^{(\ell)}$, $\ell = 1, ..., L$.
        \algstore{myalg}
  \end{algorithmic}
\end{algorithm}

\floatname{algorithm}{}
\begin{algorithm}
 \renewcommand{\thealgorithm}{}
  \caption{\textbf{Proposed Algorithm} (continued)}
  \begin{algorithmic}
      \algrestore{myalg}
      \small
\State - assume the optimal solution $p_i^*$ belongs to case 1. \label{step:impo}
\State - find $p_i^*$ from \eqref{eq_Ch_4:power} when $\lambda_1 = \lambda_2^{(\ell)} = 0$.
\If{in Step \ref{step:impo}, the assumption on the CCI constraint is not true and the assumption on the ACI constraint is true}
\State - the optimal solution $p_i^*$ belongs to case 2, i.e., find a non-negative $\lambda_1$ from \eqref{eq_Ch_4:power} such that $\sum_{i = 1}^{N} p_i^* = P_{{\rm{th}}}^{(m)} X^{(m)}$.
\EndIf
\If{in Step \ref{step:impo}, the assumption on the CCI constraint is true and the assumption on the ACI constraint is not true}
\State - the optimal solution $p_i^*$ belongs to case 3, i.e., find a non-negative $\lambda_2^{(\ell)}$ from \eqref{eq_Ch_4:power} such that $\sum_{i = 1}^{N} p_i^* \varpi_i^{(\ell)} = P_{{\rm{th}}}^{(\ell)} X^{(\ell)}$, $\ell = 1, ..., L$.
\EndIf
\If{in Step \ref{step:impo}, the assumption on the CCI constraint is not true and the assumption on the ACI constraint is not true}
\State - the optimal solution $p_i^*$ belongs to case 4, i.e., find non-negative $\lambda_1$ and $\lambda_2^{(\ell)}$ from \eqref{eq_Ch_4:power} such that $\sum_{i = 1}^{N} p_i^* = P_{{\rm{th}}}^{(m)} X^{(m)}$ and $\sum_{i = 1}^{N} p_i^* \varpi_i^{(\ell)} = P_{{\rm{th}}}^{(\ell)} X^{(\ell)}$, $\ell = 1, ..., L$, respectively.
\EndIf
\State \textbf{OUTPUT} $p_i^*$, $i$ = 1, ..., $N$.
\end{algorithmic}
\end{algorithm}

The complexity order to find $p_i^*$ is $\mathcal{O}(N\Omega)$, where $\mathcal{O}(\Omega)$ is the complexity to find the Lagrangian multipliers.
The authors in \cite{palomar2005practical} showed that the Lagrange multipliers $\lambda_{1}$ and $\lambda_{2}^{(\ell)}$, $\ell = 1, ..., L$, that satisfy the CCI and ACI constraints, respectively, can be obtained with linear complexity of the number of subcarrier $N$, i.e., $\mathcal{O}(N)$. Hence, the complexity of the proposed algorithm is $\mathcal{O}(N^2)$.

\section{Numerical Results} \label{sec_Ch_4:sim}

Without loss of generality, we assume that the OFDM SU coexists with one frequency-adjacent PU and one co-channel PU. The SU parameters are as follows: number of subcarriers $N = 128$ and subcarrier spacing $\Delta f = \frac{1.25 \: \rm{MHz}}{N} = 9.7656$ kHz. The propagation path loss parameters are as follows: exponent $= 4$, wavelength $= \frac{3 \times 10^8}{900 \times 10^6} = 0.33$ meters, distance between SU transmitter and receiver pair $= 1$ km, distance to the $\ell$th PU $d_{\ell} = 1.2$ km, distance to the $m$th PU $d_m = 5$ km, and reference distance $d_0 = 100$ m. A Rayleigh fading environment is considered, where the average channel power gains between the SU transmitter and receiver pair $\mathbb{E}\{|\mathcal{H}_i|^2\}$, between the SU transmitter and the receiver of the $\ell$th PU $\mathbb{E}\{|\mathcal{H}_{{\rm{sp}}}^{(\ell)}|^2\}$, and between the SU transmitter and the receiver of the $m$th PU $\mathbb{E}\{|\mathcal{H}_{{\rm{sp}}}^{(m)}|^2\}$ are set to 0 dB. $\sigma_n^2$ is assumed to be $10^{-15}$ W and the PU signal at the SU receiver is assumed to be an elliptically filtered white noise process of variance $\sigma_n^2$ \cite{bansal2008optimal}. Representative results are presented in this section, which were obtained through Monte Carlo trials for $10^{4}$ channel realizations. Unless otherwise mentioned, the value of the probabilities $\Psi_{{\rm{th}}}^{(m)}$ and $\Psi_{{\rm{th}}}^{(\ell)}$ is set to 0.9, $P_{{\rm{th}}}^{(m)} = 10^{-11}$ W, and $P_{{\rm{th}}}^{(\ell)} = 10^{-11}$ W. In order to better understand the MOOP approach, we consider the performance of the proposed algorithm when: 1) $\alpha_{\mathrm{CCI}} \neq 0$, $\alpha_{\mathrm{rate}} \neq 0$, and $\alpha_{\mathrm{ACI}}^{(\ell)} = 0$, 2) $\alpha_{\mathrm{ACI}}^{(\ell)} \neq 0$, $\alpha_{\mathrm{rate}} \neq 0$, and $\alpha_{\mathrm{CCI}} = 0$, and 3) $\alpha_{\mathrm{CCI}} \neq 0$, $\alpha_{\mathrm{ACI}}^{(\ell)} \neq 0$, and $\alpha_{\mathrm{rate}} \neq 0$.

\subsection{Performance of the Proposed Algorithm}
\begin{figure}
\centering
		\includegraphics[width=0.75\textwidth]{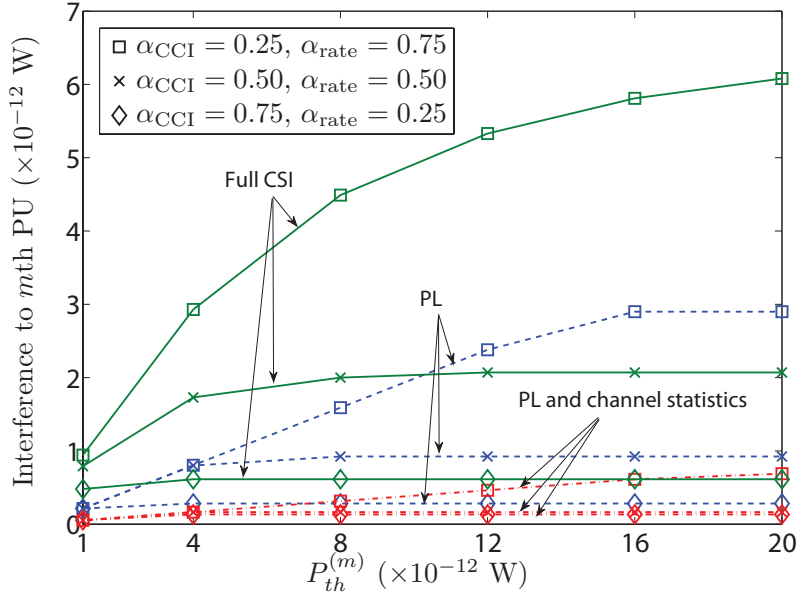}
	\caption{Interference leaked to the $m$th PU as a function of $P_{{\rm{th}}}^{(m)}$ for different values of $\alpha_{\mathrm{CCI}}$ and $\alpha_{\mathrm{rate}}$ and for different degree of CSI knowledge, at $P_{{\rm{th}}}^{(\ell)} = 10^{-11}$ W and $\Psi_{{\rm{th}}}^{(m)} = \Psi_{{\rm{th}}}^{(\ell)} = 0.9$.}
	\label{fig_Ch_4:1_min_CCI_PCCI_PCCI}
\end{figure}

\begin{figure}
\centering
		\includegraphics[width=0.750\textwidth]{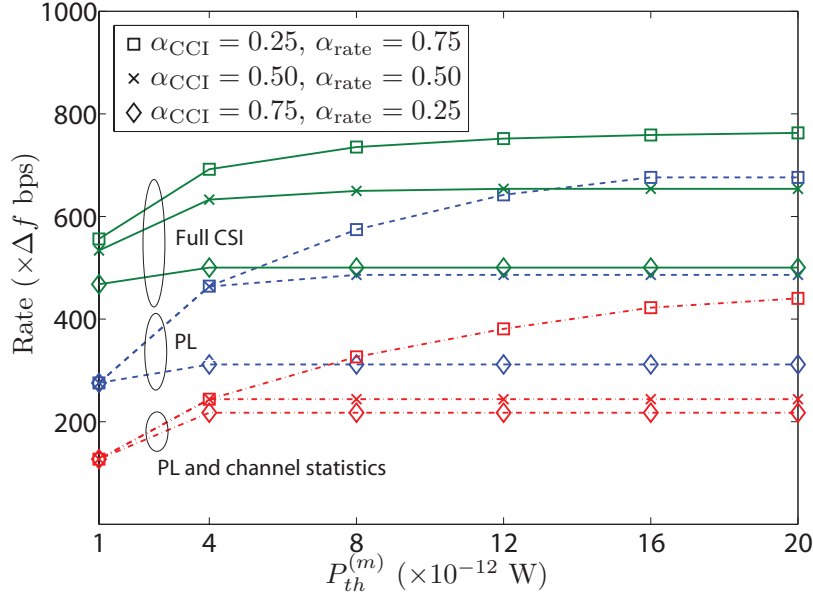}
	\caption{SU rate as a function of $P_{{\rm{th}}}^{(m)}$ for different values of $\alpha_{\mathrm{CCI}}$ and $\alpha_{\mathrm{rate}}$ and for different degree of CSI knowledge, at $P_{{\rm{th}}}^{(\ell)} = 10^{-11}$ W and $\Psi_{{\rm{th}}}^{(m)} = \Psi_{{\rm{th}}}^{(\ell)} = 0.9$.}
	\label{fig_Ch_4:1_min_CCI_rate_PCCI}
\end{figure}

\begin{figure}
\centering
		\includegraphics[width=0.75\textwidth]{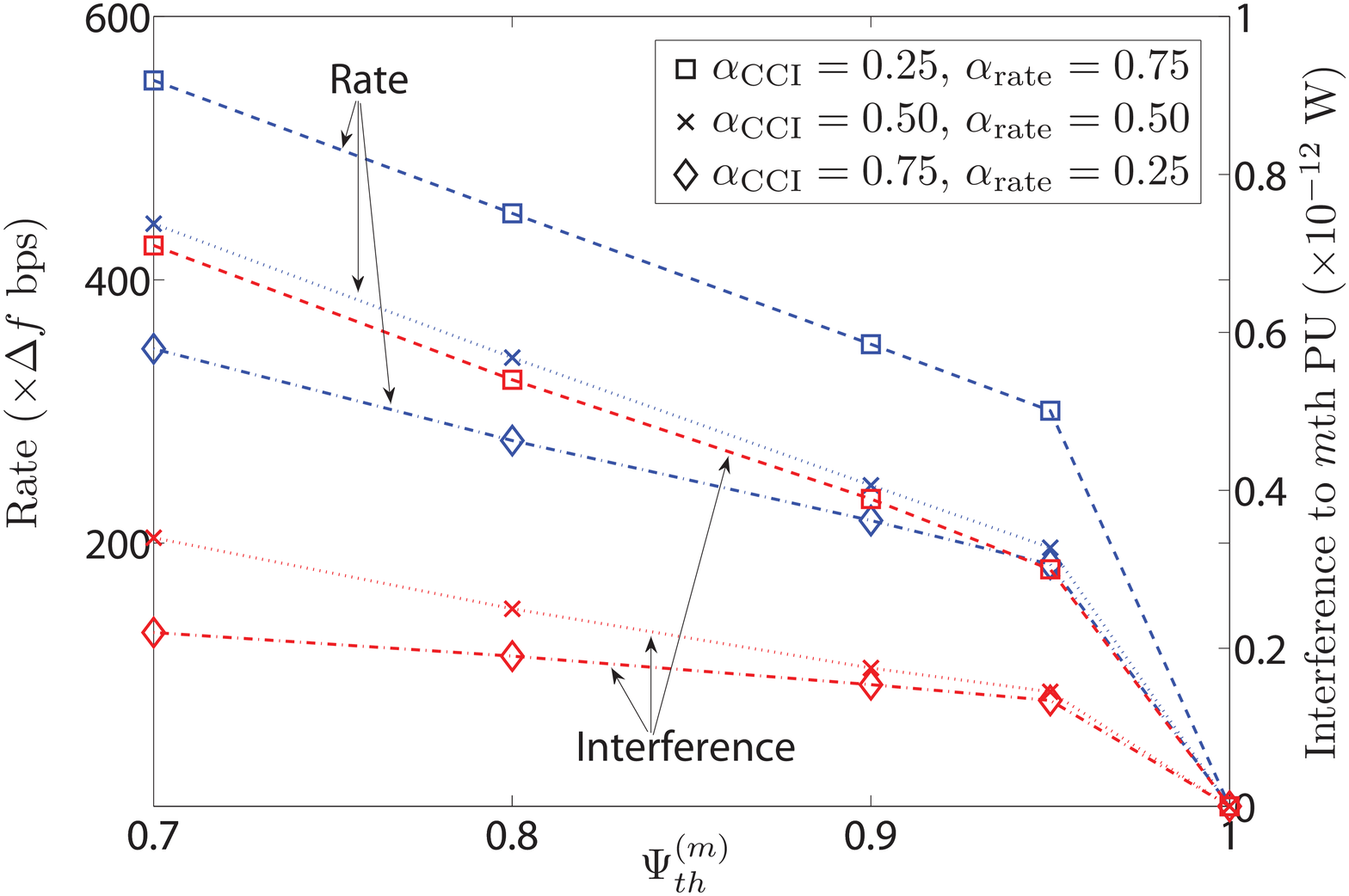}
	\caption{SU rate and interference leaked to the $m$th PU as a function of $\Psi_{{\rm{th}}}^{(m)}$ for different values of $\alpha_{\mathrm{CCI}}$ and $\alpha_{\mathrm{rate}}$, at $P_{{\rm{th}}}^{(m)} = P_{{\rm{th}}}^{(\ell)} = 10^{-11}$~W and $\Psi_{{\rm{th}}}^{(\ell)} = 0.9$.}
	\label{fig_Ch_4:Effect of gamma}
\end{figure}

Fig. \ref{fig_Ch_4:1_min_CCI_PCCI_PCCI} shows the interference leaked to the $m$th PU receiver as a function of $P_{{\rm{th}}}^{(m)}$\footnote{It is worthy to mention that the proposed algorithm performance is investigated over a large scale of $P_{{\rm{th}}}^{(m)}$ values, however, we focus here on the range up to $20\times10^{-12}$ W. This is as for higher than $20\times10^{-12}$ W the performance starts to saturate.} at $\alpha_{\mathrm{ACI}}^{(\ell)} = 0$ for different values of $\alpha_{\mathrm{CCI}}$ and $\alpha_{\mathrm{rate}}$ and for different degrees of CSI knowledge. As can be seen, increasing the value of $\alpha_{\mathrm{CCI}}$ (which is equivalent to decreasing the value of $\alpha_{\mathrm{rate}}$, as $\alpha_{\mathrm{CCI}} + \alpha_{\mathrm{rate}} = 1$ at $\alpha_{\mathrm{ACI}}^{(\ell)} = 0$) reduces the leaked interference to the $m$th PU for all the cases of CSI knowledge. This can be easily explained, as increasing $\alpha_{\mathrm{CCI}}$ gives more weight to minimizing the CCI objective and less weight to maximizing the rate objective in (\ref{eq_Ch_4:OP1}). Accordingly, increasing $\alpha_{\mathrm{CCI}}$ reduces the CCI to the $m$th PU receiver, but also the SU achievable rate. The interference leaked to the $m$th PU receiver increases linearly with increasing $P_{{\rm{th}}}^{(m)}$ for lower values of $P_{{\rm{th}}}^{(m)}$ and saturates for higher values of $P_{{\rm{th}}}^{(m)}$. This can be explained as follows. For lower values of $P_{{\rm{th}}}^{(m)}$, the interference leaked to the $m$th PU receiver is higher than the value of $P_{{\rm{th}}}^{(m)}$ and, hence, it is limited by the value of $P_{{\rm{th}}}^{(m)}$. On the other hand, for higher values of $P_{{\rm{th}}}^{(m)}$, the interference leaked to the $m$th PU receiver is less than the value of $P_{{\rm{th}}}^{(m)}$ as it is minimized by the proposed algorithm, and, hence,  it is kept constant. As expected, knowing the full CSI allows the SU to exploit this knowledge and to transmit with higher power (without violating the interference constraints at the PUs) and higher rate (as shown in the discussion of Fig.~\ref{fig_Ch_4:1_min_CCI_rate_PCCI}). On the other hand, partial CSI knowledge reduces the transmission opportunities of the SU in order not to violate the interference constraints. Note that the case of knowing only the path loss generates higher interference levels (and higher SU transmit power, hence, higher SU rates as shown in Fig. \ref{fig_Ch_4:1_min_CCI_rate_PCCI}) to existing PUs when compared to the case of knowing the path loss and the channel statistics. This is due to the high values of the predefined probabilities $\Psi_{{\rm{th}}}^{(m)}$ and $\Psi_{{\rm{th}}}^{(\ell)}$ (= 0.9); reducing these values produces higher interference levels to the PUs and higher SU rates, as it will be shown later in Fig. \ref{fig_Ch_4:Effect of gamma}.

Fig. \ref{fig_Ch_4:1_min_CCI_rate_PCCI} depicts the SU achievable rate as a function of $P_{{\rm{th}}}^{(m)}$  at $\alpha_{\mathrm{ACI}}^{(\ell)} = 0$ for different values of $\alpha_{\mathrm{CCI}}$ and $\alpha_{\mathrm{rate}}$ and for different degrees of CSI knowledge.
Similar to the discussion of Fig. \ref{fig_Ch_4:1_min_CCI_PCCI_PCCI}, the SU achievable rate saturates for higher values of $P_{{\rm{th}}}^{(m)}$. This is because the SU transmit power saturates for higher values of $P_{{\rm{th}}}^{(m)}$. As expected, increasing the value of $\alpha_{\mathrm{CCI}}$ (or decreasing the value of $\alpha_{\mathrm{rate}}$) decreases the SU achievable rate. Further, knowing the full CSI results in higher transmission rate when compared to partial CSI knowledge.

In Fig. \ref{fig_Ch_4:Effect of gamma}, the leaked interference to the $m$th PU receiver and SU achievable rate are depicted as a function of the probability $\Psi_{{\rm{th}}}^{(m)}$, respectively, at $\alpha_{\mathrm{ACI}}^{(\ell)} = 0$ for different values of $\alpha_{\mathrm{CCI}}$ and $\alpha_{\mathrm{rate}}$. As expected, increasing the value of $\Psi_{{\rm{th}}}^{(m)}$, decreases the leaked interference to the $m$th PU receiver and the SU achievable rate in order to meet such tight statistical constraints (i.e., meeting the CCI constraint with higher probability). The achieved SU rate and leaked interference to the $m$th PU receiver drop to zero for  $\Psi_{{\rm{th}}}^{(m)} = 1$, as the proposed algorithm cannot meet such stringent requirements of satisfying the active CCI constraint all the time, without knowledge of the instantaneous channel gains.

\begin{figure}
\centering
		\includegraphics[width=0.75\textwidth]{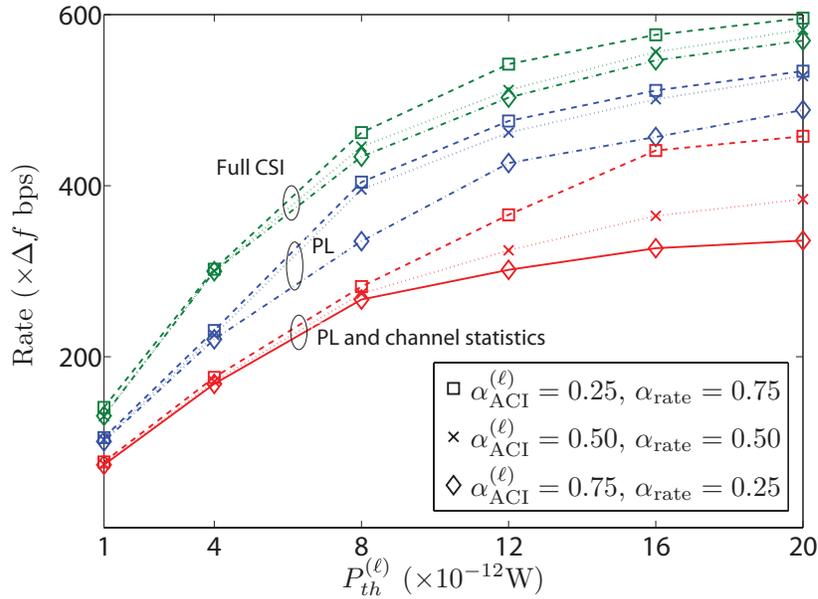}
	\caption{SU rate as a function of $P_{{\rm{th}}}^{(\ell)}$ for different values of $\alpha_{\mathrm{ACI}}^{(\ell)}$ and $\alpha_{\mathrm{rate}}$ and for different degree of CSI knowledge, at $P_{{\rm{th}}}^{(m)} = 10^{-11}$ W and $\Psi_{{\rm{th}}}^{(m)} = \Psi_{{\rm{th}}}^{(\ell)} = 0.9$.}
	\label{fig_Ch_4:2_min_ACI_rate_PACI}
\end{figure}

\begin{figure}
\centering
		\includegraphics[width=0.75\textwidth]{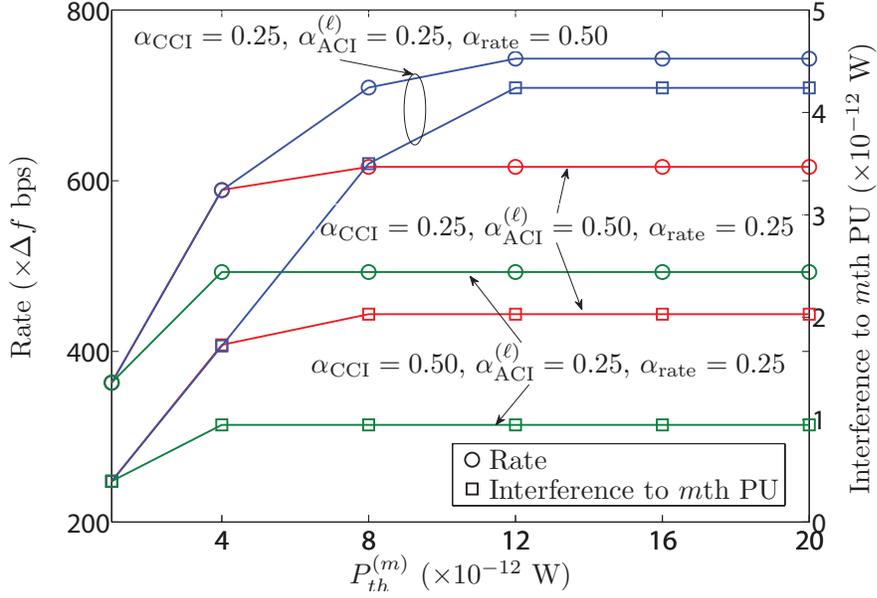}
	\caption{Effect of $\alpha_{\rm{CCI}}$, $\alpha_{\rm{ACI}}^{(\ell)}$, and $\alpha_{\rm{rate}}$ on the SU performance.}
	\label{fig_Ch_4:3obj}
\end{figure}

Fig. \ref{fig_Ch_4:2_min_ACI_rate_PACI} shows the achievable SU rate as a function of $P_{{\rm{th}}}^{(\ell)}$ at $\alpha_{\mathrm{CCI}} = 0$ for different values of $\alpha_{\mathrm{ACI}}^{(\ell)}$ and $\alpha_{\mathrm{rate}}$ and different degrees of CSI knowledge.
As can be noticed, increasing the value of $P_{{\rm{th}}}^{(\ell)}$ increases the SU rate. This occurs as increasing $P_{{\rm{th}}}^{(\ell)}$ apparently increases the transmit power and, hence, the rate increases.
As expected, increasing the value of $\alpha_{\mathrm{ACI}}^{(\ell)}$ (which is equivalent to decreasing the value of $\alpha_{\mathrm{rate}}$, as $\alpha_{\mathrm{ACI}}^{(\ell)} + \alpha_{\mathrm{rate}} = 1$ at $\alpha_{\mathrm{CCI}} = 0$) decreases the achievable SU rate. Moreover, knowing the full CSI allows the SU to achieve higher rates and transmit higher power without violating the CCI and ACI constraints.

Fig. \ref{fig_Ch_4:3obj} shows the effect of changing the weighting coefficients $\alpha_{\rm{CCI}}$, $\alpha_{\rm{ACI}}^{(\ell)}$, and $\alpha_{\rm{rate}}$ on the SU rate and interference to the $m$th PU. Similar to the previous discussions, one can notice that the SU achieves a higher transmission rate for increased $\alpha_{\rm{rate}}$ and the leaked interference to the $m$th PU is reduced for increased $\alpha_{\rm{CCI}}$. Additionally, the effect of $\alpha_{\rm{CCI}}$ on the SU rate is stronger when compared with that of $\alpha_{\rm{ACI}}^{(\ell)}$; this is because the SU rate is a function of the transmit power which is affected more by $\alpha_{\rm{CCI}}$ (related to the power itself) than $\alpha_{\rm{ACI}}^{(\ell)}$ (related to the weighted power). We should note that
increasing $\alpha_{\rm{ACI}}^{(\ell)}$ reduces the interference to the adjacent $\ell$th PU receiver; however, the results are not included due to space limitations.

\begin{figure}
\centering
		\includegraphics[width=0.750\textwidth]{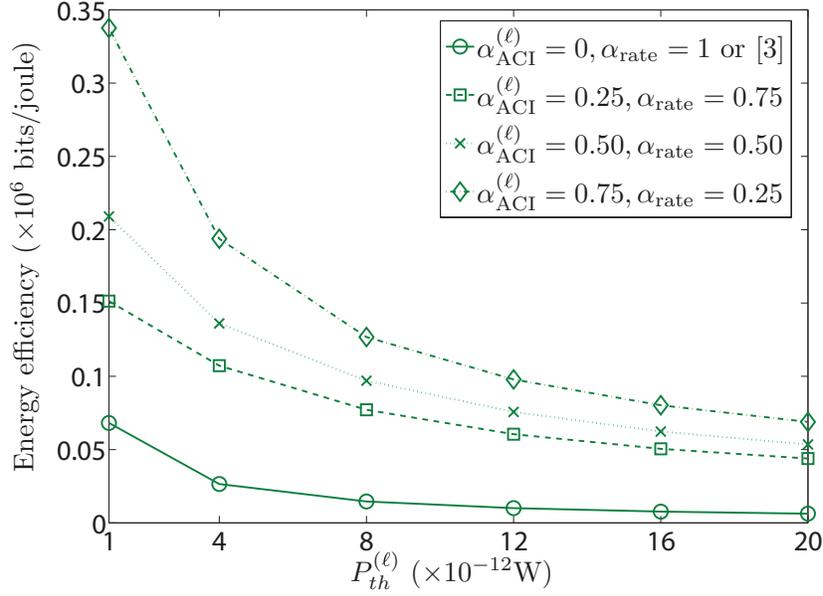}
	\caption{Comparison between the SU energy efficiency of the proposed algorithm and the algorithm in \cite{bansal2008optimal} that corresponds to $\alpha_{\mathrm{CCI}} = \alpha_{\mathrm{ACI}}^{(\ell)} = 0$.}
	\label{fig_Ch_4:2_min_ACI_efficiency_PACI_comp}
\end{figure}

\begin{figure}
\centering
		\includegraphics[width=0.750\textwidth]{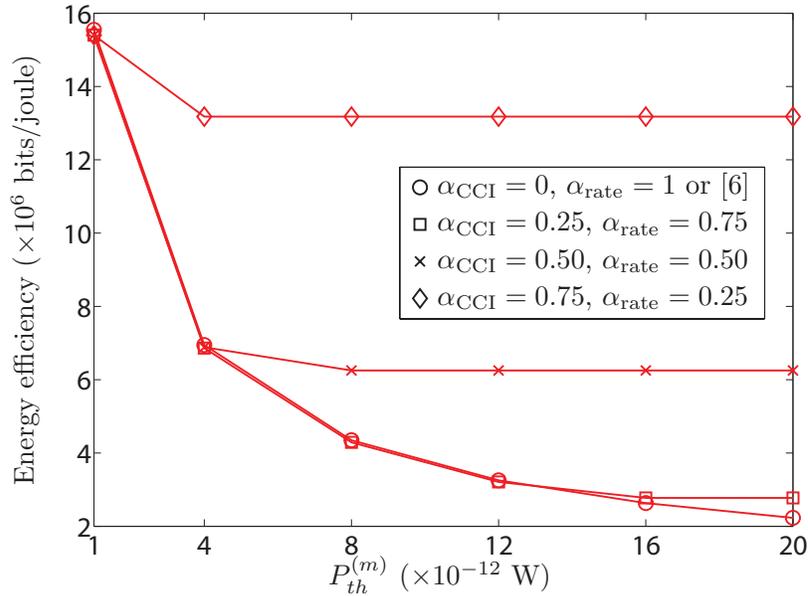}
	\caption{Comparison between the SU energy efficiency of the proposed algorithm and the algorithm in \cite{bansal2011adaptive} that corresponds to $\alpha_{\mathrm{CCI}} = \alpha_{\mathrm{ACI}}^{(\ell)} = 0$.}
	\label{fig_Ch_4:1_min_CCI_efficiency_PCCI_comp}
\end{figure}

\subsection{Performance Comparison with Algorithms in the Literature}

Fig. \ref{fig_Ch_4:2_min_ACI_efficiency_PACI_comp} compares the energy efficiency (in bits/joule) of the work in  \cite{bansal2008optimal} and the proposed algorithm, at $\alpha_{\mathrm{CCI}} = 0$ for different values of $\alpha_{\mathrm{ACI}}^{(\ell)}$ and $\alpha_{\mathrm{rate}}$ and  for the same operating conditions. As can be seen, the energy efficiency of the proposed algorithm is higher than its counterpart in \cite{bansal2008optimal} and it decreases with increasing $P_{{\rm{th}}}^{(\ell)}$. This is due to the logarithmic expression of the rate, i.e., $\log_2(1 + \gamma_i p_i)$, where increasing $P_{{\rm{th}}}^{(\ell)}$ (that corresponds to increasing the value of $p_i$) at the low range of the power results in a notable increase in the rate, while increasing the power at the high range of the power results in a negligible increase in the rate. The computational complexity of the work in \cite{bansal2008optimal} is $\mathcal{O}(N^3)$ when compared with $\mathcal{O}(N^2)$ of the proposed algorithm; hence, the improved energy efficiency of the proposed algorithm is achieved with reduced complexity.

In Fig. \ref{fig_Ch_4:1_min_CCI_efficiency_PCCI_comp}, the energy efficiency of the work in \cite{bansal2011adaptive} and the proposed algorithm, at $\alpha_{\mathrm{ACI}}^{(\ell)} = 0$ for different values of $\alpha_{\mathrm{CCI}}$ and $\alpha_{\mathrm{rate}}$, is compared for the same operating conditions. As can be noticed, the proposed algorithm is more energy efficient when compared to the work in \cite{bansal2011adaptive}.
The energy efficiency of the proposed algorithm saturates for higher values of $P_{{\rm{th}}}^{(m)}$; this is expected as the transmit power and the rate saturate for higher values of $P_{{\rm{th}}}^{(m)}$ (as can be seen from Figs. \ref{fig_Ch_4:1_min_CCI_PCCI_PCCI} and \ref{fig_Ch_4:1_min_CCI_rate_PCCI}). The complexity of the algorithm in \cite{bansal2011adaptive} is $\mathcal{O}(N^3)$ when compared with $\mathcal{O}(N^2)$ of the proposed algorithm; hence, the improved energy efficiency of the proposed algorithm is achieved with reduced complexity.

\section{Conclusions} \label{sec_Ch_4:conc}
In this paper, we considered an OFDM-based CR network and adopted a multiobjective optimization approach to investigate the tradeoff between improving the spectrum utilization (through increasing the SU transmission rate) and reducing the CCI and ACI to the PUs. This formulation is considered as a generalization of the work in the literature that focused only on maximizing the SU transmission rate. A flexible low complexity algorithm was proposed to solve the MOOP problem. Simulation results showed the flexibility of the proposed algorithm, with which the SU can tradeoff rates and interference levels optimally by changing the weighting coefficients. Further, results show the advantage of using the MOOP approach when compared to the single objective approaches in terms of improving the energy efficiency with reduced complexity.
\vspace{-10pt}

\bibliographystyle{IEEEtran}
\bibliography{IEEEabrv,mybib_file} 

\chapter{} \label{ch:TVT_EE}
\section{Abstract}
In this paper, we propose a novel algorithm to optimize the energy-efficiency (EE) of orthogonal frequency division multiplexing-based cognitive radio systems under channel uncertainties.
We formulate an optimization problem that guarantees a minimum required rate and a specified power budget for the secondary user (SU), while restricting the interference to primary users (PUs) in a statistical manner.
The optimization problem is non-convex and it is transformed to an equivalent problem using the concept of fractional programming. Unlike all related works in the literature, we consider the effect of imperfect channel-state-information (CSI) on the links between the SU transmitter and receiver pairs and we additionally consider the effect of limited sensing capabilities of the SU.  Since the interference constraints are met statistically, the SU transmitter does not require perfect CSI feedback from the PUs receivers. Simulation results show that the EE deteriorates as the channel estimation error increases. Comparisons with relevant works from the literature show that the interference thresholds at the PUs receivers can be severely exceeded and the EE is slightly deteriorated if the SU does not account for spectrum sensing errors.

\section{Introduction}
Cognitive radio (CR)  can considerably enhance the spectrum utilization efficiency by dynamically sharing the spectrum between licensed/primary users (PUs) and unlicensed/secondary users (SUs) \cite{cabric2008addressing}.
This is achieved by granting  SUs opportunistic access to the white spaces within  PUs spectrum, while controlling the interference to  PUs. Orthogonal frequency division multiplexing (OFDM) is recognized as an attractive modulation technique for CR due to its spectrum shaping flexibility, adaptivity in allocating vacant radio resources, and capability of analyzing the spectral activities of PUs \cite{bedeer2011partial, bedeer2012jointVTC, bedeer2012EBERGC, bedeer2012UBERGC, bedeer2012novelICC, bedeer2012adaptiveRWS, bedeer2013resource, bedeer2013adaptive, bedeer2013novel, bedeer2014rateCONF, bedeer2014multiobjective, bedeer2013joint, 
bedeer2014energy, bedeer2015systematic, bedeer2015rate,wang2011new}. Generally speaking, the interference introduced to  PUs bands in OFDM-based CR networks can be classified as: 1) mutual interference (co-channel interference (CCI) and adjacent channel interference (ACI)) between the SU and PUs due to the non-orthogonality of their respective transmissions \cite{wang2011new} and 2) interference due to the SU's imperfect spectrum sensing capabilities \cite{cabric2008addressing}.

Most of the existing research has focused on optimizing the transmission rate of SUs while limiting the interference introduced to PUs to predefined thresholds (see, e.g., \cite{almalfouh2011interference, bansal2011adaptive} and references therein). Recently, optimizing the energy-efficiency (EE)---defined as the total energy consumed to deliver one bit, or its inverse\footnote{The EE can be defined as the number of bits per unit energy. However, it is common to define it as the total energy consumed to deliver one bit, please see \cite{amin2012cooperative, wang2012optimal,  oto2012energy}.}
---has received increasing attention due to steadily rising energy costs and environmental concerns \cite{amin2012cooperative, wang2012optimal,  oto2012energy, xie2012energy, wangenergy, mao2013energy, mao2013energy2, amin2012opportunistic}.
Wang   \textit{et al.} in \cite{wang2012optimal} optimized the EE of an OFDM-based CR network subject to power budget and interference constraints; however, this comes at the expense of deteriorating the rate of the SU.
Oto and Akan in \cite{oto2012energy} found the optimal packet size that maximizes the EE of CR sensor networks while maintaining acceptable interference levels to the licensed PUs. In \cite{xie2012energy}, Xie \textit{et al.} investigated the problem of maximizing the EE of heterogeneous cognitive radio networks coexisting with femtocells.
Wang \textit{et al.} in \cite{wangenergy} optimized the EE of OFDM-based CR system subject to PUs interference constraints and different SUs rates.
In \cite{mao2013energy}, Mao \textit{et al.} optimized the EE of CR MIMO broadcast channels while guaranteeing certain interference threshold at the PUs receivers.
The same authors optimized the EE of OFDM-based CR systems subject to controlled interference leakage to PUs in \cite{mao2013energy2}.
To the authors' knowledge, all prior research on optimizing the EE has assumed that the SU has perfect spectrum sensing capabilities and perfect channel-state-information (CSI) for the links between the SU transmitter and receiver pairs \cite{mao2013energy, mao2013energy2, wang2012optimal, oto2012energy, xie2012energy, wangenergy}. However, in practice sensing is not fully reliable due to SU hardware limitations and variable channel conditions. Furthermore, it is also of practical importance to study the impact of channel estimation errors for the SU links on the EE optimization problem.


In this paper,
we formulate a novel EE optimization problem for the SU subject to its total transmit power budget and predefined quality-of-service (QoS) in terms of the minimum supported rate, as well as statistical constraints on the CCI and ACI to existing PUs. The optimization problem considers channel estimation errors for the links between the SU transmitter and receiver pairs, along with SU spectrum sensing errors. Furthermore, the SU does not rely on perfect CSI for the links between the SU transmitter and PUs receivers, since the interference constraints are met statistically.


The remainder of the paper is organized as follows. Section \ref{sec_Ch_5:model} introduces the system model. Section \ref{sec_Ch_5:opt} analyzes the optimization problem and outlines the proposed algorithm for its solution. Simulation results are presented in Section \ref{sec_Ch_5:sim}, while conclusions are drawn in Section \ref{sec_Ch_5:conc}.


\section{System Model} \label{sec_Ch_5:model}
\subsection{System Description}


The available spectrum is assumed to be divided into $L$ subchannels that are licensed to $L$ PUs. 
We assume that the SU periodically senses the PUs spectrum in order to identify vacant bands for its transmission. Without loss of generality, we consider that the SU senses that subchannel $m$, of bandwidth $B$, is vacant.
However, due to the varying channel conditions between the SU and PUs, the SU may not detect the presence of the $m$th PU.
This means that the SU identifies the $m$th PU band as vacant when it is truly occupied. This is referred to as a mis-detection error and it is assumed to occur with a probability $\rho^{(m)}_{\rm{md}}$. On the other hand, the SU may identify the $\ell$th PU band as occupied when it is truly vacant. This is referred to as a false-alarm error and it is assumed to occur with a probability $\rho^{(\ell)}_{\rm{fa}}$.  Mis-detection errors lead to severe co-channel interference to the $m$th PU, while false-alarm errors result in the SU wasting transmission opportunities.

\subsection{Modeling the Statistical CCI and ACI Constraints with Imperfect SU Sensing}
Using the Bayes' theorem and the law of total probability, the probability that subchannel $m$ is truly occupied under the condition that the SU identified it to be vacant can be expressed as \cite{almalfouh2011interference}
\begin{IEEEeqnarray}{c}
\beta_{\rm{ov}}^{(m)} = \frac{\rho^{(m)}_{\rm{md}}  \rho^{(m)}}{\rho^{(m)}_{\rm{md}}  \rho^{(m)} + (1 - \rho^{(m)}_{\rm{fa}}) (1 - \rho^{(m)})}, \label{eq:b_ov}
\end{IEEEeqnarray}
where $\rho^{(m)}$ is the probability that the PU transmits on subchannel $m$ and $\beta_{\rm{ov}}^{(m)}$ represents the probability that the interference due to mis-detection errors will be present in subchannel $m$, which is determined to be vacant by the SU.
Furthermore, the probability that subchannel $\ell$ is truly occupied by the PU under the condition that the SU identified it to be occupied can be written as
\begin{IEEEeqnarray}{c}
\beta_{\rm{oo}}^{(\ell)} = \frac{ (1 - \rho^{(\ell)}_{\rm{md}})  \rho^{(\ell)}}{ (1 - \rho^{(\ell)}_{\rm{md}})  \rho^{(\ell)} + \rho^{(\ell)}_{\rm{fa}} (1 - \rho^{(\ell)})}. \label{eq:b_oo}
\end{IEEEeqnarray}
Note that for perfect sensing
$\beta_{\rm{ov}}^{(m)} = 0$ and $\beta_{\rm{oo}}^{(\ell)} = 1$.

Estimating the channel gains between the SU transmitter and the PUs receivers is challenging without the PUs cooperation. Hence, we assume that the SU transmitter has only knowledge of the fading distribution type and its corresponding parameters of the channels on these links. This is a reasonable assumption for certain wireless environments. For example, a Rayleigh distribution is usually assumed for the magnitude of the fading channel coefficients in non-line-of-sight urban environments.
The constraint on the CCI from the SU to the $m$th PU is formulated as $\beta_{\rm{ov}}^{(m)} |\mathcal{H}_{\rm{sp}}^{(m)} |^2 G^{(m)}  \sum_{i = 1}^{N} p_i \leq P_{\rm{th}}^{(m)},$
where $\mathcal{H}_{\rm{sp}}^{(m)}$ and $G^{(m)}$ are the channel gain and the distance-based path loss\footnote{The SU is assumed to know the PUs location information by accessing a Radio Environment Map \cite{zhao2007applying}.} to the distant $m$th PU receiver, $p_i$ is the power allocated to subcarrier $i$, $i = 1, ..., N$, and $P_{\rm{th}}^{(m)}$ is the interference threshold at the $m$th PU receiver. Since $\mathcal{H}_{\rm{sp}}^{(m)}$ is not perfectly known at the SU transmitter, the CCI constraint is limited below the threshold $P_{\rm{th}}^{(m)}$ with at least a probability of $\Psi_{\rm{th}}^{(m)}$. This is formulated as $\textup{Pr}\left(\beta_{\rm{ov}}^{(m)} | \mathcal{H}_{\rm{sp}}^{(m)} |^2 G^{(m)}     \sum_{i = 1}^{N} p_i \leq P_{\rm{th}}^{(m)} \right) \geq \Psi_{\rm{th}}^{(m)}.$
A non-line-of-sight propagation environment is assumed; therefore, the channel gain $\mathcal{H}_{\rm{sp}}^{(m)}$ can be modeled as a zero-mean complex Gaussian random variable, and, hence, $| \mathcal{H}_{\rm{sp}}^{(m)} |^2$ follows the exponential distribution \cite{proakisdigital}. After some mathematical manipulations, the CCI statistical constraints  can be expressed as $\sum_{i = 1}^{N} p_i \leq \frac{1}{\beta_{\rm{ov}}^{(m)}} \frac{\nu^{(m)}}{G^{(m)} \left(-\ln(1 - \Psi_{\rm{th}}^{(m)})\right)} P_{\rm{th}}^{(m)}$,
where $\frac{1}{\nu^{(m)}}$ is the mean of the exponential distribution.
To further reflect the SU transmitter's power amplifier limitations and/or satisfy regulatory maximum power limits, the total SU transmit power is limited to a certain threshold $P_{\rm{th}}$ as $\sum_{i = 1}^{N} p_i \leq P_{\rm{th}}$. Therefore, the constraint on the SU total transmit power can be generalized as
\begin{IEEEeqnarray}{c} \label{eq:CCI_constraint}
\sum_{i = 1}^{N} p_i \leq \left[P_{\rm{th}},\frac{1}{\beta_{\rm{ov}}^{(m)}} \frac{\nu^{(m)}}{G^{(m)} \left(-\ln(1 - \Psi_{\rm{th}}^{(m)})\right)} P_{\rm{th}}^{(m)}\right]^-,
\end{IEEEeqnarray}
where $[x,y]^-$ represents $\min(x,y)$. The ACI is mainly due to the power spectral leakage of the SU subcarriers to the PUs receivers. This depends on the power allocated to each SU subcarrier and the spectral distance between the SU subcarriers and the PUs receivers. Similar to the CCI constraint, the statistical ACI constraint can be written as
\begin{IEEEeqnarray}{c} \label{eq:ACI_constraint}
\sum_{i = 1}^{N} p_i \: \varpi_i^{(\ell)} \leq \frac{1}{\beta_{\rm{oo}}^{(\ell)}} \frac{\nu^{(\ell)}}{G^{(\ell)} \left(-\ln(1 - \Psi_{\rm{th}}^{(\ell)})\right)} P_{\rm{th}}^{(\ell)}, \quad \ell = 1, ..., L,
\end{IEEEeqnarray}
where $\frac{1}{\nu^{(\ell)}}$ and $G^{(\ell)}$ are the mean of the exponential distribution and the distance-based path loss to the $\ell$th PU and $\varpi_i^{(\ell)} = T_{{\rm{s}},m} \: \int_{f_{i,\ell}-\frac{B_\ell}{2}}^{f_{i,\ell}+\frac{B_\ell}{2}} \textup{sinc}^2(T_{\rm{s}} f) \: df$, with $T_{{\rm{s}},m}$ as the SU OFDM symbol duration, $f_{i,\ell}$ as the spectral distance between the SU subcarrier $i$ and the $\ell$th PU  frequency band, $B_{\ell}$ as the bandwidth of the $\ell$th PU, and $\textup{sinc}(x) = \frac{\sin(\pi x)}{\pi x}$.

\subsection{Modeling the Imperfect CSI on the Link Between the SU Transmitter and Receiver}
Unlike all the previous works in the literature that assume perfect CSI for the links between the SU transmitter and receiver pairs \cite{mao2013energy, mao2013energy2, oto2012energy, xie2012energy, wangenergy, wang2012optimal}, we consider the effect of the channel estimation errors on these links.
The channel is assumed to change slowly and is modeled as a time-invariant finite impulse response system with order equal to $N_{\rm{ch}}$, ${\bf{h}} = \left[ {h(0),} \, h(1), \, \cdots , \, h(N_{\rm{ch}})  \right]^T$, where each channel tap is assumed to be complex Gaussian distributed with zero-mean and variance $\sigma _{h}^2$. To avoid the intersymbol interference, a cyclic prefix is added at the SU transmitter and removed at the receiver. The noise at the SU receiver is modeled as additive white Gaussian noise (AWGN) with zero mean and correlation matrix equal to $\sigma ^2 _n \bf{I}$, where $\bf{I}$ is the identity matrix. The training pilot symbols $\bf{b}_{\rm{pilot}}$ are added to the precoded block, where the receiver knows the pilot pattern and estimates the channel using the linear minimum mean square error estimator (LMMSE)  as ${\bf{\hat h}} = \left( {\sigma ^2 _n {\bf{R}}_h^{ - 1}  + {\bf{B}}^H {\bf{B}}} \right)^{ - 1} {\bf{B}}^H \bf{x},$
where $\bf{x}$ is the received block and $\bf{B}$ is an $N \times (N_{\rm{ch}}+1)$ column wise circulant matrix with the first column equal to $\bf{x}$ \cite{ohno2004capacity}. The subchannel estimates are computed as \cite{ohno2004capacity} $\left[ \hat H(1), \, \hat H(W), \dots , \hat H(W^{N-1}) \right]^T = \sqrt{N} {\bf{F}}_{N_{\rm{ch}}} \bf{ \hat h},$
where $W=e^{j2\pi/N}$, ${\bf{F}}_{N_{\rm{ch}}}$ is a submatrix of $\bf{F}$ corresponding to the first $N_{\rm{ch}}+1$ columns, and ${\bf{F}}$ is the $N \times N$ discrete Fourier transform matrix with the $(l,n)$ element defined as $\left[ \bf{F} \right]_{l,n} = W^{-ln}/ \sqrt{N}$. The channel capacity is expressed in terms of the channel estimate across subcarriers \cite{ohno2004capacity}, while taking the interference from the PUs into account, as
\begin{equation}  \label{eq3}
c(\mathbf{p}) = \Delta f \sum\limits_{i = 1}^N {\log _2 \left( {1 + \frac{{\left| {\hat H\left( {W^i } \right)} \right|^2 G \, p_i }}{{\sigma _{\Delta H}^2 G \, p_i  + \sigma ^2 _n + \mathcal{J}_i}}} \right)},
\end{equation}
where $\Delta f$ is the subcarrier bandwidth, $\mathbf{p} = [p_1, ..., p_N]^{\rm{T}}$ is the vector representing the power allocated to each subcarrier, $G$ is the distance-based path loss, $\mathcal{J}_i$ is the interference from the PUs to subcarrier $i$ of the SU (it depends on the SU receiver windowing function and power spectral density of the PUs \cite{weiss2004mutual}), and $\sigma _{\Delta H}^2$ is the minimum mean square error (MMSE) of the channel estimate. The latter can be expressed as $\sigma _{\Delta H}^2  = \frac{{\left( {N_{\rm{ch}} + 1} \right)\sigma _h^2  \sigma ^2 _n }}{{\sigma _n^2   +  \sigma _h^2 G P_{\rm{pilots}} }}$,
where $P_{\rm{pilots}}$ is the pilots' transmitted power \cite{ohno2004capacity}. 
\section{Optimization Problem and Proposed Algorithm} \label{sec_Ch_5:opt}

\subsection{Optimization Problem Formulation and Analysis}
Our target is to optimize the SU EE, under channel uncertainties, while guaranteeing a total transmit power budget, limiting the CCI and ACI to the $m$th and $\ell$th PUs receivers below certain thresholds with a predefined probability, and ensuring the SU QoS in terms of a minimum supported rate. In this paper, we minimize the EE defined as the total energy consumed to deliver one bit.
Accordingly, the optimization problem is formulated as
\begin{IEEEeqnarray}{c} \label{eq:MOOP_1}
\mathcal{OP}1: \quad \underset{p_i}{\textup{min}} \quad \eta_{\rm{EE}} =  \frac{\kappa \sum_{i = 1}^{N} p_i + p_{\rm{c}}} {c(\mathbf{p})} \nonumber \\
\textup{subject to} \qquad \textup{C1}: (\ref{eq:CCI_constraint}), \quad \textup{C2}: (\ref{eq:ACI_constraint}), \quad \textup{C3}: c(\mathbf{p}) \geq R_{\rm{th}},
\end{IEEEeqnarray}
where $\kappa$ is a constant that depends on the power amplifier efficiency, $p_{\rm{c}}$ is the circuitry  power consumption, and $R_{\rm{th}}$ is the minimum required SU rate. The objective function in (\ref{eq:MOOP_1}) is non-convex; hence, $\mathcal{OP}1$ is non-convex and the global optimal solution is not guaranteed. The non-convex optimization problem in (\ref{eq:MOOP_1}) can be transformed to an equivalent optimization problem using the concept of fractional programming \cite{dinkelbach1967nonlinear}. Let us define a new objective function as
\begin{IEEEeqnarray}{c}
\Phi(\mathbf{p}, q) = \kappa \sum_{i = 1}^{N} p_i + p_{\rm{c}} - q \: c(\mathbf{p}),
\end{IEEEeqnarray}
where $q$ is a non-negative parameter/constant (and not a variable). We define a new optimization problem $\mathcal{OP}2$ as
\begin{IEEEeqnarray}{c}
\mathcal{OP}2: \quad \underset{p_i}{\textup{min}} \quad  \Phi(\mathbf{p}, q), \qquad \qquad \qquad
\textup{subject to} \qquad \textup{C1---C3}. \label{eq:OP1}
\end{IEEEeqnarray}
One can show that $\mathcal{OP}2$ is quasi-convex (the proof is not provided due to space limitations), and, hence, the global optimality is guaranteed. It was shown in \cite{dinkelbach1967nonlinear} that at a certain value of the parameter $q$, denoted as $q^*$, the optimal solution of $\mathcal{OP}2$ is also the optimal solution to $\mathcal{OP}1$. Hence, finding the optimal power allocation $\mathbf{p}^*$ of $\mathcal{OP}1$ can be realized by finding the optimal power allocation $\mathbf{p}^*(q)$ of $\mathcal{OP}2$; then update the value of $q$ until it reaches $q^*$ \cite{dinkelbach1967nonlinear}.
Following \cite{dinkelbach1967nonlinear}, let us define $\Phi_{\rm{min}}(q) = \underset{p_i}{\min} \{ \Phi(\mathbf{p}, q) | \mathbf{p} \in \mathcal{S}\}$ to be the minimum of $ \Phi(\mathbf{p}, q)$, where $\mathcal{S}$ is the non-empty feasible region of $\mathcal{OP}1$ and $\mathcal{OP}2$ and $q^*$ is the minimum of $\eta_{\rm{EE}}(\mathbf{p})$, i.e., $q^* = \eta_{\rm{EE}}(\mathbf{p}^*) = \frac{\kappa \sum_{i = 1}^{N} p_i^* + p_{\rm{c}}} {c(\mathbf{p}^*)}$ . If $\Phi_{\rm{min}}(q^*) = 0$, then the power that corresponds to $q^* = \eta_{\rm{EE}}(\mathbf{p}^*)$ is the optimal solution of $\mathcal{OP}1$ \cite{dinkelbach1967nonlinear}.
$\mathcal{OP}2$ can be solved by applying the Karush-Kuhn-Tucker (KKT) conditions \cite{Boyd2004convex},
where the Lagrangian function is expressed as
\begin{IEEEeqnarray}{RCL}
\mathcal{L}(\mathbf{p},\mathbf{y},\boldsymbol \lambda) & = & \kappa \sum_{i = 1}^{N} p_i + p_{\rm{c}} - q \: c(\mathbf{p}) \nonumber \\
& & + \lambda_{1} \left[\sum_{i = 1}^{N} p_i - \left[P_{\rm{th}},\frac{1}{\beta_{\rm{ov}}^{(m)}} \frac{\nu^{(m)}}{G^{(m)} \left(-\ln(1 - \Psi_{\rm{th}}^{(m)})\right)} P_{\rm{th}}^{(m)}\right]^- + y_{1}^2\right] \nonumber \\
& & + \sum_{\ell = 1}^{L} \lambda_{2}^{(\ell)} \left[ \sum_{i = 1}^{N} p_i \: \varpi_i^{(\ell)} - \frac{1}{\beta_{\rm{oo}}^{(\ell)}} \frac{\nu^{(\ell)}}{G^{(\ell)} \left(-\ln(1 - \Psi_{\rm{th}}^{(\ell)})\right)} P_{\rm{th}}^{(\ell)} + y_{2}^{(\ell)^2} \right]  \nonumber \\
& & + \lambda_{3} \left[R_{\rm{th}} -  c(\mathbf{p}) + y_{3}^2 \right],
\end{IEEEeqnarray}
where $\boldsymbol \lambda = [\lambda_{1}, \lambda_{2}^{(\ell)}, \lambda_{3}]^{\rm{T}}$ and $\mathbf{y} = [y_1^2, y_2^{{(\ell)}^2}, y_3^2]^{\rm{T}}$, $\ell = 1, ..., L$, are the vectors of the Lagrange multipliers and slack variables, respectively. A stationary point can be found when $\nabla \mathcal{L}(\mathbf{p},\mathbf{y},\boldsymbol \lambda) = 0$, which yields

\begin{subequations}
\begin{IEEEeqnarray}{rcl}
\frac{\partial \mathcal{L}}{\partial p_i} &{}={}&  \frac{- \frac{\Delta f}{\ln(2)} (q + \lambda_{3})  | {\hat H\left( {W^i } \right)} |^2 G\: ( \sigma_n^2 + \mathcal{J}_i)}{\sigma _{\Delta H}^2 G^2 (\sigma _{\Delta H}^2 + | {\hat H\left( {W^i } \right)}|^2) p_i^2 + G \: ( \sigma_n^2 + \mathcal{J}_i) (2 \sigma _{\Delta H}^2 + | {\hat H\left( {W^i } \right)} |^2) p_i + \: ( \sigma_n^2 + \mathcal{J}_i)^2} \nonumber \\ & & + \kappa  + \lambda_{1} + \sum_{\ell = 1}^{L} \lambda_{2}^{(\ell)} \varpi_i^{(\ell)} = 0, \label{eq:L_first}\\
\frac{\partial \mathcal{L}}{\partial \lambda_{1}} &{}={}& \sum_{i = 1}^{N} p_i - \left[P_{\rm{th}},\frac{1}{\beta_{\rm{ov}}^{(m)}}
\frac{\nu^{(m)}}{G^{(m)} (-\ln(1 - \Psi_{\rm{th}}^{(m)}))} P_{\rm{th}}^{(m)}\right]^- + y_{1}^2 = 0, \\
\frac{\partial \mathcal{L}}{\partial \lambda_{2}^{(\ell)}} &{}={}& \sum_{i = 1}^{N} p_i \: \varpi_i^{(\ell)} - \frac{1}{\beta_{\rm{oo}}^{(\ell)}} \frac{\nu^{(\ell)}}{G^{(\ell)} (-\ln(1 - \Psi_{\rm{th}}^{(\ell)}))} P_{\rm{th}}^{(\ell)} + y_{2}^{(\ell)^2} = 0, \\
\frac{\partial \mathcal{L}}{\partial \lambda_{3}} &{}={}& R_{\rm{th}} -  c(\mathbf{p}) + y_{3}^2 = 0,\\
\nonumber
\end{IEEEeqnarray}
\begin{IEEEeqnarray}{rcl}
\frac{\partial \mathcal{L}}{\partial y_{1}} &{}={}& 2 \lambda_{1} \, y_{1} = 0, \label{eq:OP_1_6}\\
\frac{\partial \mathcal{L}}{\partial y_{2}^{(\ell)}} &{}={}& 2 \lambda_{2}^{(\ell)} \, y_{2}^{(\ell)} = 0, \label{eq:OP_1_7}\\
\frac{\partial \mathcal{L}}{\partial y_{3}} &{}={}& 2 \lambda_{3} \, y_{3} = 0,
 \label{eq:L_end}
\end{IEEEeqnarray}
\end{subequations}
It can be seen that (\ref{eq:L_first})--(\ref{eq:L_end}) represent $N + 2 L + 4$ equations in the $N + 2 L + 4$ unknown components of the vectors $\mathbf{p}, \mathbf{y}$, and $\boldsymbol \lambda$. From (\ref{eq:L_first}), the optimal power allocation per subcarrier is given as
\begin{IEEEeqnarray}{c}
p_i^* = \left[\chi_i \left(-1 + \left( 1 - \left(\frac{( \sigma_n^2 + \mathcal{J}_i)}{G} - \frac{\frac{\Delta f}{\ln(2)} (q + \lambda_{3}) | {\hat H\left( {W^i } \right)} |^2}{\kappa + \lambda_{1} + \sum_{\ell = 1}^{L} \lambda_{2}^{(\ell)} \varpi_i^{(\ell)}} \right) \frac{2}{\chi_i \left( 2\sigma _{\Delta H}^2  + | {\hat H\left( {W^i } \right)} |^2  \right)}\right)^{1/2} \right)\right]^+, \nonumber \\ \label{eq:optimal}
\end{IEEEeqnarray}
where $[x]^+$ represents $\max(0,x)$ and the value of $\chi_i$ is calculated as $\chi_i = \frac{{( \sigma_n^2 + \mathcal{J}_i) \left( {2\sigma _{\Delta H}^2  + \left| {\hat H\left( {W^i } \right)} \right|^2 } \right)}}{{2\sigma _{\Delta H}^2 \left( {\sigma_{\Delta H}^2  + \left| {\hat H\left( {W^i } \right)} \right|^2 } \right)G}}.$
In (\ref{eq:optimal}), the values of the Lagrangian multipliers $\lambda_{1}$, $\lambda_{2}^{(\ell)}$, and $\lambda_{3}$ are determined based on whether the constraints on the CCI/total transmit power, ACI, and rate are active or inactive, respectively (a constraint on the form $\Gamma(x) \leq \Gamma_{\rm{th}}$ is said to be inactive if $\Gamma(x) < \Gamma_{\rm{th}}$, while it is active if $\Gamma(x) = \Gamma_{\rm{th}}$).
Equation (\ref{eq:OP_1_6}) implies that either $\lambda_{1} = 0$ or $y_{1} = 0$, (\ref{eq:OP_1_7}) implies that either $\lambda_{2}^{(\ell)} = 0$ or $y_{2}^{(\ell)} = 0$, and (\ref{eq:L_end}) implies that either $\lambda_{3} = 0$ or $y_{3} = 0$. Hence, eight possible cases exist, as follows:

---\textit{Cases 1 \& 2}: setting $\lambda_{1} = 0$, $\lambda_{2}^{(\ell)} = 0$, and $\lambda_{3} = 0$ (case 1)/$y_{3} = 0$ (case 2) results in the optimal solution for inactive CCI/total transmit power constraint, inactive ACI constraints, and inactive/active rate constraint, respectively.


---\textit{Case 3 \& 4}: setting $y_{1} = 0$, $\lambda_{2}^{(\ell)} = 0$, and $\lambda_{3} = 0$ (case 3)/$y_{3} = 0$ (case 4) results in the optimal solution for active CCI/total transmit power constraint, inactive ACI constraint, and inactive/active rate constraint, respectively.


---\textit{Case 5 \& 6}: setting $\lambda_{1} = 0$, $y_{2}^{(\ell)} = 0$, and $\lambda_{3} = 0$ (case 5)/$y_{3} = 0$ (case 6) results in the optimal solution for inactive CCI/total transmit power constraint, active ACI constraint, and inactive/active rate constraint, respectively.


---\textit{Case 7 \& 8}: setting $y_{1} = 0$, $y_{2}^{(\ell)} = 0$, and $\lambda_{3} = 0$ (case 7)/$y_{3} = 0$ (case 8) results in the optimal solution for active CCI/total transmit power constraint, active ACI constraint, and inactive/active rate constraint, respectively.

\subsection{Proposed Algorithm and Complexity Analysis}

The proposed algorithm can be formally stated as follows:
\floatname{algorithm}{}
\begin{algorithm}
\renewcommand{\thealgorithm}{}
\caption{\textbf{Proposed Algorithm}}
\begin{algorithmic}[1]
\small
\State \textbf{INPUT} $P_{\rm{th}}$, $P_{\rm{th}}^{(m)}$, $P_{\rm{th}}^{(\ell)}$, $R_{\rm{th}}$, $\nu^{(m)}$, $\nu^{(\ell)}$, $G^{(m)}$, $G^{(\ell)}$, $\Psi_{\rm{th}}^{(m)}$, $\Psi_{\rm{th}}^{(\ell)}$, $\beta_{\rm{ov}}^{(m)}$, $\beta_{\rm{oo}}^{(\ell)}$, $G$, $\sigma _n^2$, $\hat H\left( {W^i }\right)$, $\sigma _{\Delta H}^2$, $\Delta f$, $N$, $\delta > 0$, $q = q_{\rm{initial}}$ and $\Phi_{\rm{min}} = -\infty$.
\While{$ \Phi_{\rm{min}}(q) < - \delta$}
\State - assume the optimal solution $p_i^*$ belongs to case 1, i.e., $\sum_{i = 1}^{N} p_i^* < \left[P_{\rm{th}},\frac{1}{\beta_{\rm{ov}}^{(m)}} \frac{\nu^{(m)}}{G^{(m)} \left(-\ln(1 - \Psi_{\rm{th}}^{(m)})\right)} P_{\rm{th}}^{(m)}\right]^-$, $\sum_{i = 1}^{N} p_i^* \: \varpi_i^{(\ell)} < \frac{1}{\beta_{\rm{oo}}^{(\ell)}} \frac{\nu^{(\ell)}}{G^{(\ell)} \left(-\ln(1 - \Psi_{\rm{th}}^{(\ell)})\right)} P_{\rm{th}}^{(\ell)}$, and $c(\mathbf{p}) > R_{\rm{th}}$. \label{step:1}
\State - find $p_i^*$ from (\ref{eq:optimal}) when $\lambda_{1} = \lambda_{2}^{(\ell)} = \lambda_{3} = 0$.
\If{in Step \ref{step:1}, the assumption on the CCI/total transmit power constraint is true, the assumption on the ACI constraint is true, and the assumption on the rate constraint is not true}.
\State - the optimal solution belongs to case 2, i.e., find non-negative $\lambda_{3}$ from (\ref{eq:optimal}) such that $c(\mathbf{p}) = R_{\rm{th}}$.
\State - if the assumption on the CCI/total transmit power and ACI constraints are violated, then $p_i^* = 0$.
\ElsIf{in Step \ref{step:1}, the assumption on the CCI/total transmit power constraint is not true, the assumption on the ACI constraint is true, and the assumption on the rate constraint is true}
\State - the optimal solution belongs to case 3, i.e., find non-negative $\lambda_{1}$ from (\ref{eq:optimal}) such that $\sum_{i = 1}^{N} p_i^* = \left[P_{\rm{th}},\frac{1}{\beta_{\rm{ov}}^{(m)}} \frac{\nu^{(m)}}{G^{(m)} \left(-\ln(1 - \Psi_{\rm{th}}^{(m)})\right)} P_{\rm{th}}^{(m)}\right]^-$.
\State - if the assumption on the rate constraint is violated, then $p_i^* = 0$.
\ElsIf{in Step \ref{step:1}, the assumption on the CCI/total transmit power constraint is not true, the assumption on the ACI constraint is true, and the assumption on the rate constraint is not true}
        \algstore{myalg}
  \end{algorithmic}
\end{algorithm}

\floatname{algorithm}{}
\begin{algorithm}
 \renewcommand{\thealgorithm}{}
  \caption{\textbf{Proposed Algorithm} (continued)}
  \begin{algorithmic}
      \algrestore{myalg}
      \small
\State - the optimal solution belongs to case 4, i.e., find non-negative $\lambda_{1}$ and $\lambda_{3}$ from (\ref{eq:optimal}) such that $\sum_{i = 1}^{N} p_i^* = \left[P_{\rm{th}},\frac{1}{\beta_{\rm{ov}}^{(m)}} \frac{\nu^{(m)}}{G^{(m)} \left(-\ln(1 - \Psi_{\rm{th}}^{(m)})\right)} P_{\rm{th}}^{(m)}\right]^-$ and $c(\mathbf{p}) = R_{\rm{th}}$.
\State - if the assumption on the ACI constraint is violated, then $p_i^* = 0$.
\ElsIf{in Step \ref{step:1}, the assumption on the CCI/total transmit power constraint is true, the assumption on the ACI constraint is not true, and the assumption on the rate constraint is true}
\State - the optimal solution belongs to case 5, i.e., find non-negative $\lambda_{2}^{(\ell)}$ from (\ref{eq:optimal}) such that $\sum_{i = 1}^{N} p_i^* \: \varpi_i^{(\ell)} = \frac{1}{\beta_{\rm{oo}}^{(\ell)}} \frac{\nu^{(\ell)}}{G^{(\ell)} \left(-\ln(1 - \Psi_{\rm{th}}^{(\ell)})\right)} P_{\rm{th}}^{(\ell)}$.
\State - if the assumption on the rate constraint is violated, then $p_i^* = 0$.
\ElsIf{in Step \ref{step:1}, the assumption on the CCI/total transmit power constraint is true, the assumption on the ACI constraint is not true, and the assumption on the rate constraint is not true}
\State - the optimal solution belongs to case 6, i.e., find non-negative $\lambda_{2}^{(\ell)}$ and $\lambda_{3}$ from (\ref{eq:optimal}) such that $\sum_{i = 1}^{N} p_i^* \: \varpi_i^{(\ell)} = \frac{1}{\beta_{\rm{oo}}^{(\ell)}} \frac{\nu^{(\ell)}}{G^{(\ell)} \left(-\ln(1 - \Psi_{\rm{th}}^{(\ell)})\right)} P_{\rm{th}}^{(\ell)}$ and $c(\mathbf{p}) = R_{\rm{th}}$.
\State - if the assumption on the CCI/total transmit power constraint is violated, then $p_i^* = 0$.
\ElsIf{in Step \ref{step:1}, the assumption on the CCI/total transmit power constraint is not true, the assumption on the ACI constraint is not true, and the assumption on the rate constraint is true}
\State - the optimal solution belongs to case 7, i.e., find non-negative $\lambda_{1}$ and $\lambda_{2}^{(\ell)}$  from (\ref{eq:optimal}) such that  $\sum_{i = 1}^{N} p_i^* = \left[P_{\rm{th}},\frac{1}{\beta_{\rm{ov}}^{(m)}} \frac{\nu^{(m)}}{G^{(m)} \left(-\ln(1 - \Psi_{\rm{th}}^{(m)})\right)} P_{\rm{th}}^{(m)}\right]^-$ and $\sum_{i = 1}^{N} p_i^* \: \varpi_i^{(\ell)} = \frac{1}{\beta_{\rm{oo}}^{(\ell)}} \frac{\nu^{(\ell)}}{G^{(\ell)} \left(-\ln(1 - \Psi_{\rm{th}}^{(\ell)})\right)} P_{\rm{th}}^{(\ell)}$.
\State - if the assumption on the rate constraint is violated, then $p_i^* = 0$.
\ElsIf{in Step \ref{step:1}, the assumption on the CCI/total transmit power constraint is not true, the assumption on the ACI constraint is not true, and the assumption on the rate constraint is not true}
\State - the optimal solution belongs to case 8, i.e., find non-negative $\lambda_{1}$, $\lambda_{2}^{(\ell)}$, and $\lambda_{3}$  from (\ref{eq:optimal}) such that  $\sum_{i = 1}^{N} p_i^* = \left[P_{\rm{th}},\frac{1}{\beta_{\rm{ov}}^{(m)}} \frac{\nu^{(m)}}{G^{(m)} \left(-\ln(1 - \Psi_{\rm{th}}^{(m)})\right)} P_{\rm{th}}^{(m)}\right]^-$, $\sum_{i = 1}^{N} p_i^* \: \varpi_i^{(\ell)} = \frac{1}{\beta_{\rm{oo}}^{(\ell)}} \frac{\nu^{(\ell)}}{G^{(\ell)} \left(-\ln(1 - \Psi_{\rm{th}}^{(\ell)})\right)} P_{\rm{th}}^{(\ell)}$, and $c(\mathbf{p}) = R_{\rm{th}}$.
\Else
\State - $p_i^* = 0$.
\EndIf
\State - update $\Phi_{\rm{min}}(q) = \underset{p_i}{\min} \{ \Phi(\mathbf{p}, q) \} | \mathbf{p} \in \mathcal{S}\}$
\State - Calculate $q = \frac{\kappa \sum_{i = 1}^{N} p_i^* + p_{\rm{c}}} {c(\mathbf{p})}$.
\EndWhile
\State \textbf{OUTPUT} $q^* = q$ and $p_i^*$, $i$ = 1, ..., $N$.
\end{algorithmic}
\end{algorithm}


\vspace{30pt}
Efficient algorithms are presented in \cite{palomar2005practical} to find the Lagrange multipliers $\lambda_{1}$ and $\lambda_{2}^{(\ell)}$, and $\lambda_{3}$ that satisfy the CCI/total transmit power, ACI, and rate constraints, respectively, with complexity order of $\mathcal{O}(N)$. Accordingly, the complexity order of the proposed algorithm can be $\mathcal{O}(N_qN^2)$, where $N_q$ is the number of executions of the while loop. The average (over the number of channel realizations) value for $N_q$ is 4 for $\delta = 10^{-8}$ and 4.46 for $\delta = 10^{-14}$; both values are significantly lower than the number of subcarriers $N$. Hence, the complexity of the proposed algorithm is of the order  $\mathcal{O}(N^2)$.

\section{Numerical Results} \label{sec_Ch_5:sim}

Without loss of generality, we assume that the OFDM SU coexists with one frequency-adjacent PU and one co-channel PU. The SU parameters are chosen as follows: number of subcarriers $N = 128$ and subcarrier spacing $\Delta f = \frac{1.25 \: \rm{MHz}}{N} = 9.7656$ kHz. The propagation path loss parameters are as follows: distance between SU transmitter and receiver pair $= 1$ km, distance to the $\ell$th PU $d_{\ell} = 1.2$ km, distance to the $m$th PU $d_m = 1.5$ km, reference distance $= 100$ m, exponent $= 4$, and wavelength $= \frac{3 \times 10^8}{900 \times 10^6} = 0.33$ meters. A Rayleigh fading environment is considered with $N_{\rm{ch}} = 5$, where the average channel power gains between the SU transmitter and the receiver of the $\ell$th PU $\mathbb{E}\{|\mathcal{H}_{sp}^{(\ell)}|^2\}$ and between the SU transmitter and the receiver of the $m$th PU $\mathbb{E}\{|\mathcal{H}_{sp}^{(m)}|^2\}$ are set to 0 dB. $\sigma_n^2$ is assumed to be $4 \times 10^{-16}$ W, the PUs signal is assumed to be an elliptically filtered white noise process \cite{weiss2004mutual} of variance $4 \times 10^{-16}$ W, $p_c = P_{\rm{th}} = 2$ W, $\kappa = 7.8$, $\delta = 10^{-8}$, $\Psi_{\rm{th}}^{(m)} = \Psi_{\rm{th}}^{(\ell)} = 0.9$, and $P_{\rm{th}}^{(m)} = P_{\rm{th}}^{(\ell)} = 10^{-13}$ W. Representative results are presented in this section, which were obtained through Monte Carlo trials for $10^{4}$ channel realizations. Unless otherwise mentioned, imperfect spectrum sensing is assumed. Following \cite{almalfouh2011interference} and in order to favor the PUs protection,  $\rho_{\rm{md}}^{(m)}$ is uniformly distributed over the interval [0.01,  0.05], and it is lower than $\rho_{\rm{fa}}^{(m)}$, which is uniformly distributed over the interval [0.01,  0.1].
$\rho^{(m)}$ and $\rho^{(\ell)}$ are uniformly distributed between [0, 1] and the EE, measured in J/bits, is the total energy consumption to deliver one bit.

\begin{figure}[!t]
\centering
\includegraphics[width=0.75\textwidth]{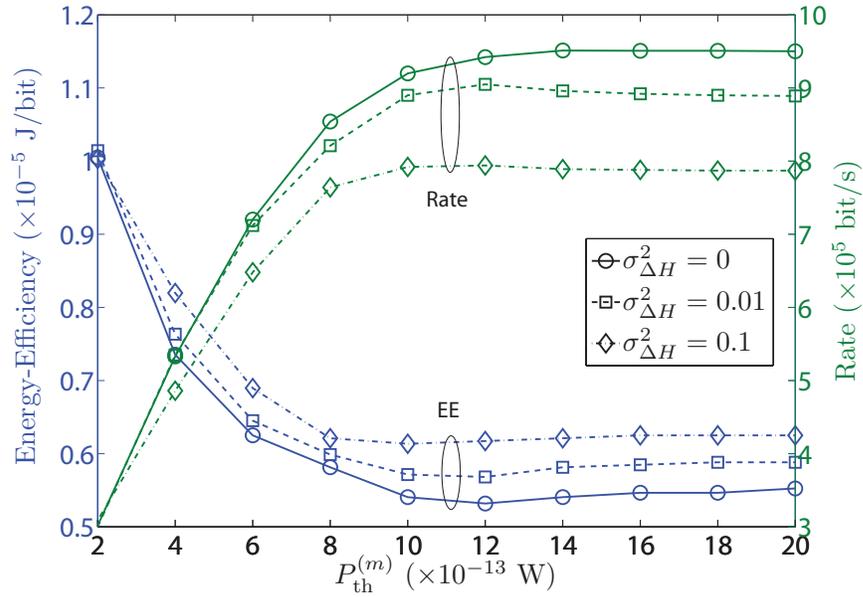}
\caption{Effect of $\sigma_{\Delta H}^2$ on the SU performance.}
\label{fig:MMSE}
\end{figure}

In Fig. \ref{fig:MMSE}, the EE (in J/bits) and the transmission rate (in bits/sec) of the SU  are depicted as a function of $P_{\rm{th}}^{(m)}$, for $R_{\rm{th}} = 0$ and different values of $\sigma_{\Delta H}^2$. As can be seen, the EE decreases and the rate increases as $P_{\rm{th}}^{(m)}$ increases, and both saturate for higher values of $P_{\rm{th}}^{(m)}$. This is as for lower values of $P_{\rm{th}}^{(m)}$ the total transmit power is limited, and increasing $P_{\rm{th}}^{(m)}$ increases the transmit power, and, hence, enables the proposed algorithm to improve both the EE and rate of the SU. The EE keeps improving until the optimal power budget is reached, after which a further increase in $P_{\rm{th}}^{(m)}$  does not improve the EE, and, hence, the rate is kept constant. As the value of $\sigma_{\Delta H}^2$ increases, i.e., the estimation error increases, both the EE and the rate  deteriorate accordingly.

\begin{figure}[!t]
\centering
\includegraphics[width=0.75\textwidth]{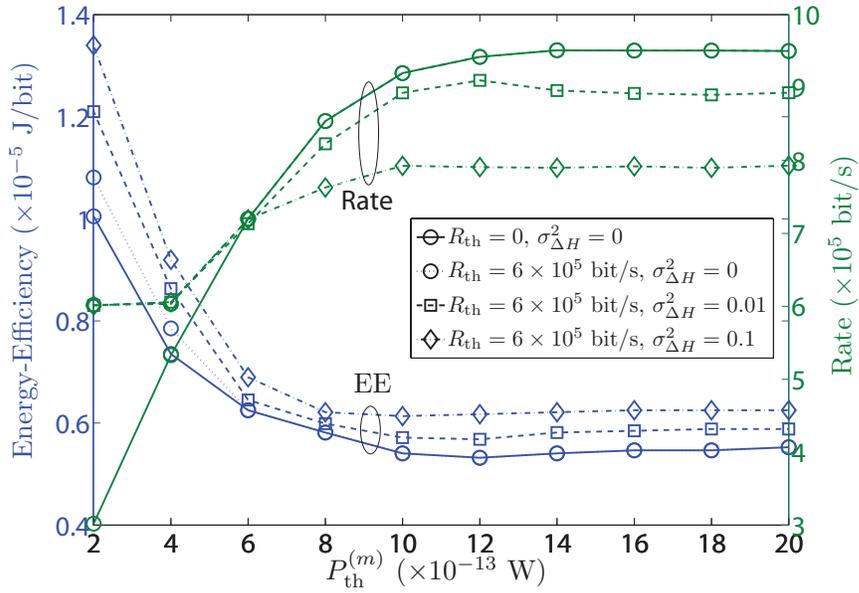}
\caption{Effect of $R_{\rm{th}}$ and $\sigma_{\Delta H}^2$ on the SU performance.}
\label{fig:PCCI}
\end{figure}

Fig. \ref{fig:PCCI} depicts the SU EE and rate as a function of $P_{\rm{th}}^{(m)}$, for different values for $R_{\rm{th}}$ and $\sigma_{\Delta H}^2$. As expected, for $\sigma_{\Delta H}^2 = 0$, increasing $R_{\rm{th}}$ from 0 to $6 \times 10^5$ bits/sec guarantees the SU rate at low values of $P_{\rm{th}}^{(m)}$ (i.e., when the rate drops below $6 \times 10^5$ bits/sec); however, this comes at the expense of increasing the EE. On the other hand, for $R_{\rm{th}} = 6 \times 10^5$ bits/sec, increasing the estimation error deteriorates both the rate and the EE of the SU at high values of $P_{\rm{th}}^{(m)}$; for low values of the $P_{\rm{th}}^{(m)}$, the SU maintains its required rate but this is at the expense of increasing the EE.

\begin{figure}[!t]
\centering
\includegraphics[width=0.75\textwidth]{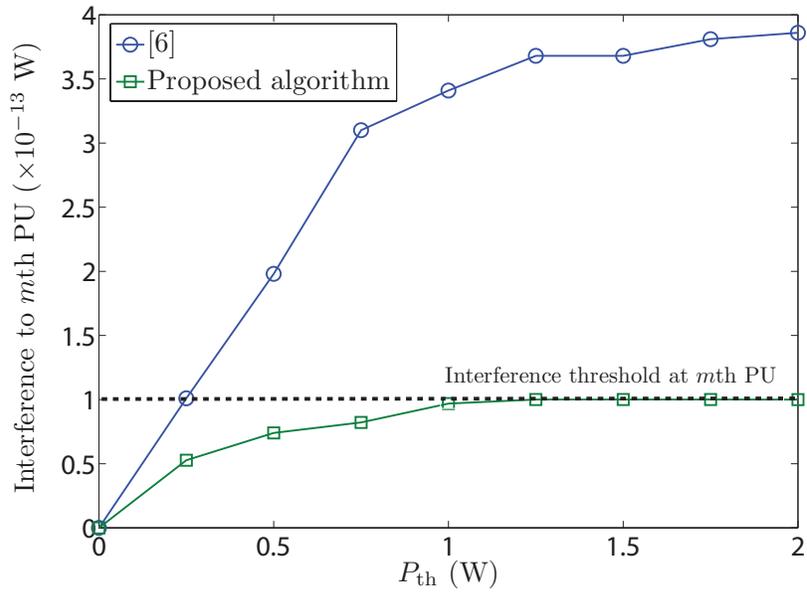}
\caption{Comparison with the work in \cite{wang2012optimal} to show the effect of perfect and imperfect sensing assumptions on the interference leaked to the $m$th PU.}
\label{fig:comp_interference}
\end{figure}

\begin{figure}[!t]
\centering
\includegraphics[width=0.75\textwidth]{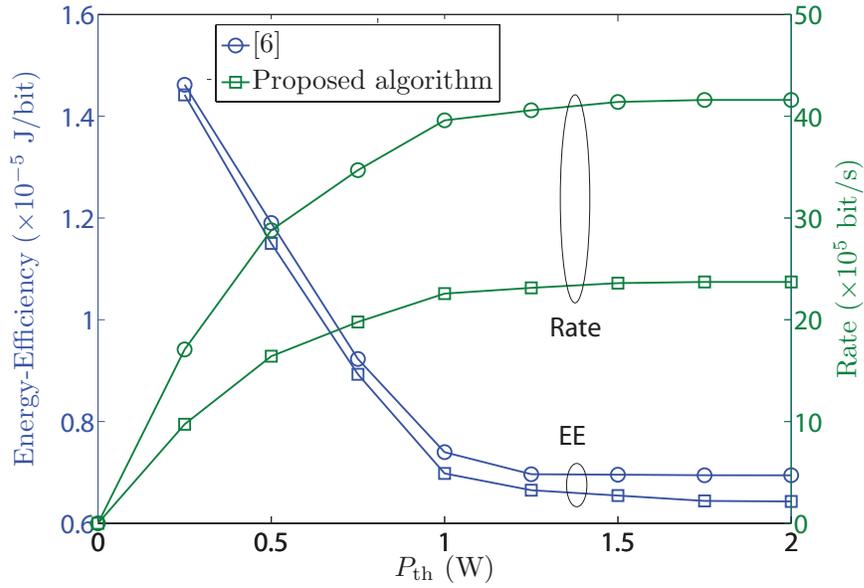}
\caption{Comparison with the work in \cite{wang2012optimal} to show the effect of perfect and imperfect sensing assumptions on the EE and the rate of SU.}
\label{fig:comp_EE_SE}
\end{figure}

In order to show the effect of assuming perfect spectrum sensing, Figs. \ref{fig:comp_interference} and \ref{fig:comp_EE_SE} compare the interference introduced into the $m$th PU band, and the EE and rate, respectively, for the proposed algorithm and the work in \cite{wang2012optimal} that assumes perfect sensing capabilities for the SU. We set $\sigma_{\Delta H}^2 = 0$ and $R_{\rm{th}} = 0$ in the proposed algorithm, in order to match the conditions in \cite{wang2012optimal}.
As can be seen in Fig. \ref{fig:comp_interference}, if the sensing errors are not taken into consideration when optimizing the EE as in \cite{wang2012optimal} (i.e., the SU is assumed to sense the PUs bands perfectly, which is not true in practice),
then the interference leaked in the $m$th PU band exceeds the threshold (note that this is due to the increase of the transmit power for the case of perfect spectrum sensing assumptions). On the other hand, if the sensing errors are considered in the optimization problem (i.e., the SU is assumed to sense the PUs bands with a certain probability of error), then the interference to the $m$th PU band is below the threshold.
In Fig. \ref{fig:comp_EE_SE} and as expected, the SU rate is higher if perfect spectrum sensing is assumed because the transmit power is higher. Additionally, the EE (in J/bits) is higher when compared to its counterpart that considers spectrum sensing errors due to increasing the transmit power as discussed in Fig.  \ref{fig:comp_interference}.

\section{Conclusions} \label{sec_Ch_5:conc}
In this paper, we proposed an optimal power loading algorithm that optimizes the EE of an OFDM-based CR system under different channel uncertainties. The algorithm considers the channel estimation errors for the links between the SU transmitter and receiver pairs and also the effect of the imperfect sensing capabilities of the SU. Further, the algorithm does not require perfect CSI for the links from the PUs receivers to the SU transmitter. Simulation results showed that increasing the channel estimation errors deteriorates the EE. Further, they showed that assuming that the SU has perfect sensing capabilities deteriorates the EE and violates the interference constraints at the PUs receivers.
Additionally, the results demonstrated that the proposed algorithm guarantees a minimum QoS for the SU at the expense of deteriorating the EE. 
\bibliographystyle{IEEEtran}
\bibliography{IEEEabrv,mybib_file} 

\chapter{Conclusions and Future Work} \label{ch:conc}
In this final chapter, we summarize the contributions presented in this dissertation and
discuss several potential extensions to our work.

\section{Conclusions}
The following conclusions can be drawn from this dissertation:
\begin{itemize}
  \item We illustrated that the MOOP approach is a strong candidate for the optimal link adaptation problem when compared to single objective optimization approaches. The adopted MOOP approach showed significant performance improvements in terms of the achieved throughput/rate and transmit power, when compared with other works in the literature that separately maximized the throughput/rate (while constraining the transmit power) or minimized the transmit power (while constraining the throughput), at the cost of no additional complexity. Moreover, the MOOP showed better performance in terms of the energy efficiency.
  \item The MOOP approach allowed the MCM system to tune for various levels of throughput/rate and transmit power, without resolving different single objective optimization problems. This was achieved through changing the weighting coefficient associated with each objective.
  \item The improved performance of the MOOP approach did not come at the cost of additional computational complexity. For the formulated MOOP problems, we proposed low complexity algorithms that do not necessarily require perfect CSI on the links between the SU transmitter and receiver pair and/or on the links between the SU transmitter and the PUs receivers. We additionally quantified the performance loss due to partial CSI on the links.
  \item We showed that the interference constraints at the PUs receivers can be severely violated if the SU is assumed to have perfect spectrum sensing capabilities.
  \item We showed that adding a fading margin is crucial to compensate for the violation of the interference constraints at the PUs receivers due to imperfect CSI on the links between the SU transmitter and the PUs receivers.
\end{itemize}

\section{Future Work}
There are various directions to extend our work, which can be briefly outlined as follows:
\begin{itemize}
	\item As discussed earlier, the MOOP approach showed superior performance over traditional single objective optimization approaches. It would be worth researching to extend the problem formulation to the scenario of adaptive modulation and coding, i.e., optimally allocate a specific code and its code rate in addition to the modulation type/order and the power per subcarrier.
	\item Extending the formulated MOOP problems to include multiple SUs is interesting and important. This can be done in both the downlink and uplink scenarios. In both cases, the SUs will employ orthogonal frequency division multiplying access (OFDMA) in order to efficiently access the spectrum. Accordingly, the optimization problem will optimally allocate the subcarriers in addition to bits and power to SUs in order to improve the SUs network performance.
	\item The MOOP approach can be extended and applied in energy harvesting networks. The energy harvesting receivers  should divide the received signal in order to harvest energy and decode the information as well. The MOOP approach seems to be a strong candidate for the energy harvesting receivers to balance between the decoded information rate and the amount of harvested energy.
\end{itemize}

\bibliographystyle{IEEEtran}
\bibliography{IEEEabrv,mybib_file} 


\end{document}